\RequirePackage{fix-cm}
\documentclass[twocolumn]{svjour3}          % twocolumn
\smartqed  % flush right qed marks, e.g. at end of proof
\usepackage{graphicx}
 \usepackage{varioref}%  smart page, figure, table, and equation referencing
 \usepackage{graphics}%  use the graphics package for simple commands
 \usepackage{graphicx}% or use the graphicx package for more complicated commands
 \usepackage{epsfig}% or use the epsfig package if you prefer to use the old commands
 \usepackage{wrapfig}%   wrap figures/tables in text (i.e., Di Vinci style)
 \usepackage{threeparttable}% tables with footnotes
 \usepackage{bigdelim} %% essencial %% para uso de tabelas
 \usepackage{multirow} %% essencial %% para uso de tabelas
 \usepackage{dcolumn}%   decimal-aligned tabular math columns
 \usepackage{nomencl}%   nomenclature generation via makeindex
 \usepackage{subfigure}% subcaptions for subfigures
 \usepackage{subfigmat}% matrices of similar subfigures, aka small mulitples
 \usepackage{fancyvrb}%  extended verbatim environments
 \usepackage{fancyhdr}
 \usepackage{lettrine}%  dropped capital letter at beginning of paragraph
 \usepackage{amssymb}% The amssymb package provides various useful mathematical symbols
 \usepackage[colorlinks]{hyperref}%  hyperlinks [must be loaded after dropping]
 \journalname{Brazilian Journal of Physics}
\begin{document}

\title{Aerothermodynamic Analysis of a Reentry Brazilian Satellite}

\titlerunning{Reentry Brazilian Satellite}        % if too long for running head

\author{Wilson F. N. Santos}

\authorrunning{Santos, W.F.N.} % if too long for running head

\institute{Combustion and Propulsion Laboratory \\
           National Institute for Space Research \at
           Cachoeira Paulista-SP, 12630-000 BRAZIL \\
              Tel.: +55-12-31869265\\
              Fax: +55-12-31011992\\
              \email{wilson@lcp.inpe.br}}

\date{Received: date / Accepted: date}

\maketitle

\begin{abstract}
This work deals with a computational investigation on the small ballistic reentry Brazilian vehicle SARA (acronyms for SAt\'{e}lite de Reentrada Atmosf\'{e}rica). Hypersonic flows over the vehicle SARA at zero-degree angle of attack in a chemical equilibrium and thermal non-equilibrium are modeled by the Direct Simulation Monte Carlo (DSMC) method, which has become the main technique for studying complex multidimensional rarefied flows, and that properly accounts for the non-equilibrium aspects of the flows. The emphasis of this paper is to examine the behavior of the primary properties during the high altitude portion of SARA reentry. In this way, velocity, density, pressure and temperature field are investigated for altitudes of 100, 95, 90, 85 and 80 km. In addition, comparisons based on geometry are made between axisymmetric and planar two-dimensional configurations. Some significant differences between these configurations were noted on the flowfield structure in the reentry trajectory. The analysis showed that the flow disturbances have different influence on velocity, density, pressure and temperature along the stagnation streamline ahead of the capsule nose. It was found that the stagnation region is a thermally stressed zone. It was also found that the stagnation region is a zone of strong compression, high wall pressure. Wall pressure distributions are compared with those of available experimental data and good agreement is found along the spherical nose for the altitude range investigated.
\end{abstract}

\section{Introduction}
\label{sec:1}
The microgravity field has become nowadays one of the most promising new areas for the commercialization of space. The development of scientific and technological experiments carried out under conditions of reduced gravity has been proposed by the Brazilian Space Agency by employing a recoverable orbital system. The orbital system, built in a platform with a capsule shape, is designed in order to stay in orbit during the time needed for the execution of the experiments. After that, the capsule is sent back to the Earth and recovered.

The aerothermodynamic aspects during the ballistic reentry flight, in the development of a capsule platform, offer exciting challenges to the aerodynamicists. In the Earth atmosphere reentry, the capsule undergoes not only different velocity regimes -- hypersonic, supersonic and subsonic -- but also different flow regimes -- free molecular flow, transition and continuum -- and flight conditions that may difficult their aerodynamic design. Therefore, the capsule aerodynamic design, of great importance for the flight through several atmosphere layers, has to consider important aspects such as stability, drag and heating loads. In this context, a combination of engineering tools, experimental analysis, and numerical methods is used in the design of high altitude reentry capsule aerodynamics. Nevertheless, due to difficulties and high costs associated to the experimental work at high speed flows, a numerical analysis becomes imperative.

Many experimental and numerical studies~\cite{Carlson,Gilmore,Gnoffo,Gupta,Ivanov,Longo,Moss,Savino,Vash,Weiland,Wilmoth,Wood} have been dedicated to the aerothermodynamic of vehicles reentering the Earth atmosphere. Nevertheless, for the particular case of SARA capsule, only a few studies are available in the current literature~\cite{Sharipov,Pimentel,Tchuen,Toro,Toro2,Kozak}. For the purpose of this introduction, it will be sufficient to describe only some of these studies.

Sharipov~\cite{Sharipov} has investigated the flowfield structure over the SARA capsule by employing the DSMC method. It was assumed the monatomic gas argon as the working fluid and freestream Mach numbers of 5, 10, and 20. Even considering that the real gas effect in a reentry capsule is not well represented by a monotonic gas, that investigation might be considered as the first contribution to the aerothermodynamic analysis of the SARA capsule in high altitudes.

Pimentel et al.~\cite{Pimentel} have performed inviscid hypersonic flow simulations over the SARA capsule modeled by employing the planar two-dimensional (2-D) and the axisymmetric Euler equations. Results were presented for an altitude of 80 km, Mach numbers of 15 and 18, and angle of attack of 0 and 10 degrees. They also considered air as working fluid composed of five species (N$_{2}$, O$_{2}$, O, N, and NO) with their reactions of dissociation and recombination. Pressure and temperature contour maps were presented for 2-D and axisymmetric flow.

Finally, by using axisymmetric Navier-Stokes equations, Tchuen et al.~\cite{Tchuen} have investigated the flowfield structure over the SARA capsule by considering hypersonic flow at zero angle of attack in chemical and thermal non-equilibrium. It was assumed air as working fluid composed of seven species (N$_{2}$, O$_{2}$, O, N, NO, NO$^{+}$, and $e^{-}$) associated with their reactions of dissociation and recombination. Results for pressure, skin friction, and heat transfer coefficients were presented for different combinations of Mach numbers of 10, 20 and 25 with altitudes of 75 and 80 km.

The interest in the majority of these SARA studies has basically gone into considering the flow in the continuum flow regime. Conversely, in the transition flow regime, there is a little understanding of the physical aspects of a hypersonic flow past to the SARA capsule related to the severe aerothermodynamic environment in the reentry trajectory. In this fashion, the purpose of the present account is to investigate the flowfield structure of a hypersonic flow over the SARA capsule in the transition flow regime, i.e., between the free collision flow and the continuum flow regime. The primary goal is to assess the sensitivity of the primitive properties, such as, velocity, density, pressure, and temperature due to changes on the altitude representative of a typically reentry trajectory of the SARA capsule. Therefore, the present study focuses on the low-density region in the upper atmosphere, where the non-equilibrium conditions are such that the traditional Computational Fluid Dynamics (CFD) calculations are inappropriate to yield accurate results. In such a circumstance, the Direct Simulation Monte Carlo (DSMC) method will be employed to calculate the planar two-dimensional and the axisymmetric hypersonic flow over the SARA capsule.

It is important to mention that, for the particular case of SARA, this is the first work to present the flowfield structure in the transition flow regime by considering two species (N$_{2}$ and O$_{2}$) as working fluid, and rotation and vibration internal modes of energy.

\begin{figure}[b!]
  \begin{center}
   \includegraphics[width=7.0cm,height=6.0cm]{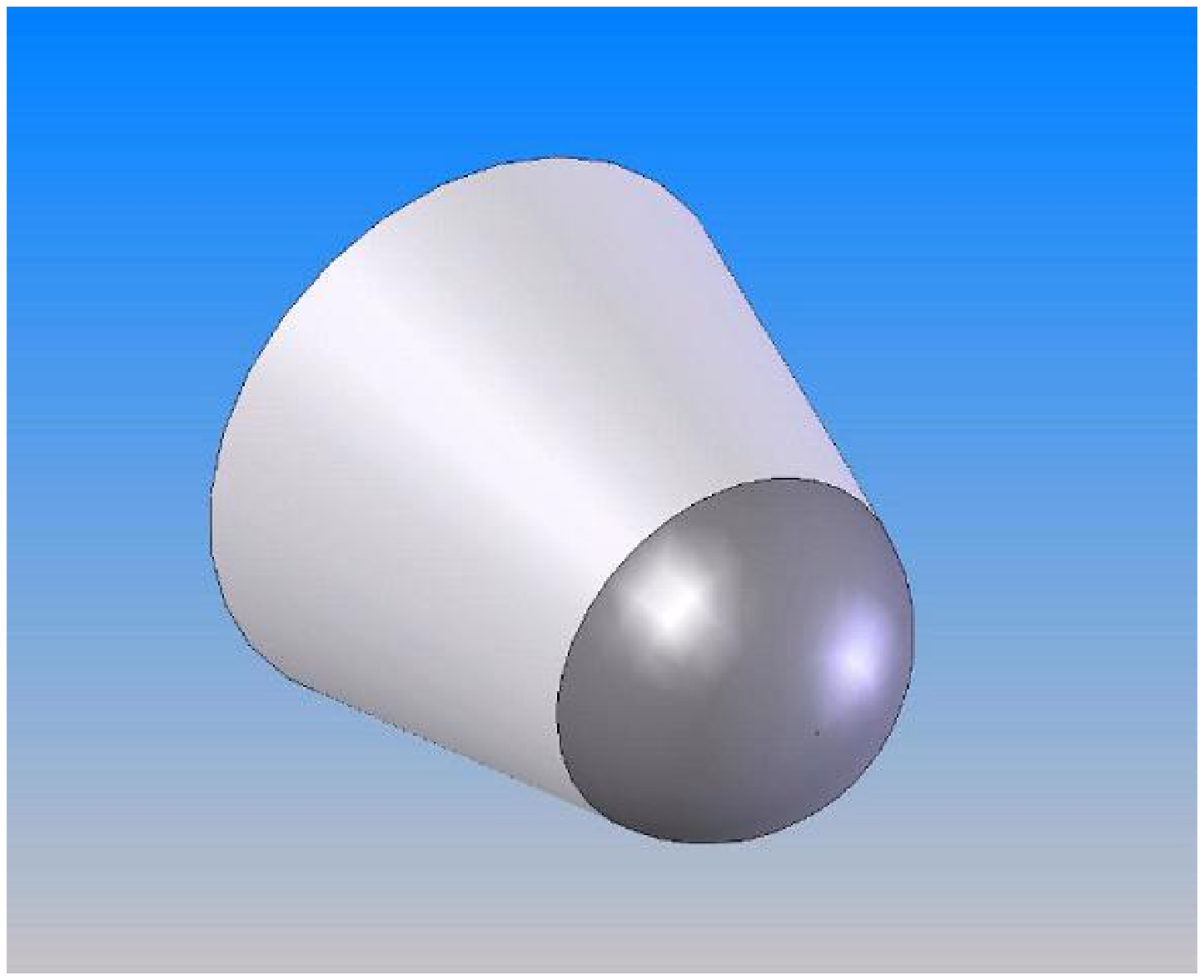}
   \includegraphics[width=7.0cm,height=6.0cm]{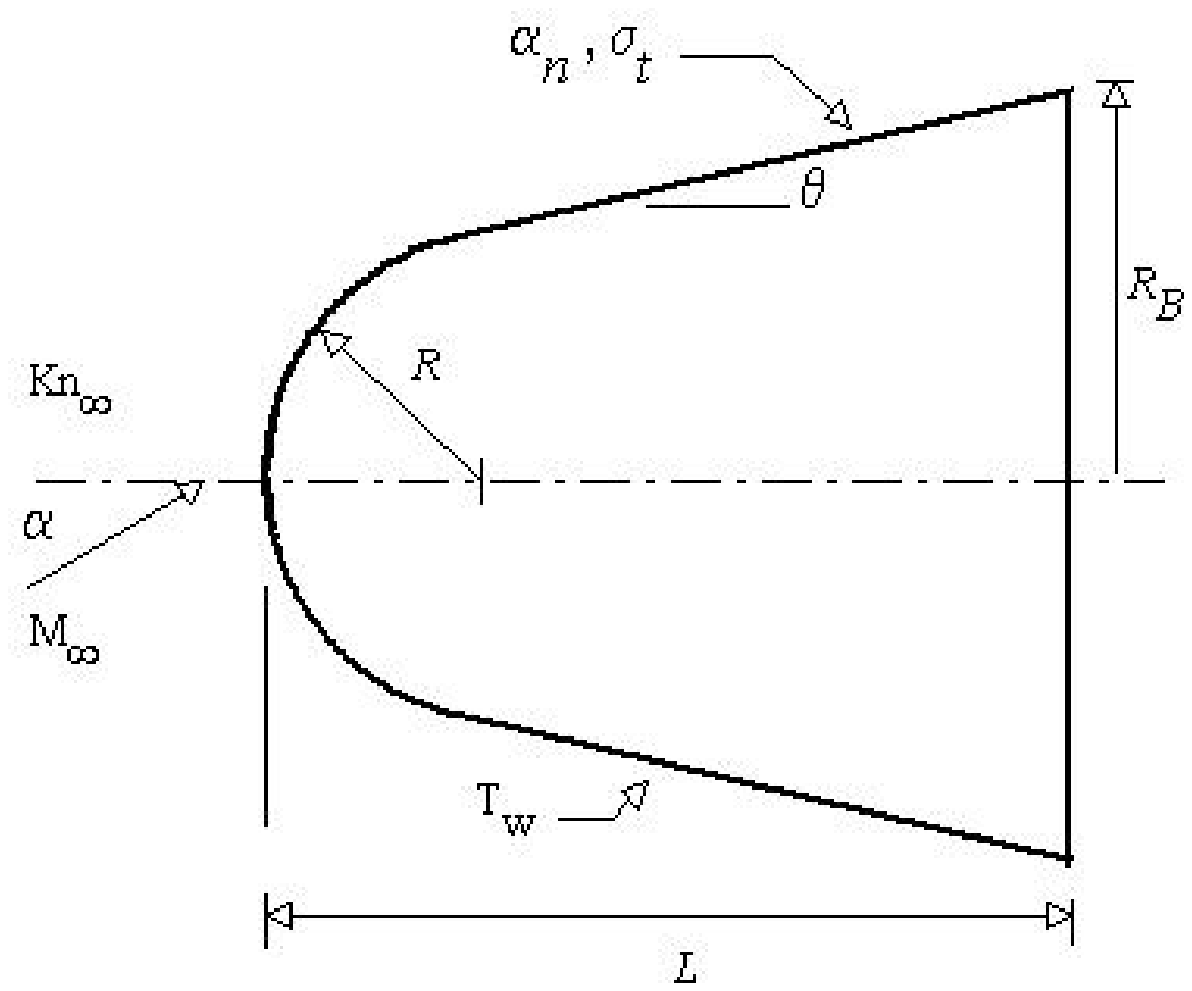}
   \caption{Drawing illustrating a schematic view of the capsule (top), and the important parameters (bottom).}
  \label{BJoPP02F01}
  \end{center}
\end{figure}

\section{Geometry Definition}
\label{sec:2}
The SARA reentry capsule is an axisymmetric  design consisting of a spherical nose with a 11.4-degree half-angle conical afterbody. The nose radius $R$ is 0.2678 m, the afterbody base has a radius $R_B$ of 0.5035 m, and the total length $L$ is 1.410 m. Figure~\ref{BJoPP02F01} illustrates schematically the capsule shape and the main important physical and geometric parameters related to the hypersonic flow on the capsule. The main physical parameters are defined as follows: $M_\infty$ is the freestream Mach number, $Kn_\infty$ stands for the Knudsen number, $\alpha$ is the angle of attack, $T_w$ is the wall temperature, and finally, $\alpha_n$ and $\sigma_t$ represent the parameters related to the gas-surface interaction. In this fashion, the flowfield structure around the capsule may be affected by the effects of compressibility, rarefaction, gas-surface interaction, etc.

\section{Freestream and Flow Conditions}
\label{sec:3}
Flow conditions represent those experienced by the capsule in the reentry trajectory from 100 to 80 km of altitude. This range of altitude is associated with the transition flow regime, which is characterized by the overall Knudsen number the order of or larger than $10^{-2}$.

Freestream flow conditions used for the numerical simulation of flow past the capsule are those given by Hirschel~\cite{Hirschel} and summarized in Tables~\ref{tab1} and~\ref{tab2}, and the gas properties~\cite{Bird94} are shown in Table~\ref{tab3}. Referring to this set of tables, $T_{\infty}$, $p_{\infty}$, $\rho_{\infty}$, $n_{\infty}$, $\lambda_{\infty}$, and $V_{\infty}$ stand respectively for temperature, pressure, density, number density, molecular mean free path, and velocity, and $\chi$, $m$, $d$ and $\omega$ account respectively for mass fraction, molecular mass, molecular diameter and viscosity index.

\begin{table}[b!]
\caption{Freestream flow conditions~\cite{Hirschel}.}
\begin{center}
\renewcommand{\arraystretch}{1.}
\setlength\tabcolsep{4.0pt}
\begin{tabular}{cccc}
\hline\hline\noalign{\smallskip} Altitude (km) & $T_{\infty}$(K) &
$p_{\infty}$(N/m$^{2}$) & $\rho_{\infty}$(kg/m$^{3}$)\\
\noalign{\smallskip} \hline \noalign{\smallskip}
80  & 180.7 & $1.03659$ & $1.999\times10^{-5}$\\
85  & 180.7 & $0.41249$ & $7.956\times10^{-6}$\\
90  & 180.7 & $0.16438$ & $3.171\times10^{-6}$\\
95  & 195.5 & $0.06801$ & $1.212\times10^{-6}$\\
100 & 210.0 & $0.03007$ & $4.989\times10^{-7}$\\
\hline\hline
\end{tabular}
\label{tab1}
\end{center}
\end{table}

\begin{table}[b!]
\caption{Freestream flow conditions~\cite{Hirschel} (cont`d).}
\begin{center}
\renewcommand{\arraystretch}{1.}
\setlength\tabcolsep{4.0pt}
\begin{tabular}{ccccccc}
\hline\hline\noalign{\smallskip} Altitude (km) & $n_{\infty}$(m$^{-3}$) &$\lambda_{\infty}$(m) &$V_{\infty}$(m/s)\\
\noalign{\smallskip} \hline \noalign{\smallskip}
80  & $4.1562\times10^{21}$ & $3.085\times10^{-3}$ & 7820\\
85  & $1.6539\times10^{20}$ & $7.751\times10^{-3}$ & 7864\\
90  & $6.5908\times10^{19}$ & $1.945\times10^{-2}$ & 7864\\
95  & $2.5197\times10^{19}$ & $5.088\times10^{-2}$ & 7866\\
100 & $1.0372\times10^{19}$ & $1.236\times10^{-1}$ & 7862\\
\hline\hline
\end{tabular}
\label{tab2}
\end{center}
\end{table}

\begin{table}[t!]
\caption{Gas properties~\cite{Bird94}}
\begin{center}
\renewcommand{\arraystretch}{1.}
\setlength\tabcolsep{3pt}
\begin{tabular}{ccccc}
\hline\hline\noalign{\smallskip}
 & $\chi$ & $m$ (kg) & $d$ (m) & $\omega$\\
\noalign{\smallskip} \hline \noalign{\smallskip}
$O_{2}$ & $0.237$ & $5.312\times10^{-26}$ & $4.01\times10^{-10}$ & $0.77$ \\
$N_{2}$ & $0.763$ & $4.650\times10^{-26}$ & $4.11\times10^{-10}$ & $0.74$ \\
\noalign{\smallskip} \hline\hline\noalign{\smallskip}
\end{tabular}
\label{tab3}
\end{center}
\end{table}

\begin{figure}[b!]
 \begin{center}
 \includegraphics[width=8.0cm,height=7.0cm]{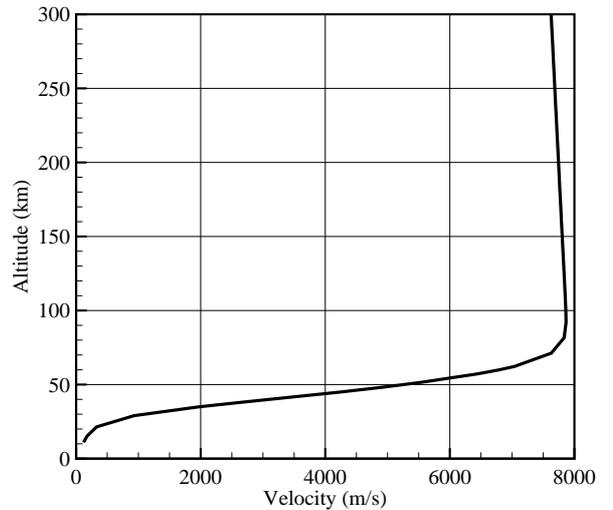}
 \end{center}
\caption{The velocity-altitude map for the SARA capsule.}
\label{BJoPP02F02}
\end{figure}

The velocity-altitude map for the SARA capsule~\cite{Toro2,Pessoa} is demonstrated in Fig.~\ref{BJoPP02F02}. This velocity-altitude map was generated based on predefined conditions, such as an initial velocity of 7626.30 m/s, an initial altitude of 300 km, a drag coefficient of 0.80, a lift coefficient of 0.0, a reference area of 0.785 m$^2$, a total mass of 150 kg, and finally,  a gravity acceleration of 9.81 m/s$^2$.

According to Fig.~\ref{BJoPP02F02}, for altitudes of 100, 95, 90, 85, and 80 km, the freestream velocity $V_{\infty}$ is 7862, 7866, 7864, 7864, and 7820 m/s, respectively. These values correspond to a freestream Mach number $M_{\infty}$ of 27.1, 28.1, 29.2, 29.2, and 29.0, respectively. In the present account, the capsule surface was kept at a constant wall temperature $T_{w}$ of 800 K for all cases investigated. This temperature is chosen to be representative of the surface temperature near the stagnation point of a reentry capsule. According to Machado and Boas~\cite{Machado}, in the stagnation region of the SARA capsule, temperature may reach a maximum value around 950 K.

The overall Knudsen number $Kn$ is defined as the ratio of the molecular mean free path $\lambda$ in the freestream gas to a characteristic dimension of the flowfield. In the present study, the characteristic dimension was defined as being the nose radius $R$. For the altitudes investigated, 100, 95, 90, 85, and 80 km, the overall Knudsen numbers correspond to $Kn_{R}$ of 0.4615, 0.1899, 0.0726, 0.0289, and 0.0115, respectively. Finally, the Reynolds number $Re_{R}$ correspond to 92, 224, 609, 3442 and 15249 for altitudes of 100, 95, 90, 85 and 80 km, respectively, based on conditions in the undisturbed stream with the nose radius $R$ as the characteristic length.

\section{Computational Method and Procedure}
\label{sec:4}
The governing equation in the transition flow regime is the Boltzmann equation~\cite{Cercignani}. The Boltzmann equation is a nonlinear integral-differential equation, with one dependent variable given by the particle distribution function. The particle properties are determined in a six-dimensional space, named the phase space, composed by three dimensions for particle coordinates, and three dimensions for particle velocities.

In order to circumvent the difficulty of a direct solution of the Boltzmann equation, the Direct Simulation Monte Carlo (DSMC) method is one of the alternative approaches for solving the Boltzmann equation by simulating the behavior of individual particles. The DSMC method, pioneered by Bird~\cite{Bird94}, has become the appropriate choice for problems involving complex multidimensional flows of rarefied hypersonic aerothermodynamics. In addition, CFD methods, which rely on continuum relations to compute the flowfield structure, will not provide accurate results for these flows in the upper atmosphere, since the assumptions made in developing the differential equations, on which CFD methods are based, break down on rarefied conditions.

In the DSMC method, the gas is modeled at the microscopic level by simulated particles. Each simulated particle represents a very large number of real molecules or atoms. These representative molecules are tracked as they move, collide and undergo boundary interactions in simulated physical space. The molecular motion, which is considered to be deterministic, and the intermolecular collisions, which are considered to be stochastic, are uncoupled over the small time step used to advance the simulation and computed sequentially.

In the present account, collisions are modeled by using the variable hard sphere (VHS) molecular model~\cite{Bird81} and the no time counter (NTC) collision sampling technique~\cite{Bird89}. The VHS model employs the simple hard sphere angular scattering law. Therefore, all directions are equally possible for post-collision velocity in the center-of-mass frame of reference. Nevertheless, the collision cross section depends on the relative speed of
colliding molecules according to some power law. The exponent is calculated by matching the viscosity of the simulated gas to that of its real counterpart.

The mechanics of the energy exchange processes between kinetic and internal modes for rotation and vibration are controlled by the Borgnakke-Larsen statistical model~\cite{Borgnakke}. The essential characteristic of this model is that a fraction of collisions is treated as completely inelastic, and the remainder of the molecular collisions is regarded as elastic. For a given collision, the probabilities are designated by the inverse of the relaxation numbers, which correspond to the number of collisions necessary, on average, for a molecule to relax. In this study, the relaxation numbers of 5 and 50 were used for the rotation and vibration, respectively. Simulations are performed using a non-reacting gas model consisting of only two chemical species, N$_{2}$ and O$_{2}$.

Finally, the freestream coefficient of viscosity $\mu_\infty$ and the freestream mean free path $\lambda_\infty$ are evaluated from a consistent definition~\cite{Bird94,Bird83} by using the VHS collision model with temperature exponents (viscosity index) of 0.74 and 0.77 for N$_{2}$ and O$_{2}$, respectively.

\section{Computational Flow Domain and Grid}
\label{sec:5}
For the numerical treatment of the problem, the flowfield around the capsule is divided into an arbitrary number of regions, which are subdivided into computational cells. The cells are further subdivided into subcells, two subcells/cell in each coordinate direction. The cell provides a convenient reference for the sampling of the macroscopic gas properties. On the other hand, the collision partners are selected from the same subcell in order to establish the collision rate. Therefore, the physical space network is used to facilitate the choice of molecules for collisions as well as for the sampling of the macroscopic flow properties, such as temperature, pressure, density, etc.

\begin{figure}[t!]
 \begin{center}
 \includegraphics[width=8.0cm,height=7.0cm]{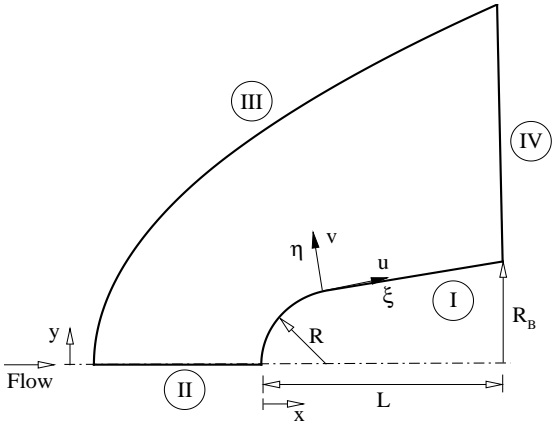}
 \end{center}
 \caption{Drawing illustrating the computational domain.}
\label{BJoPP02F03}
\end{figure}

The computational domain used for the calculation is made large enough so that the capsule disturbances do not reach the upstream and side boundaries, where freestream conditions are specified. In this manner, the computational domain changed according to the rarefaction degree of the flow on the capsule. A schematic view of the computational domain is depicted in Fig.~\ref{BJoPP02F03}. Advantage of the flow symmetry is taken into account, and molecular simulation is applied to one-half of a full configuration. According to this plot, side I is defined by the capsule surface. Diffuse reflection with complete thermal accommodation is the condition applied to this side. In a diffuse reflection model, the molecules are reflected equally in all directions. In addition, the final velocity of the molecules is randomly assigned according to a half-range Maxwellian distribution based on the wall temperature. Side II is a plane of symmetry, where all flow gradients normal to the plane are zero. At the molecular level, this plane is equivalent to a specular reflecting boundary. Side III is the freestream side through which simulated molecules enter and exit. Finally, the flow at the downstream outflow boundary, side IV, is predominantly supersonic and vacuum condition was assumed at this boundary~\cite{Guo}. As a result, simulated molecules can only exit at this boundary.

The numerical accuracy in DSMC method depends on the cell size chosen, on the time step as well as on the number of particles per computational cell. In the DSMC algorithm, the linear dimensions of the cells should be small in comparison with the length scale of the macroscopic flow gradients normal to streamwise directions. Therefore, the cell dimensions should be the order of or smaller than the local mean free path~\cite{Alexander98,Alexander00}. Furthermore, the time step should be chosen to be sufficiently small in comparison with the local mean collision time~\cite{Garcia,Hadjiconstantinou}. In general, the total simulation time, discretized into time steps, is based on the physical time of the real flow. Finally, the number of simulated particles has to be large enough to make statistical correlations between particles insignificant.

These effects were investigated in order to determine the number of cells and the number of particles required to achieve grid independent solutions. The grid generation scheme used in this study follows that procedure presented by Bird~\cite{Bird94}. Along the body surface (side I) and the outer boundary (side III), point distributions are generated in such way that the number of points on each side is the same; $\xi$-direction in Fig.~\ref{BJoPP02F03}. Then, the cell structure is defined by joining the corresponding points on each side by straight lines and then dividing each of these lines into segments which are joined to form the system of quadrilateral cells; $\eta$-direction in Fig.~\ref{BJoPP02F03}. The distribution can be controlled by a number of different distribution functions that allow the concentration of points in regions where high flow gradients or small mean free paths are expected. Grid independence was tested by running the calculations with a coarse, standard and a fine grid. A discussion of grid effects on the aerodynamic surface quantities is described in the Appendix.

\section{Computational Results and Discussion}
\label{sec:6}
This section focuses on the effects that take place in the primary properties due to variations on the altitude of the capsule. Primary properties of particular interest in the transition flow regime are velocity, density, pressure and temperature. In this scenario, this section discusses and compares differences in these quantities due to rarefaction effects.

\subsection{Velocity Field}
\label{sec:6.1}
\begin{figure}[t!]
 \begin{center}
  \includegraphics[width=7.0cm,height=6.0cm]{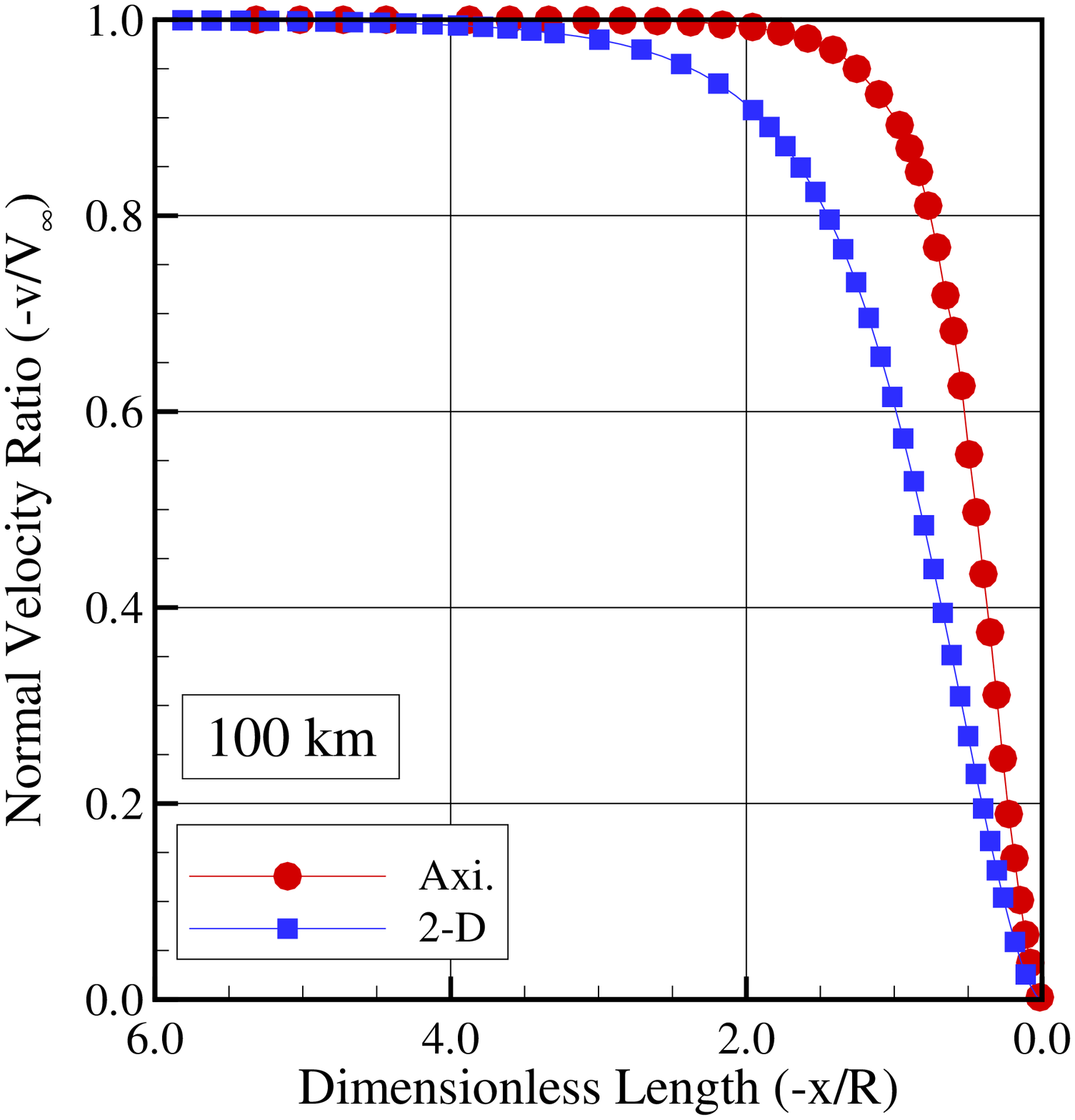}
  \includegraphics[width=7.0cm,height=6.0cm]{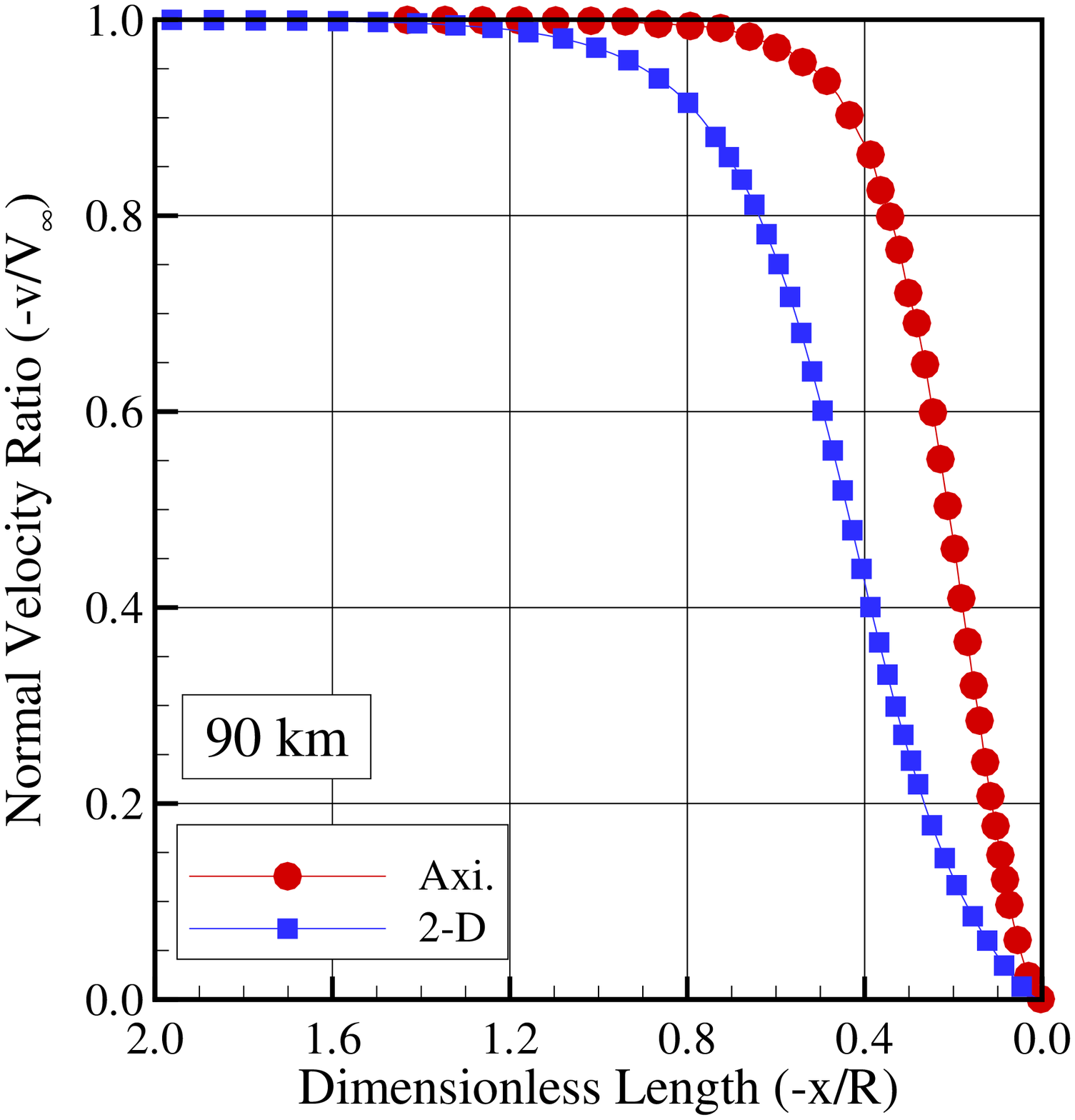}
  \includegraphics[width=7.0cm,height=6.0cm]{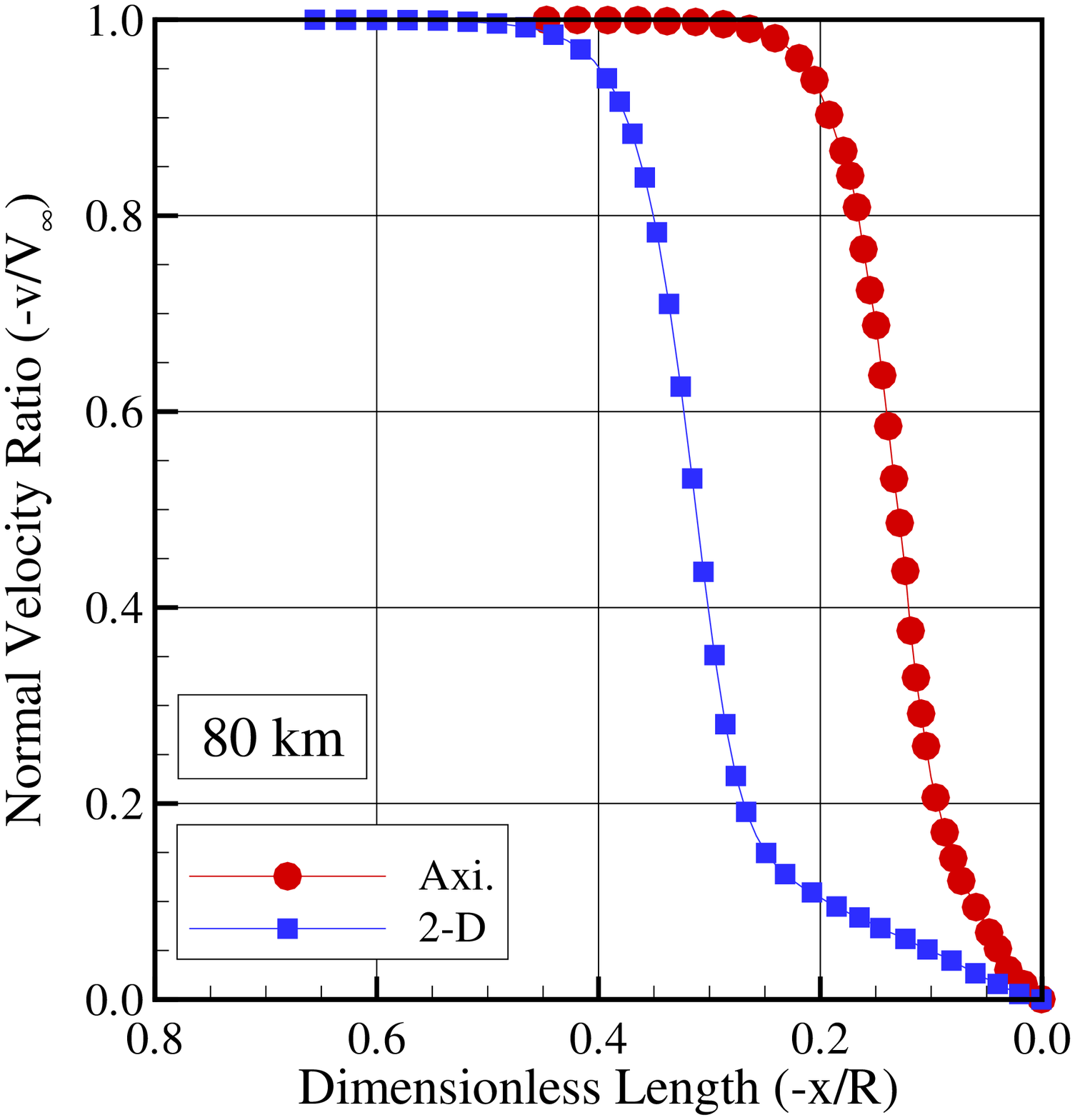}
 \end{center}
 \caption{Normal velocity ($-v/V_{\infty}$) profiles along the stagnation
          streamline for altitudes of (a) 100 km, (b) 90 km, and (c) 80 km.}
 \label{BJoPP02F04}
\end{figure}

Normal velocity profiles along the stagnation streamline and their dependence on rarefaction are illustrated in
Figs.~\ref{BJoPP02F04}(a-c) for altitudes of 100, 90 and 80 km, respectively. In this set of plots, the normal velocity $v$ is normalized by the freestream velocity $V_{\infty}$, and the distance $x$ along the stagnation streamline is normalized by the nose radius $R$. In addition, for comparison purpose, the velocity profiles for the 2-D geometry are also included in the plots. Also, in order to emphasize points of interest, a different scale is used in the abscissa axis. It is important to note that $V_{\infty}$ is slightly different for each altitude (see Tab.~\ref{tab1}) and, therefore, the comparison is made in terms of ratio. Also, the velocity profiles for the 95 km and 85 km cases are intermediate to the other cases and, therefore, they will not be shown.

\begin{figure}[t!]
 \begin{center}
  \includegraphics[width=7.0cm,height=6.0cm]{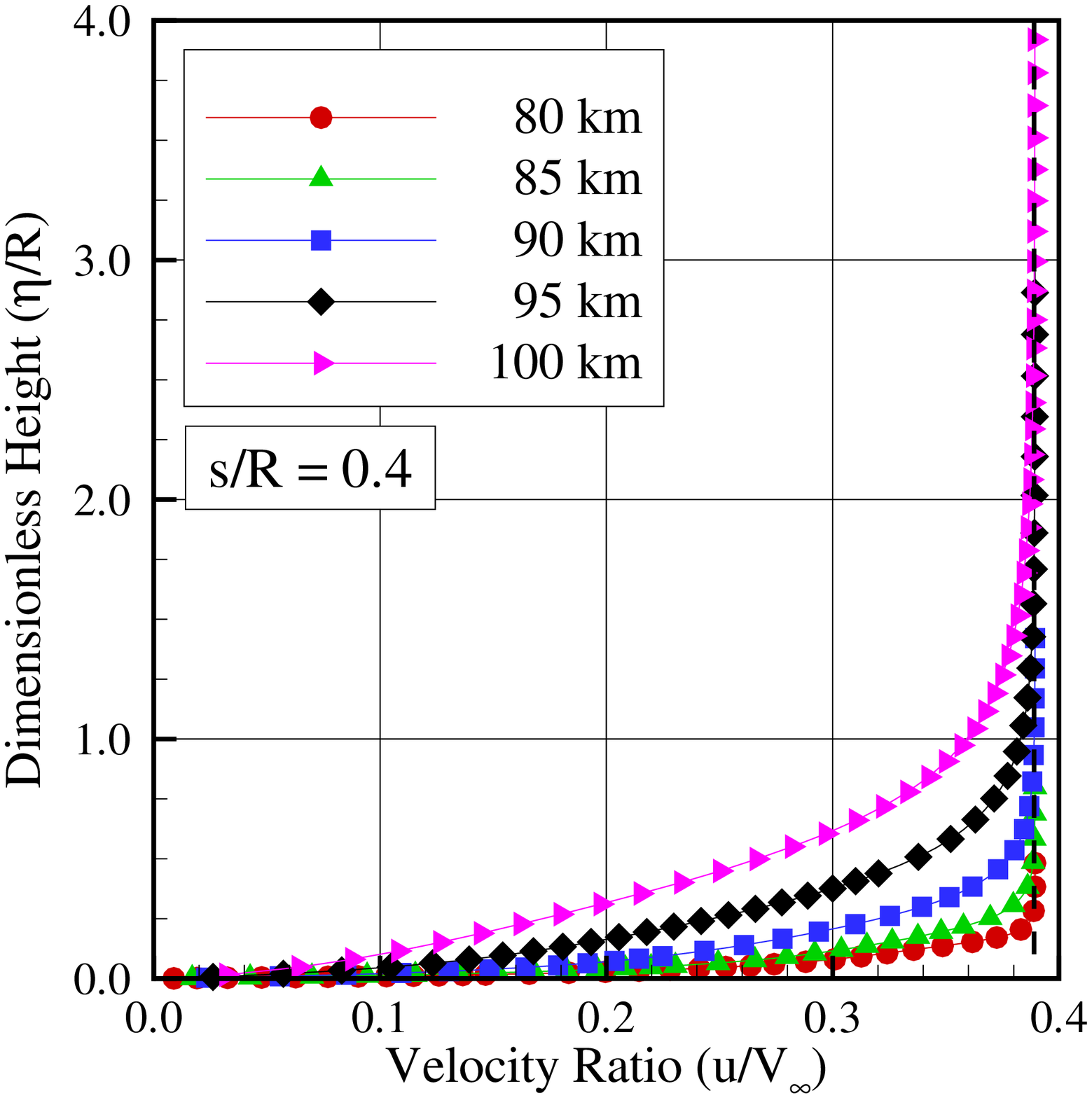}
  \includegraphics[width=7.0cm,height=6.0cm]{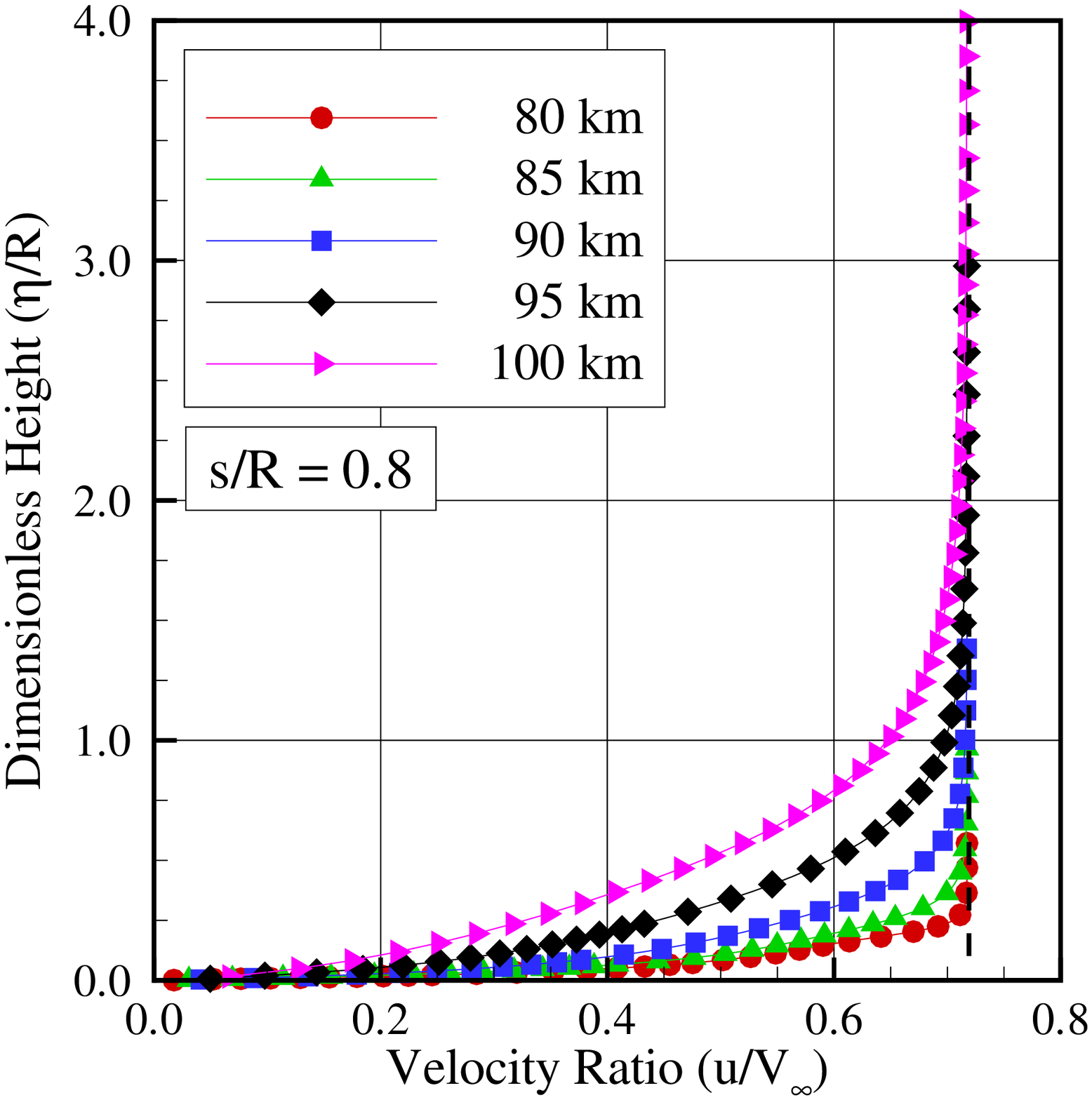}
  \includegraphics[width=7.0cm,height=6.0cm]{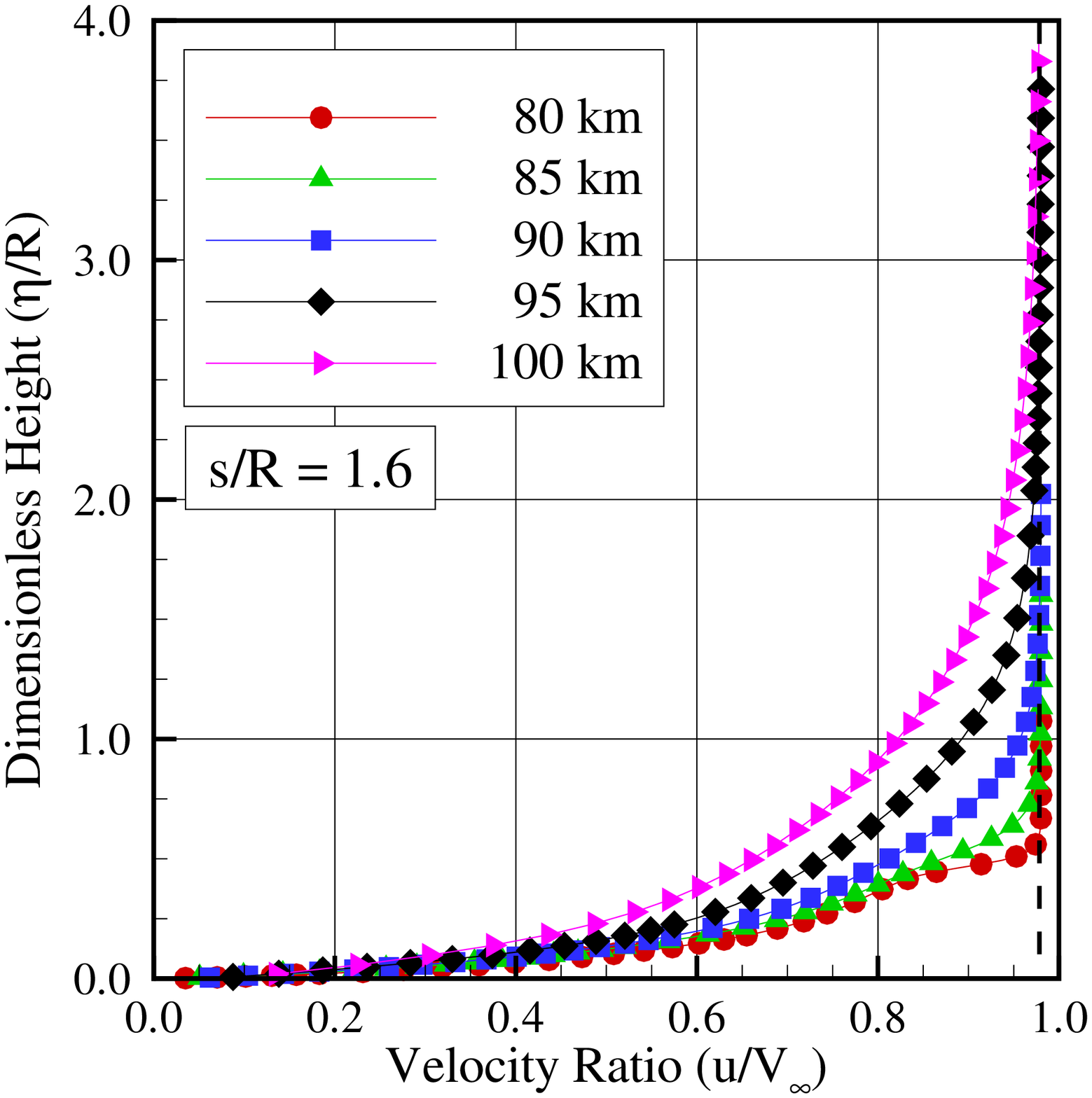}
 \end{center}
 \caption{Tangential velocity ($u/V_{\infty}$) profiles along the capsule surface
          for sections corresponding to $s/R$ of (a) 0.4, (b) 0.8 and (c) 1.6.}
 \label{BJoPP02F05}
\end{figure}

According to these plots, it is seen that the rarefaction effect influences the flowfield far upstream. The extent of this effect decreases with decreasing the Knudsen number $Kn_{R}$, i.e., by decreasing the altitude. As the altitude decreases, particles reflecting from the capsule surface diffuse less upstream due to the high density at the vicinity of the capsule nose, as will be seen subsequently. As a result, the extent of the flowfield disturbance due to the presence of the capsule decreases and becomes smaller in terms of the body dimension $R$. It should be mentioned in this context that, far from the capsule nose, the velocity ratio $v/V_{\infty}$ is equal to one, and at the stagnation point equal to zero. The region defined from the section in which the velocity ratio starts decreasing from 1, say $v/V_{\infty}$ = 0.99, up to the stagnation point, $v/V_{\infty}$ = 0, is defined herein as the upstream disturbance region, i.e., the region affected by the presence of the capsule. For instance, for the axisymmetric geometry, the extent of the upstream disturbance, based in a velocity reduction of 1\% ($v/V_{\infty}$ = 0.99), corresponds to sections $x/R$ of 1.86, 1.20, 0.75, 0.45, and 0.26 for altitudes of 100, 95, 90, 85, and 80 km, respectively. Nevertheless, for the 2-D geometry, it changes to around 3.51, 2.12, 1.21, 0.71, and 0.45 for altitudes of 100, 95, 90, 85, and 80 km, respectively. Consequently, the extension of the upstream disturbance is more pronounced for a 2-D flow than for an axisymmetric flow. It should be remarked that this effect is less pronounced for an axisymmetric geometry because of the ``relieving effect'', which is a characteristic of all three-dimensional flows. For the flow over an axisymmetric capsule, the ``addition" of a third dimension provides the molecules with extra space to move through it. In contrast, for the 2-D geometry, the molecules move only around the top or the bottom surface.

The outer extent of the flowfield disturbance over the capsule surface is demonstrated in Figs.~\ref{BJoPP02F05}(a-c) as a function of the altitude. In this set of plots, the tangential velocity $u$ is normalized by the freestream velocity $V_{\infty}$, and the height $\eta$ in the off-body direction ($\eta$-direction in Fig.~\ref{BJoPP02F03}(a)) is normalized by the nose radius $R$. In an effort to emphasize points of interest, this set of plots presents data for three stations along the body surface defined by $s/R$ of 0.4, 0.8, and 1.6, where $s$ is the arc length along the body surface measured from the stagnation point. The first two sections are located on the spherical nose, and the last one on the afterbody surface.

Interesting features can be drawn from this set of tangential velocity profiles. As the body slope decreases, the tangential velocity adjacent to the body surface increases. This is to be expected since the flow experiences an expansion as it moves downstream along the capsule surface. It is observed that the flow accelerates faster along the surface with the altitude rise. It should be noted that the tangential velocity $u_{\infty}$, defined as $\eta \rightarrow \infty$, is represented by the dashed line and shown for each station. Because of the body curvature, $u_{\infty}$( = $V_{\infty}\cos\theta$) varies as a function of the body slope.

Another interesting characteristic in these plots is the similarity of the velocity profiles along the body surface. This is an indication that the velocity profiles may be expressed in terms of functions that, in appropriate coordinates, may be independent of one of the coordinate directions. However, no attempts have been done to find such functions.

\subsection{Density Field}
\label{sec:6.2}
\begin{figure}[t!]
 \begin{center}
   \includegraphics[width=7.0cm,height=6.0cm]{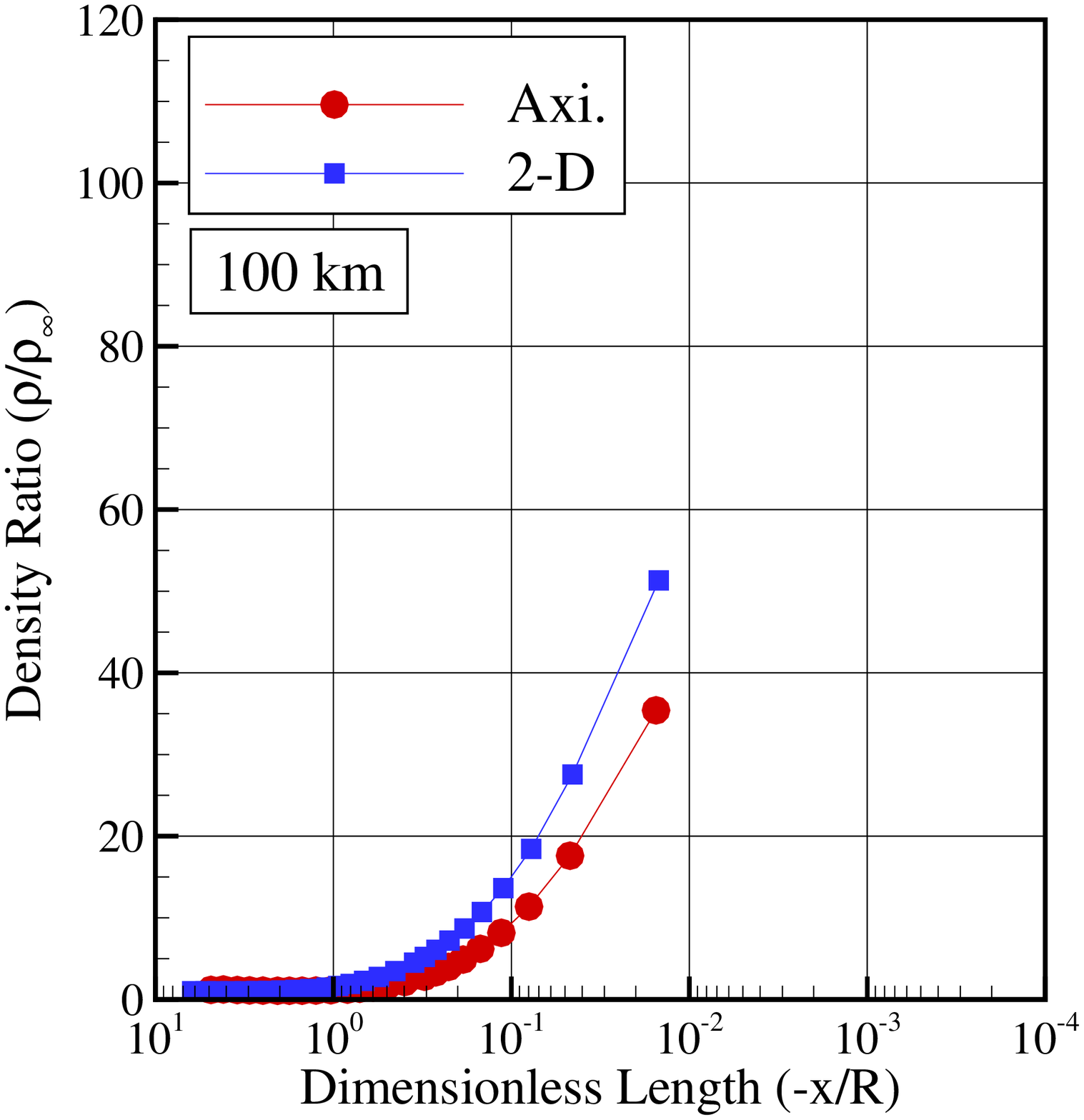}
   \includegraphics[width=7.0cm,height=6.0cm]{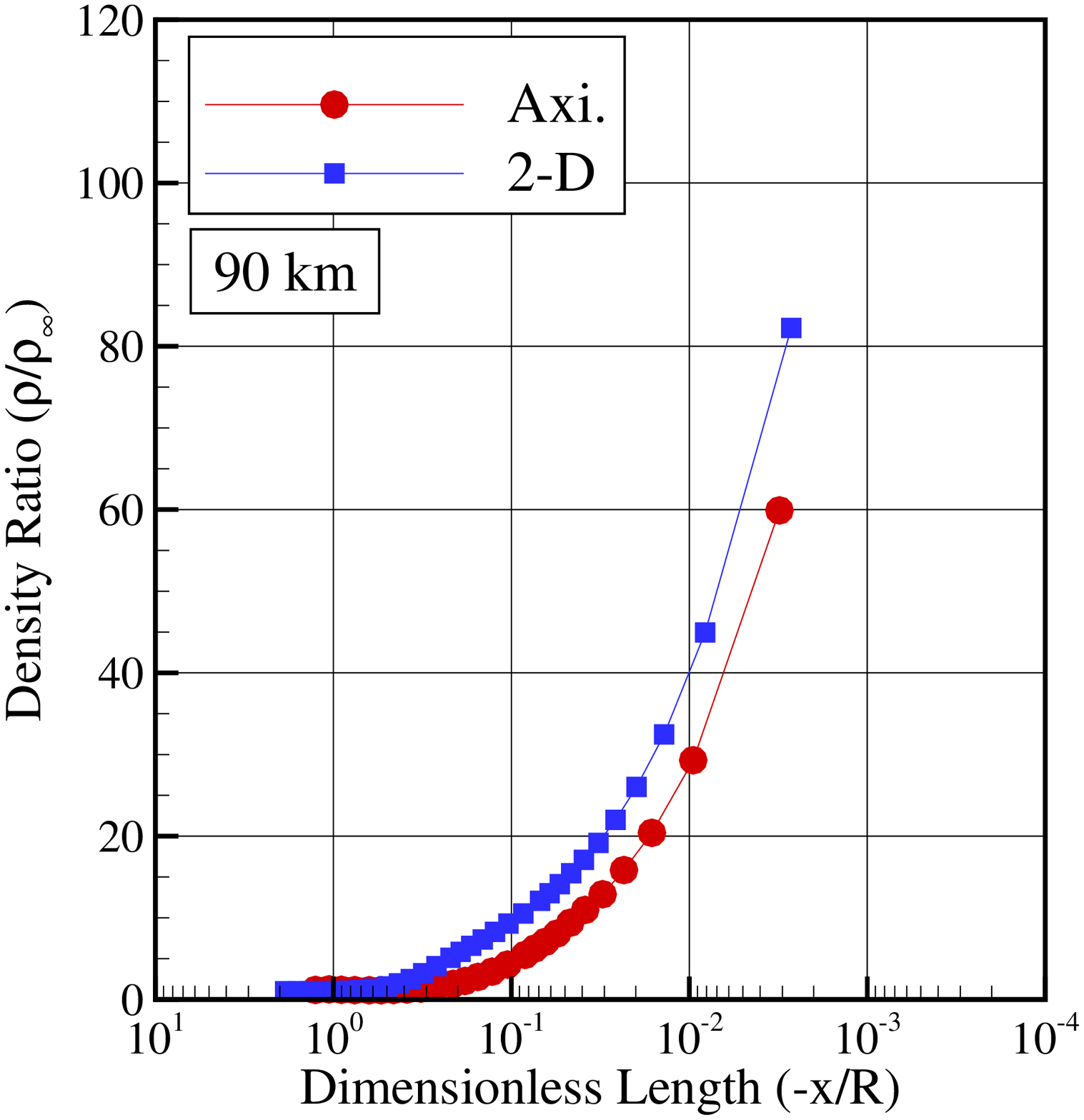}
   \includegraphics[width=7.0cm,height=6.0cm]{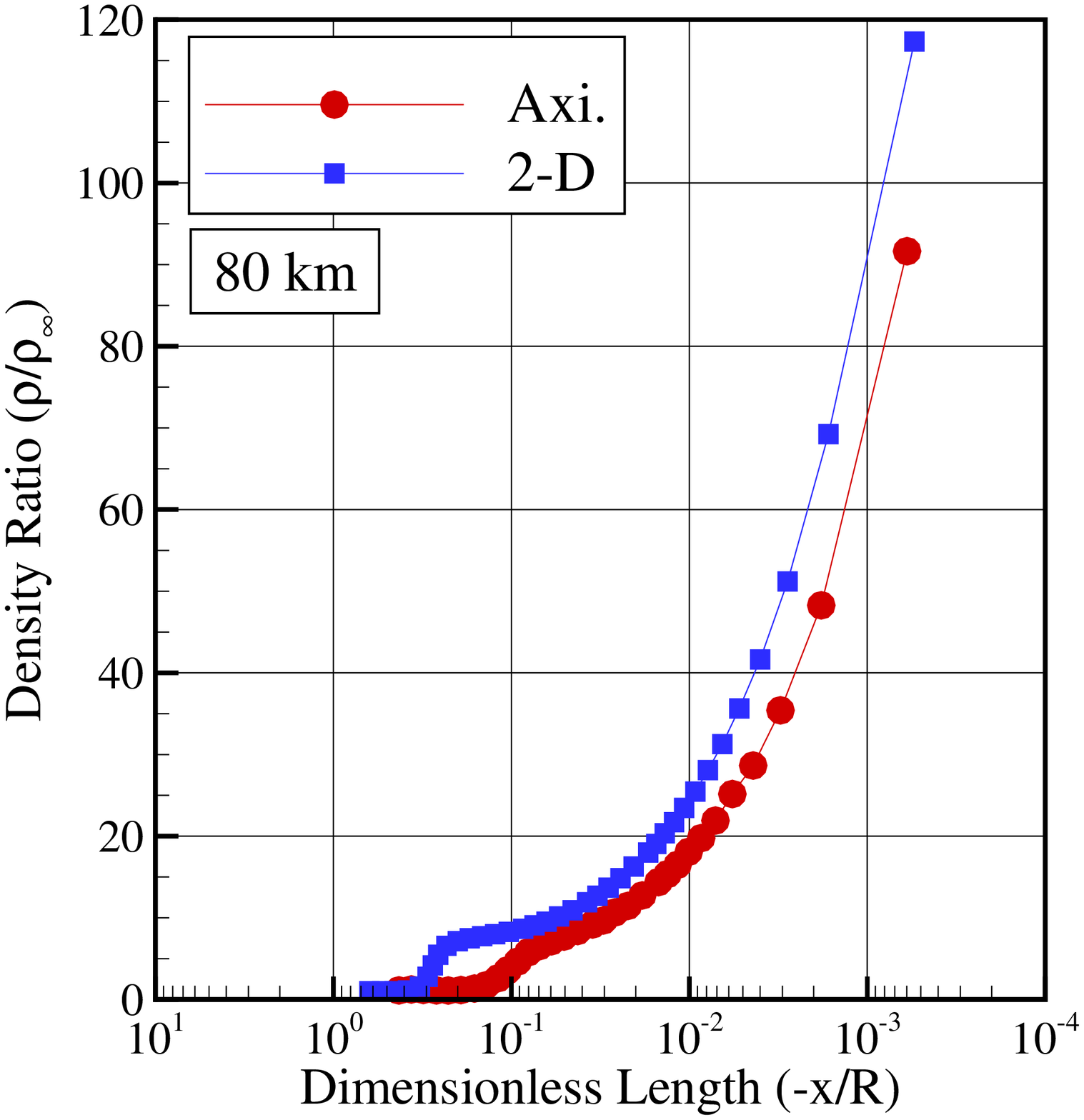}
 \end{center}
 \caption{Density ratio ($\rho/\rho_{\infty}$) profiles along the stagnation
          streamline for altitudes of (a) 100 km, (b) 90 km, and (c) 80 km.}
 \label{BJoPP02F06}
\end{figure}

The impact of rarefaction on density profiles along the stagnation streamline is displayed in Figs.~\ref{BJoPP02F06}(a-c) for altitudes of 100, 90 and 80 km, respectively. In this set of figures, density $\rho$ is normalized by the freestream density $\rho_{\infty}$. Again, as a base of comparison, density profiles for the 2-D geometry are also included in the figures.

The predictions of density for the altitude range investigated basically show no sign of a discrete shock wave. Except for the 80 km case, it is clearly seen a continuous rise in density from the freestream to the nose of the capsule, an indication of the diffuse nature of the shock wave, which corresponds to a characteristic of highly rarefied flows. By decreasing the altitude, both the freestream gas density and temperature change. As a result, the molecules interact much more with each other and collisions among them are more frequent. Therefore, density increases and the profile becomes steeper near the stagnation point. In addition, with the density rise near the stagnation point, the local mean free path decreases resulting in a lower local Knudsen number. Moreover, it is noticed that, unlike normal velocity, density has little effect on the extent of the upstream disturbance caused by the presence of the capsule, for the cases investigated. Much of the density increase in the shock layer occurs in a region smaller than one nose radius $R$, and after the temperature has reached its postshock value, as will be seen subsequently.

Still referring to Figs.~\ref{BJoPP02F06}(a-c), it can be recognized that density rises to well above the continuum inviscid limit for the cases investigated. As a point of reference, the Rankine-Hugoniot relations give a postshock density that corresponds to the ratio $\rho/\rho_{\infty}$ of 5.96 for freestream Mach number of 27. Near the stagnation point, $x/R \approx$ 0, a substantial density increase occurs, which is a characteristic of a cold-wall entry flow. In a typical entry flow, the body surface temperature is low compared to the stagnation temperature. This leads to a steep density gradient near to the body surface. For the present simulation, the ratio of wall temperature to stagnation temperature is around 0.033, which corresponds to a cold-wall flow.

\begin{figure}[t!]
 \begin{center}
   \includegraphics[width=7.0cm,height=6.0cm]{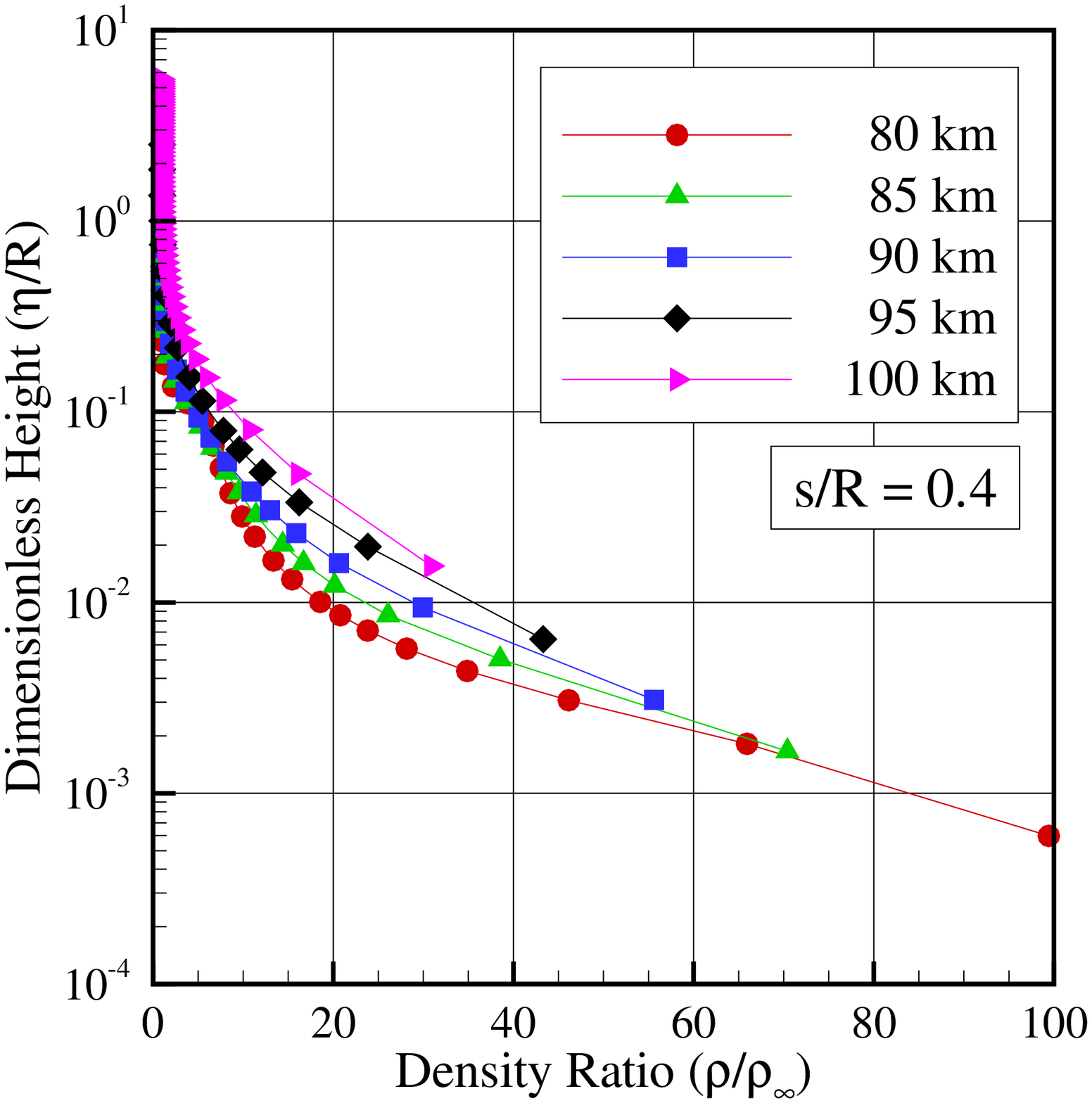}
   \includegraphics[width=7.0cm,height=6.0cm]{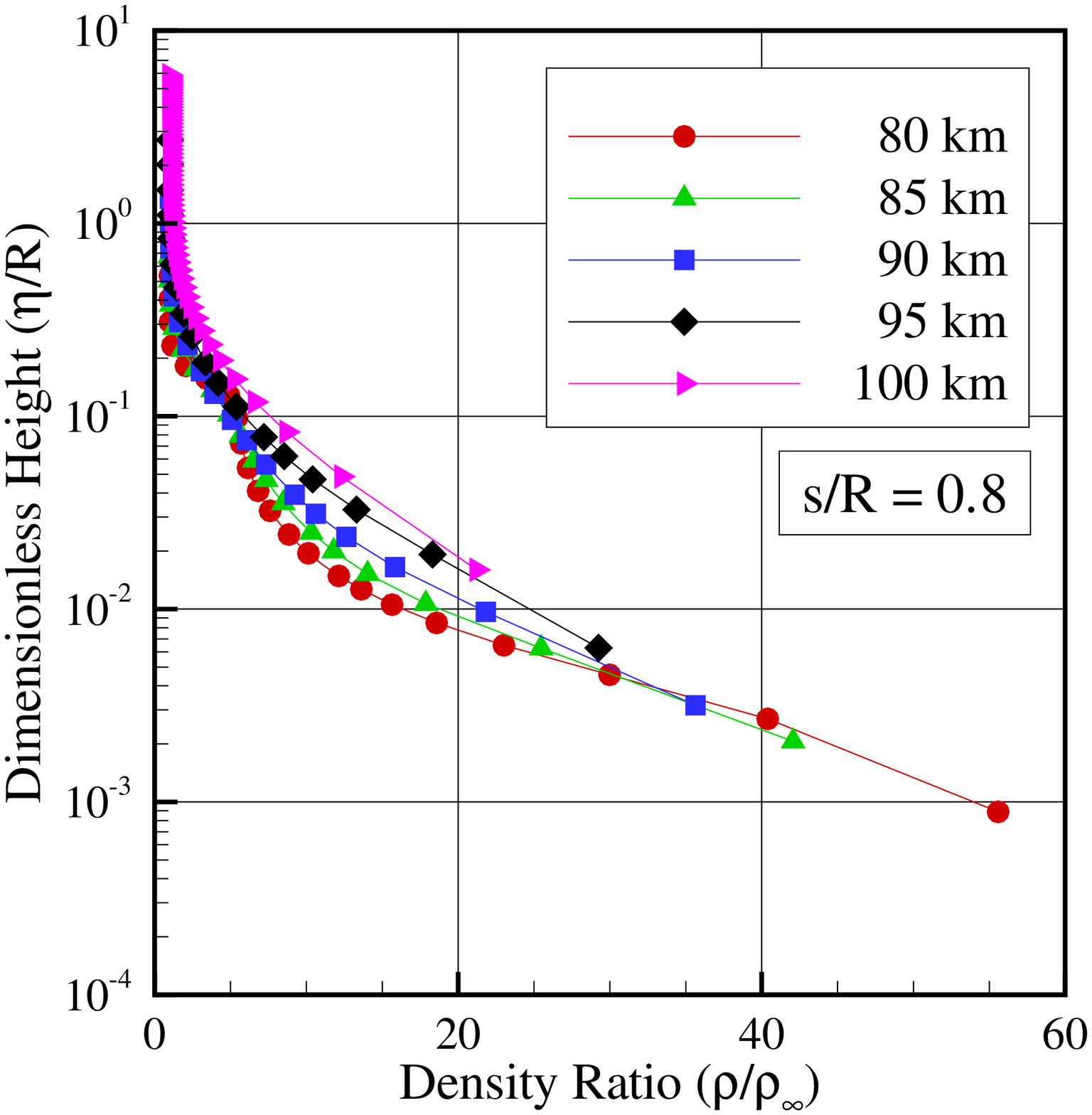}
   \includegraphics[width=7.0cm,height=6.0cm]{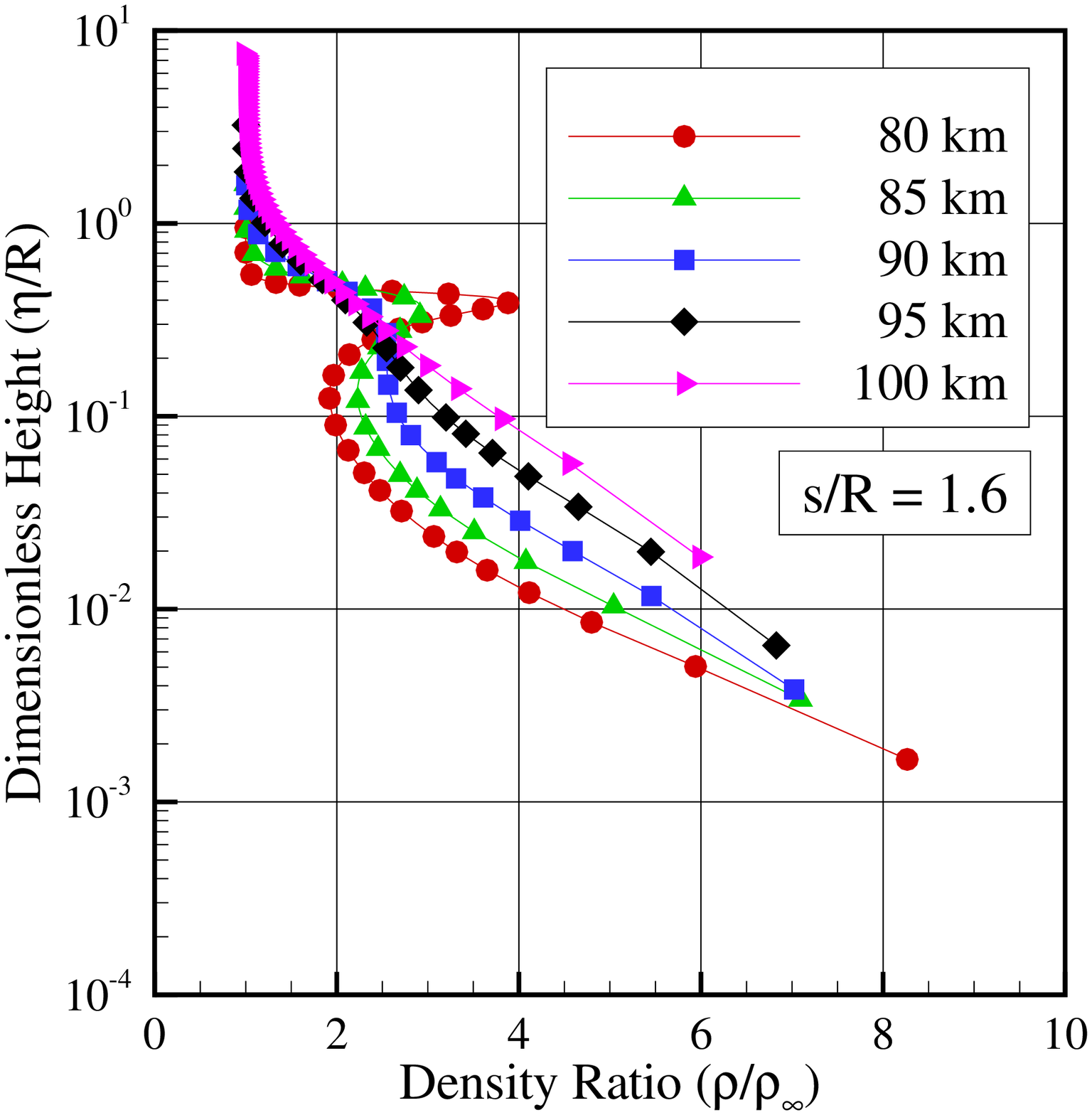}
 \end{center}
 \caption{Density ratio ($\rho/\rho_{\infty}$) profiles along the capsule surface
          for sections corresponding to $s/R$ of (a) 0.4, (b) 0.8 and (c) 1.6.}
 \label{BJoPP02F07}
\end{figure}

The deceleration of the freestream air molecules leads to an increase in the internal energy, such as translational, rotational and vibrational modes. Since the vibrational phenomenon absorbs energy, the temperature does not increase to its perfect gas value, and consequently, the density rise across the shock wave is greater than that for a perfect gas flow at the same freestream Mach number. For air behaving as a perfect gas at hypersonic speeds $\rho/\rho_{\infty} \longrightarrow$ 6, while for atmospheric hypersonic flight $\rho/\rho_{\infty}$ is the order of 10 to 14 or more~\cite{Probstein}, because of the elevated temperatures and resultant activation of more degrees of freedom. As a reference, Cuda Jr. and Moss~\cite{Cuda} have reported a simulation on hypersonic flow over blunt wedges at an altitude of 70 km, and they found a value for the stagnation density two orders of magnitude greater than the freestream value.

Variation of local density along the body normal direction, expressed as a ratio to the freestream value, is depicted in Figs.~\ref{BJoPP02F07}(a-c) as a function of the altitude. This group of diagrams presents data at three afterbody stations that correspond to the dimensionless arc length $s/R$ of 0.4, 0.8, and 1.6. According to these diagrams, it is clearly noticed that the density also experiences significant changes in the direction perpendicular to the wall as the flow moves downstream along the capsule surface. In the direction perpendicular to the wall, and for stations close to the stagnation region, the density is high adjacent to the wall and rapidly decreases inside a layer of thickness smaller than one nose radius $R$, where the density approaches the freestream value for the altitudes investigated. This characteristic is observed when the body surface is very much colder than the stagnation temperature of the oncoming gas. As a result, the gas near the body tends to be much denser and cooler than the gas in the rest of the boundary layer. Conversely, for a station far from the stagnation region, $s/R$~=~1.6, the density ratio is maximum at the wall, it decreases in the off-body direction $\eta$ and then it increases to a peak value inside the shock wave. Afterwards, the density decreases dramatically and reaches the freestream density value as $\eta/R \rightarrow \infty$.

Still referring to Fig.~\ref{BJoPP02F07}, it is very encouraging to observe that density is affected by changes in the altitude, as would be expected. According to Fig.~\ref{BJoPP02F07}(a), which corresponds to the section $s/R$ = 0.4, the density variation is in excess of one order of magnitude as compared to the freestream density for the altitude investigated. In this region, close to the stagnation region, the compression combined with a relatively cool wall produces a maximum density ratio $\rho/\rho_{\infty}$ that lies in the range from 40 to 90. Because of the flow expansion along the afterbody surface, the density ratio adjacent to the surface decreases dramatically to the range of 6 to 8 for the section corresponding to $s/R$ = 1.6.

\subsection{Pressure Field}
\label{sec:6.3}
Part of the large amount of kinetic energy present in a hypersonic freestream is converted by molecular collisions into high thermal energy surrounding the body and by flow work into increased pressure. In this manner, the stagnation line is a zone of strong compression, where pressure increases from the freestream to the stagnation point due to the shock wave that forms ahead of the capsule.

\begin{figure}[t!]
 \begin{center}
   \includegraphics[width=7.0cm,height=6.0cm]{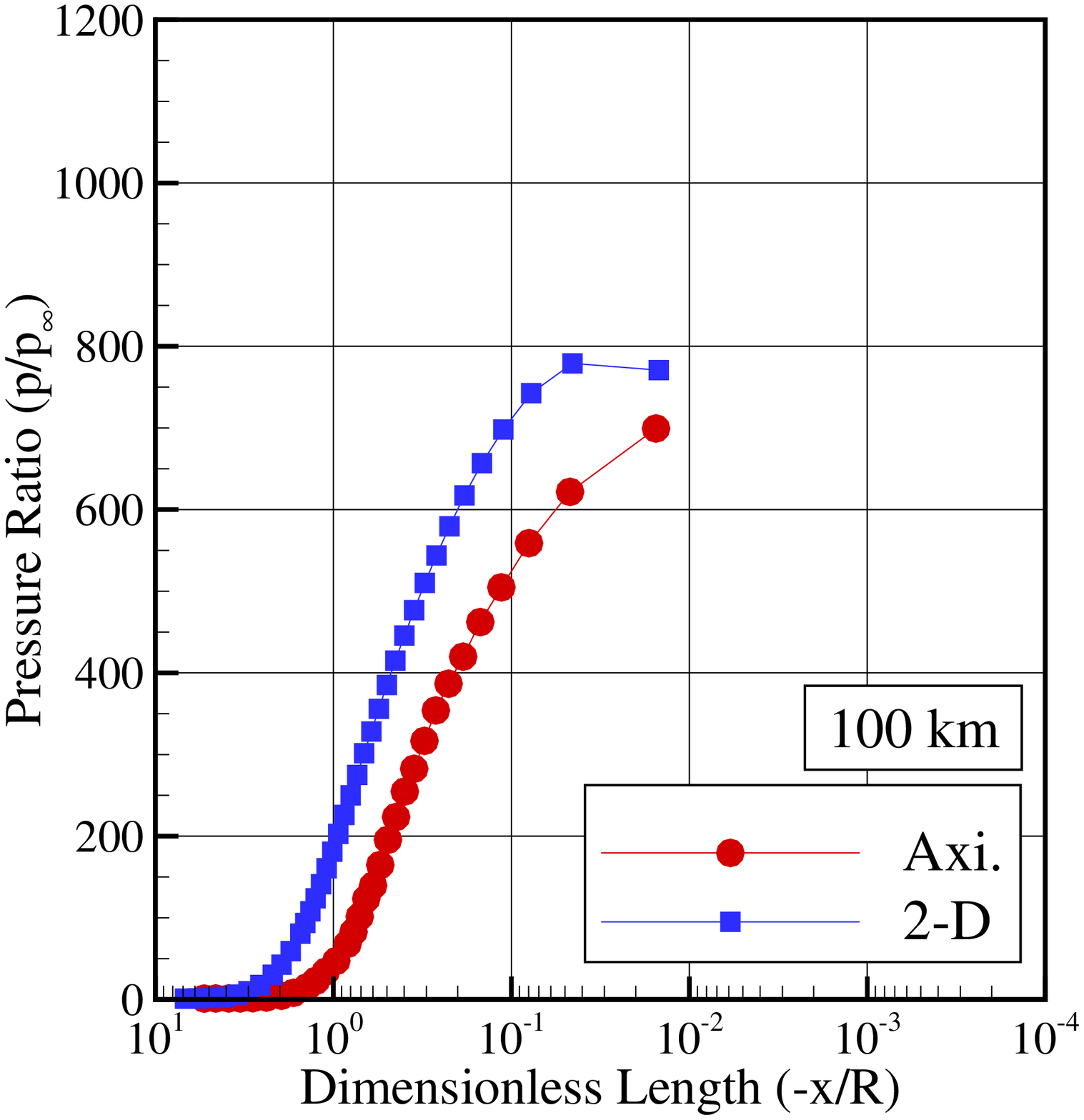}
   \includegraphics[width=7.0cm,height=6.0cm]{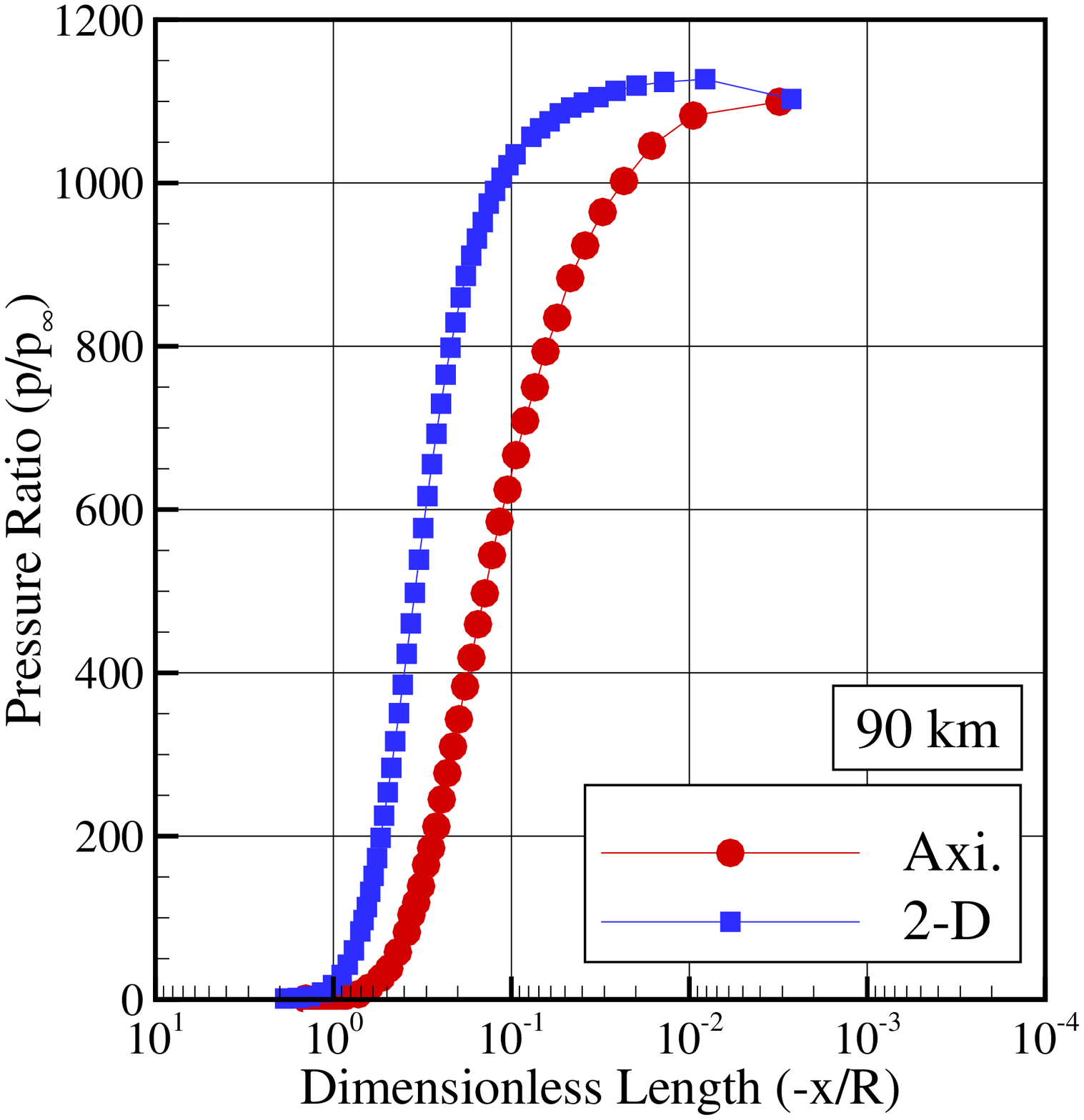}
   \includegraphics[width=7.0cm,height=6.0cm]{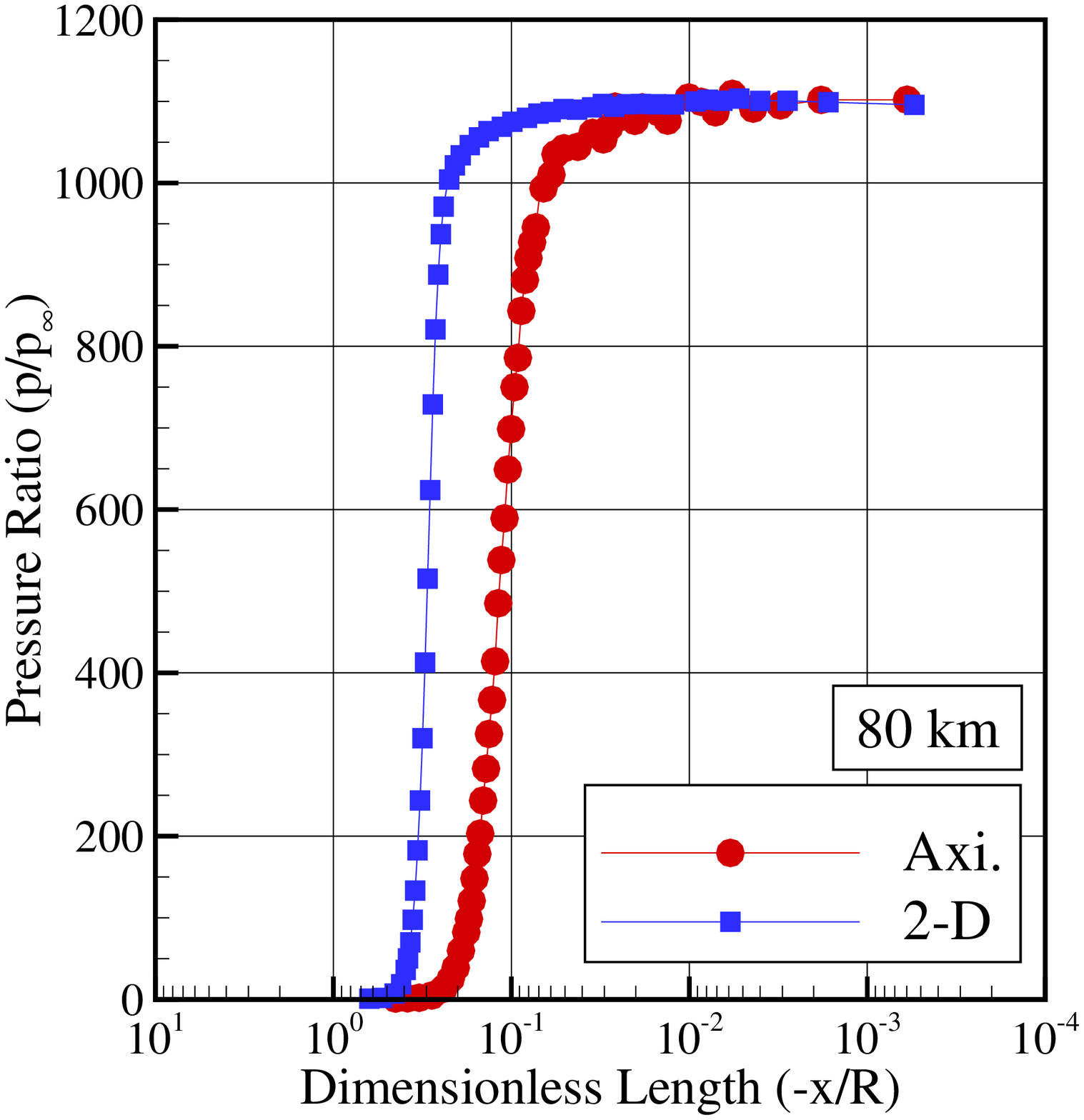}
 \end{center}
 \caption{Pressure ratio ($p/p_{\infty}$) profiles along the stagnation
          streamline for altitudes of (a) 100 km, (b) 90 km, and (c) 80 km.}
 \label{BJoPP02F08}
\end{figure}

Pressure profiles along the stagnation streamline are depicted in Figs.~\ref{BJoPP02F08}(a-c) as a function of the altitude. In this set of plots, pressure $p$ is normalized by the freestream pressure $p_{\infty}$. It may be recognized from these plots that basically there is a continuous rise in pressure from the freestream up to the stagnation point where the maximum value is attained. Near the stagnation point, a substantial pressure increase occurs with decreasing the altitude, i.e., with decreasing the Knudsen number $Kn_{R}$. It is apparent from these plots that the general shape of the pressure distribution profiles is preserved when the altitude decreases from 100 km to 80 km.

The extent of the upstream flowfield disturbance for pressure is significantly different from that presented by velocity and density. The domain of influence for pressure is higher than that for velocity and density, and lower than that presented for temperature. Similar to the density, much of the pressure increase in the shock layer occurs after the translational kinetic temperature has reached its postshock value, as will be shown subsequently.

\begin{figure}[t!]
 \begin{center}
   \includegraphics[width=7.0cm,height=6.0cm]{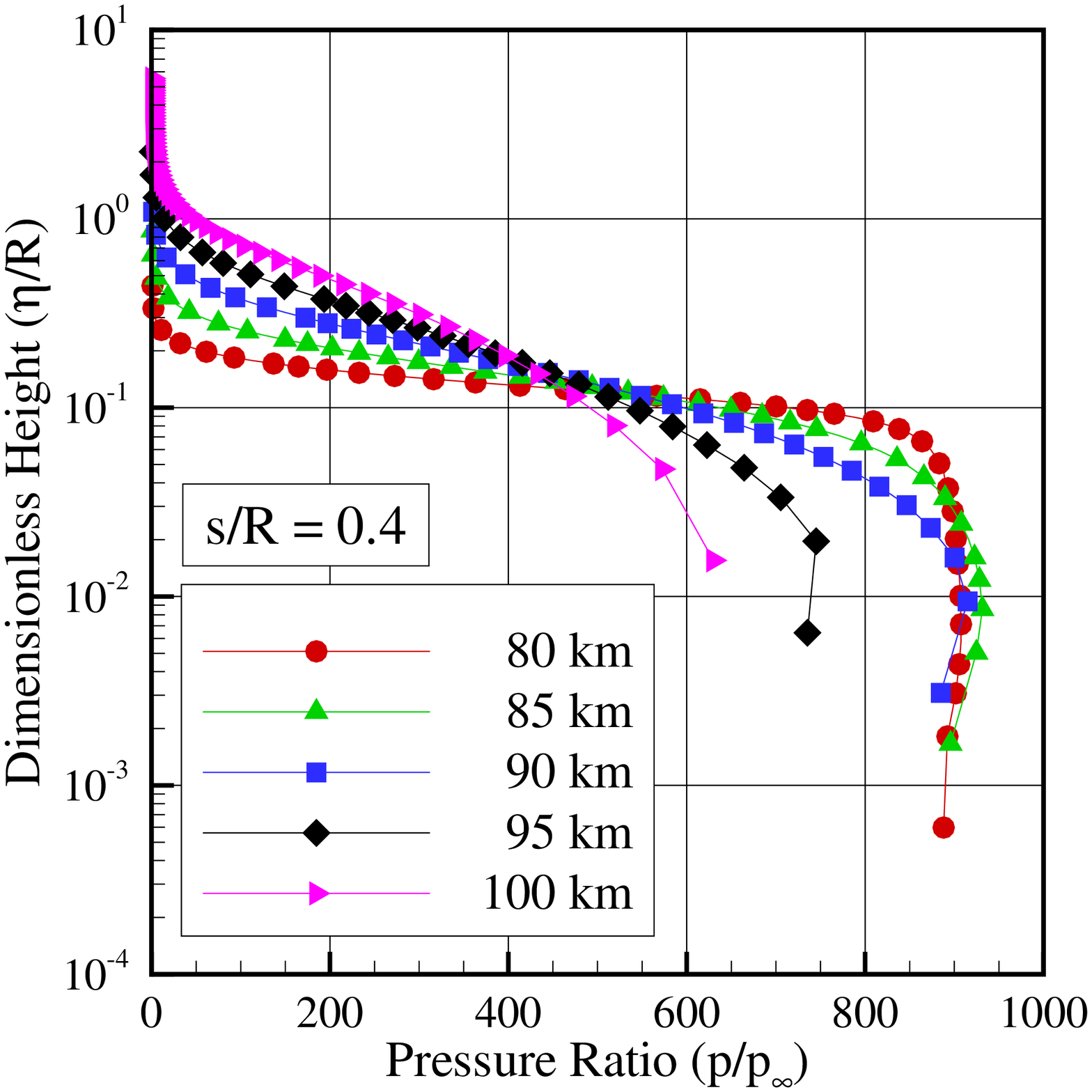}
   \includegraphics[width=7.0cm,height=6.0cm]{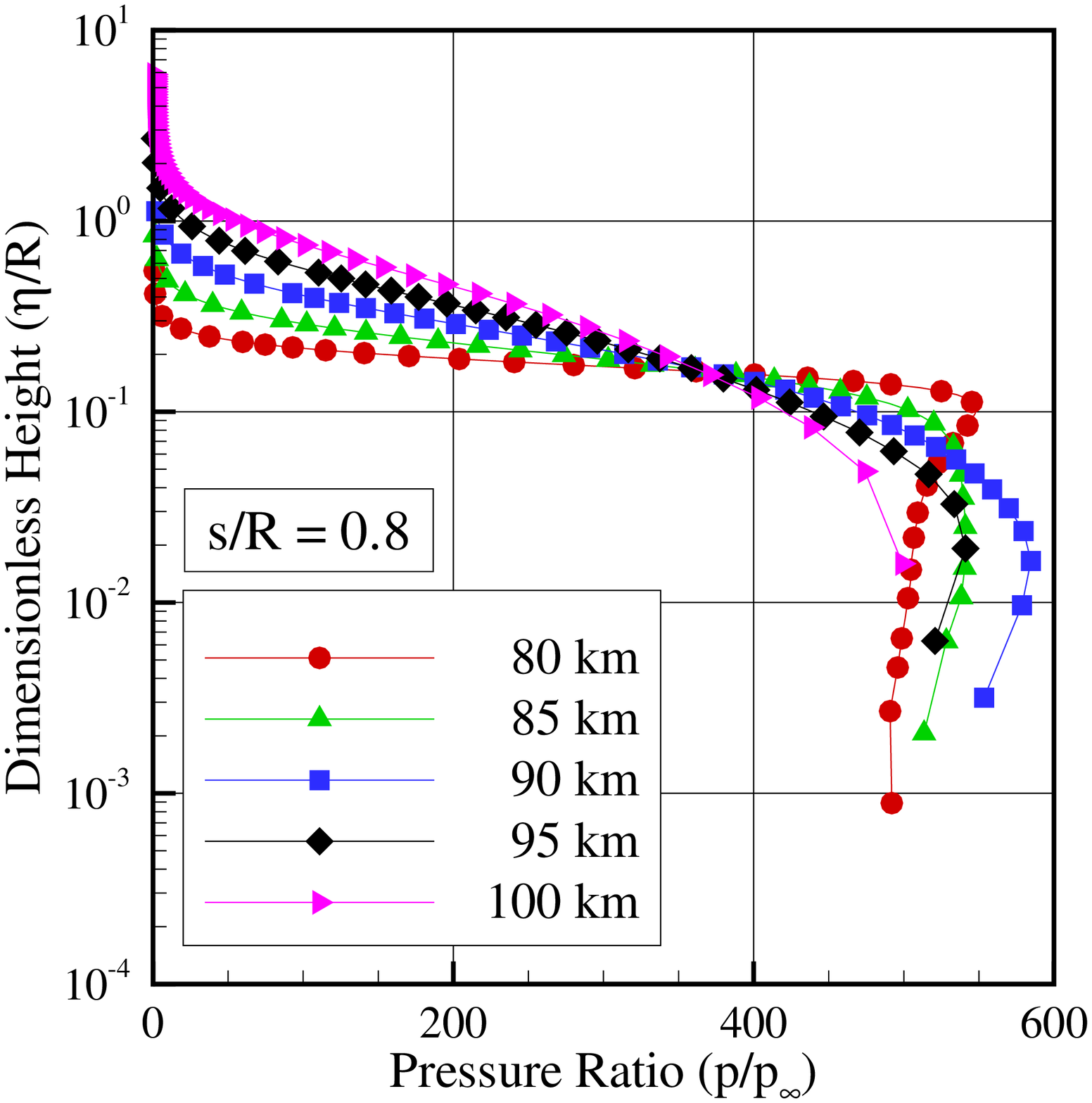}
   \includegraphics[width=7.0cm,height=6.0cm]{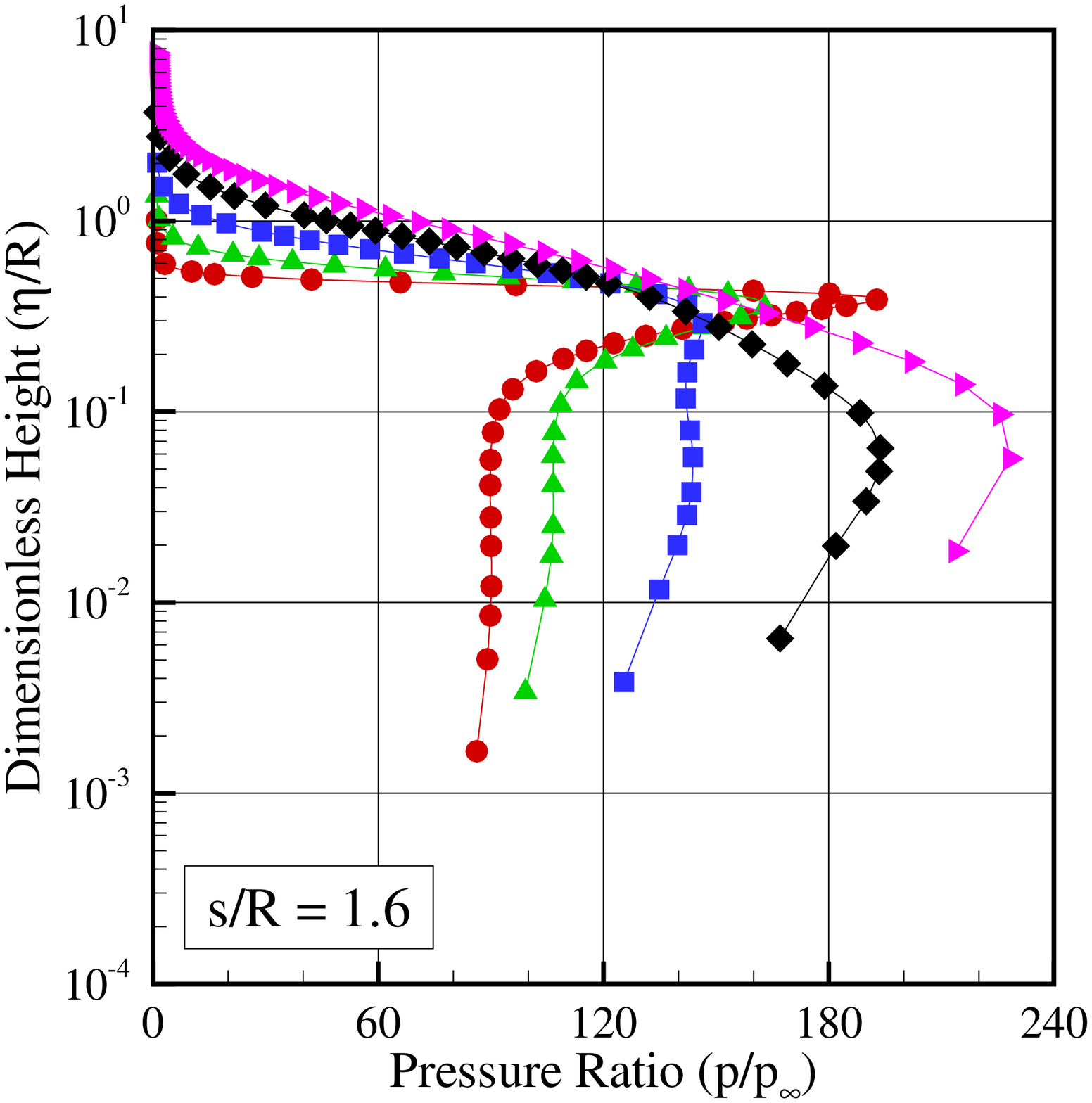}
 \end{center}
 \caption{Pressure ratio ($p/p_{\infty}$) profiles along the capsule surface
          for sections corresponding to $s/R$ of (a) 0.4, (b) 0.8 and (c) 1.6.}
 \label{BJoPP02F09}
\end{figure}

Local pressure, expressed as a ratio to the freestream value, for three stations located on the capsule surface is demonstrated in Figs.~\ref{BJoPP02F09}(a-c) as a function of the altitude. It is apparent from these profiles that pressure is affected with decreasing altitude, as was mentioned earlier. For the station corresponding to $s/R$ = 0.4, Fig.~\ref{BJoPP02F09}(a), the pressure variation is almost three orders of magnitude larger than the freestream pressure for an altitude of 80 km. In this region, at the vicinity of the stagnation point, the compression produces a maximum pressure that is around 900 times the freestream pressure for an altitude of 80 km. Due to the flow expansion along the body surface, the pressure adjacent to the surface decreases significantly. For the particular case of 80 km, the pressure ratio $p/p_{\infty}$ decreases to around 90 for the station corresponding to $s/R$ = 1.6, as shown in Fig.~\ref{BJoPP02F09}(c), a reduction of basically one order of magnitude from station  $s/R$  of 0.4 to 1.6.

\subsection{Temperature Field}
\label{sec:6.4}
The strong shock wave that forms ahead of the capsule at hypersonic flow converts part of the kinetic energy of the freestream air molecules into thermal energy. This thermal energy downstream of the shock wave is partitioned into increasing the translational kinetic energy of the air molecules, and into exciting other molecular energy states such as rotation and vibration.

\begin{figure}[t!]
 \begin{center}
   \includegraphics[width=7.0cm,height=6.0cm]{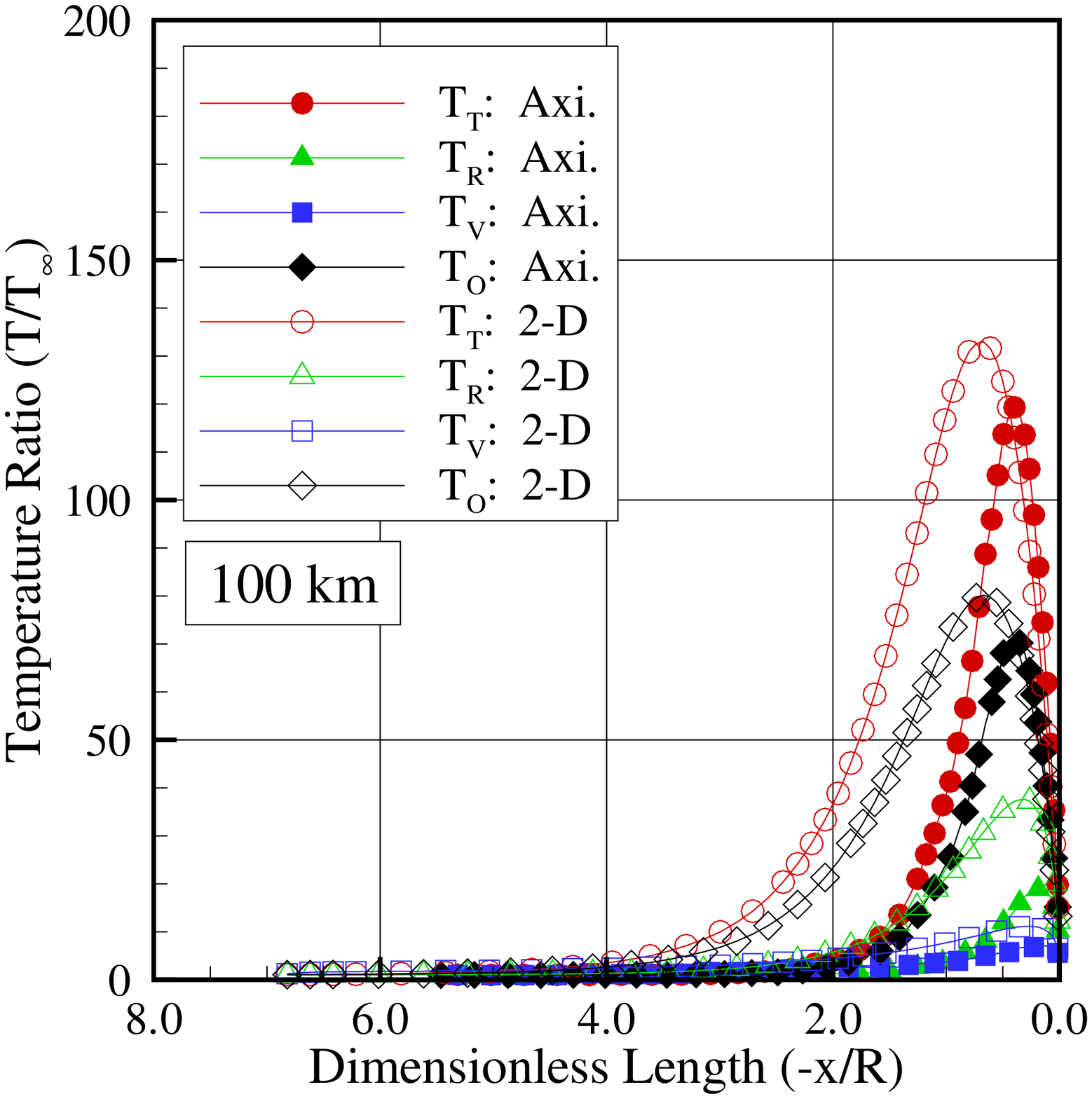}
   \includegraphics[width=7.0cm,height=6.0cm]{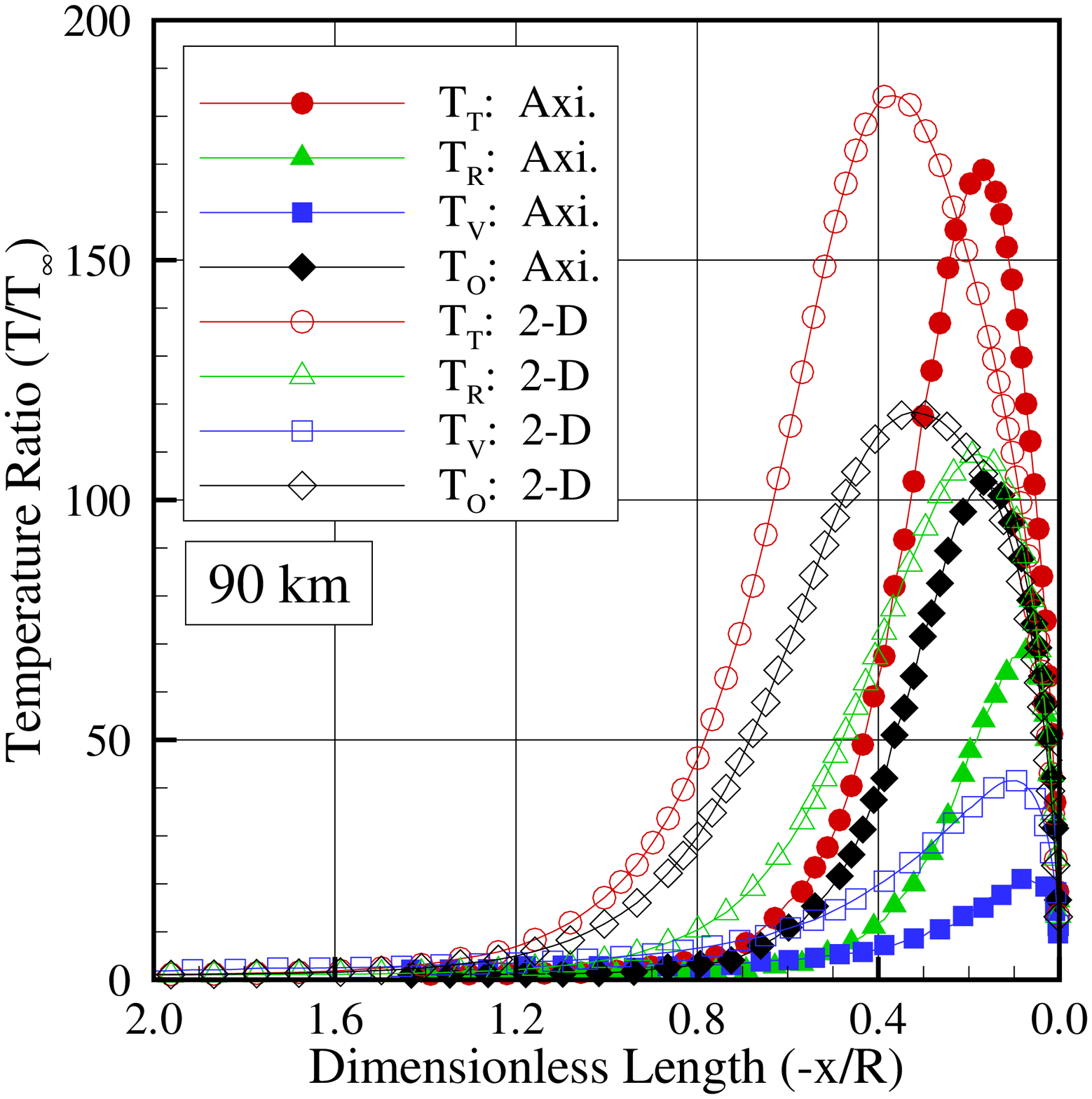}
   \includegraphics[width=7.0cm,height=6.0cm]{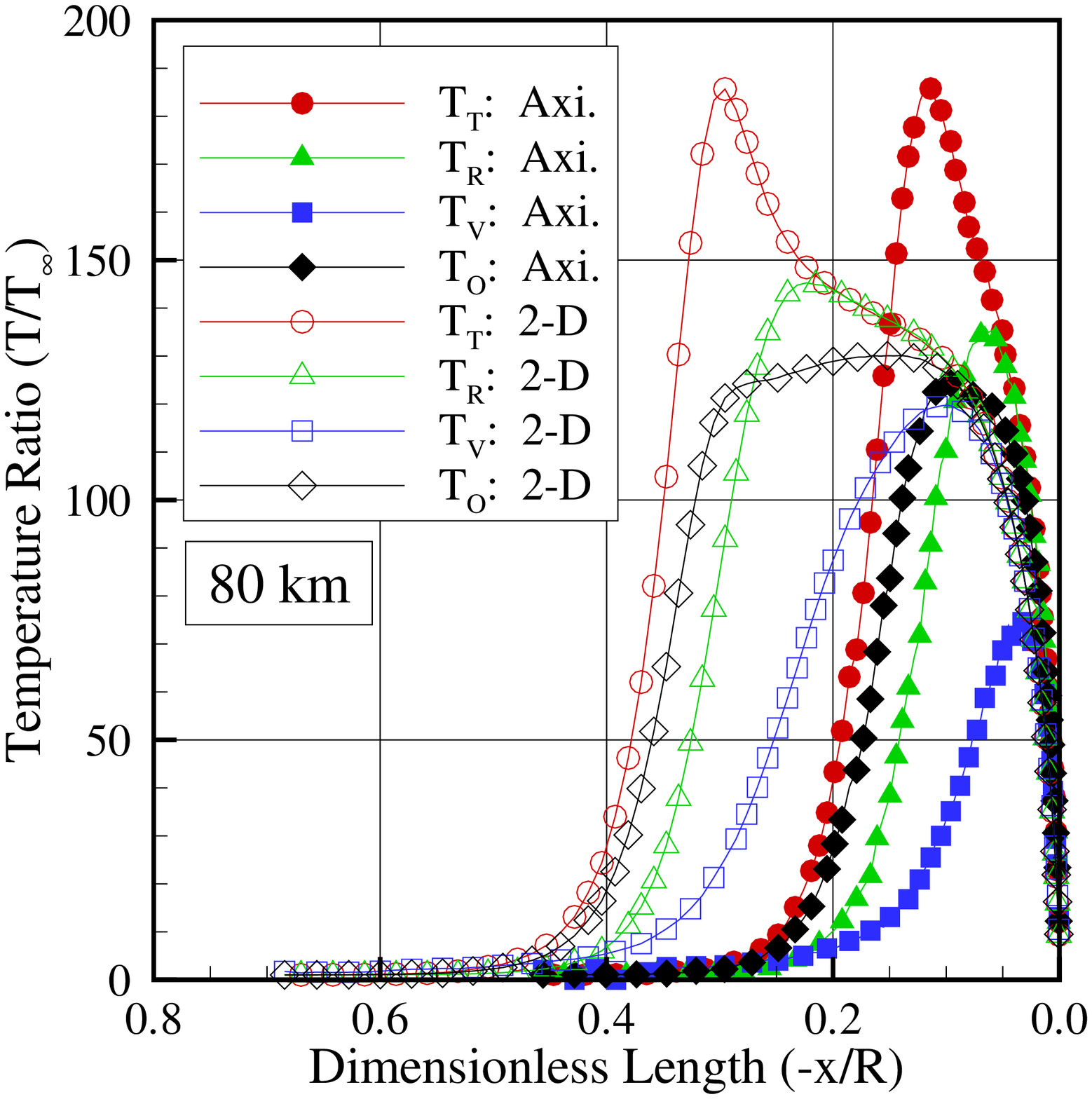}
 \end{center}
 \caption{kinetic temperature ratio ($T/T_{\infty}$) profiles along the stagnation
          streamline in a linear scale for altitudes of (a) 100 km, (b) 90 km, and (c) 80 km.}
 \label{BJoPP02F10}
\end{figure}

\begin{figure}[t!]
 \begin{center}
   \includegraphics[width=7.0cm,height=6.0cm]{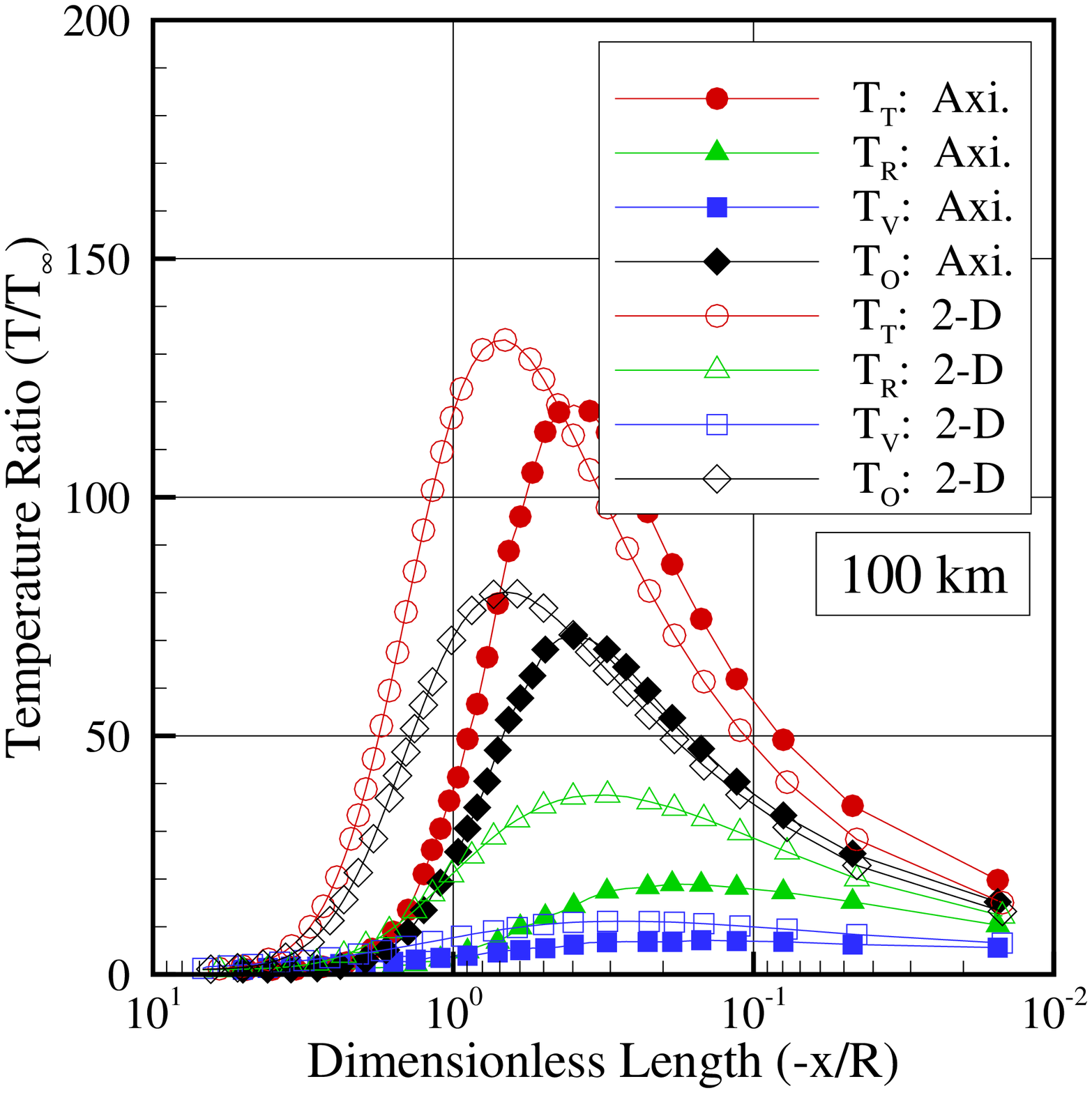}
   \includegraphics[width=7.0cm,height=6.0cm]{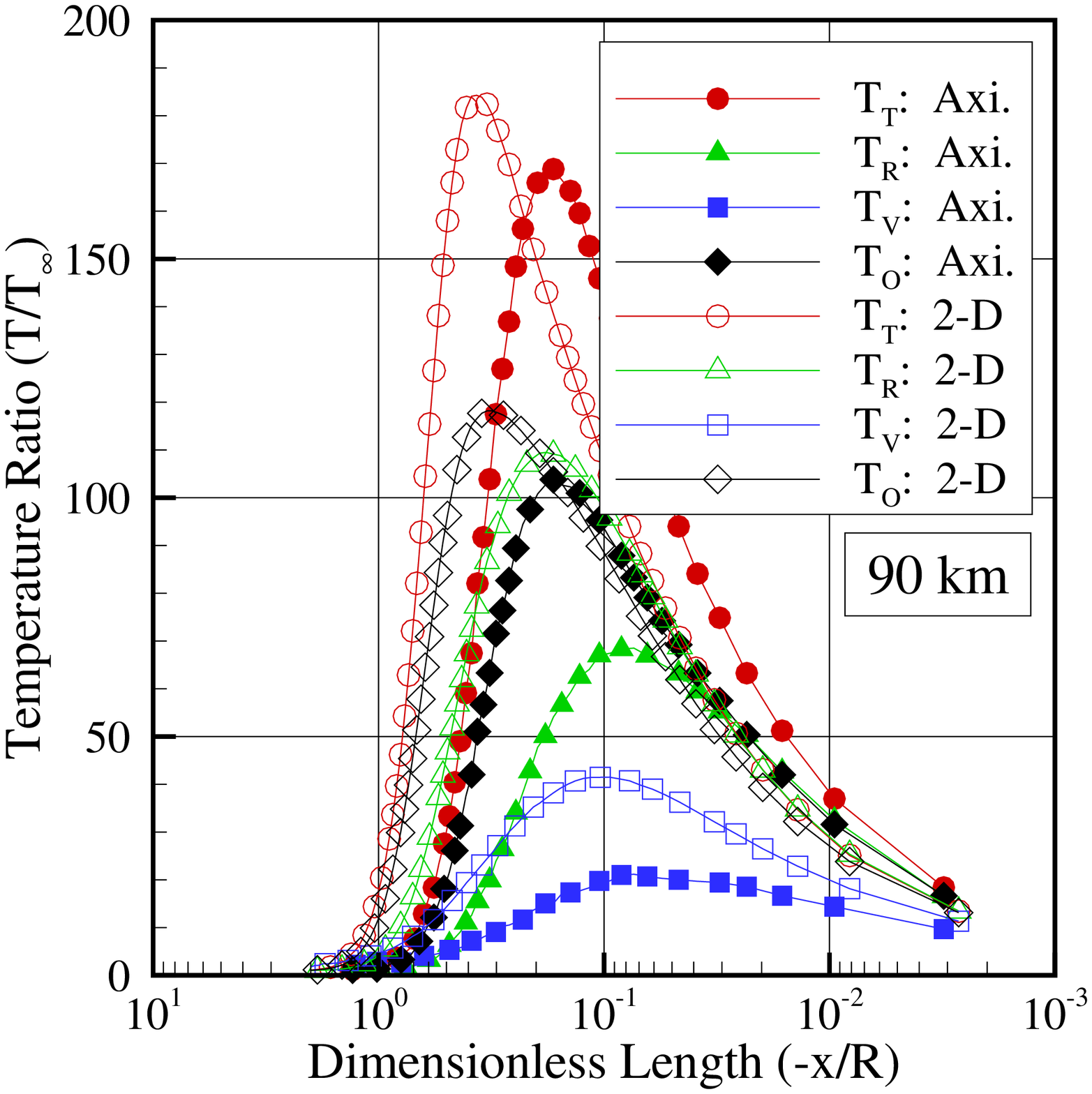}
   \includegraphics[width=7.0cm,height=6.0cm]{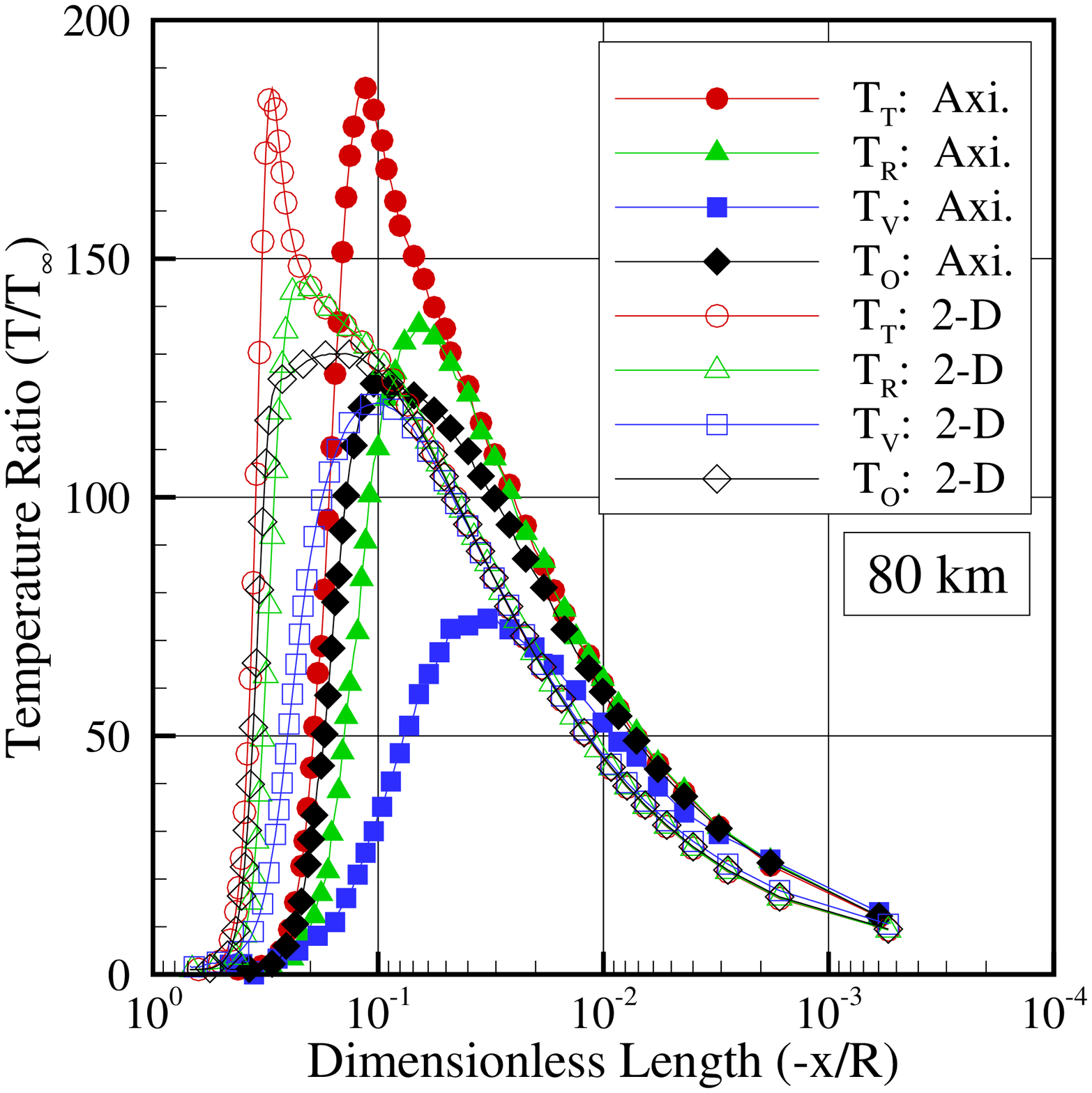}
 \end{center}
 \caption{kinetic temperature ratio ($T/T_{\infty}$) profiles along the stagnation
          streamline in a logarithm scale for altitudes of (a) 100 km, (b) 90 km, and (c) 80 km.}
 \label{BJoPP02F11}
\end{figure}

Representative kinetic temperature profiles along the stagnation streamline are demonstrated in Figs.~\ref{BJoPP02F10} and \ref{BJoPP02F11} for 100, 90 and 80 km of altitude. Figures~\ref{BJoPP02F10}(a-c) exhibit the temperature distribution with a linear scale in the abscissa in order to emphasize the extension of the upstream disturbance, while Figs.~\ref{BJoPP02F11}(a-c) depict the same temperature distribution in a logarithm scale in order to emphasize the temperature behavior at the vicinity of the stagnation point. In addition, in this set of pictures, temperature ratio stands for the translational temperature $T_{T}$, rotational temperature $T_{R}$, vibrational temperature $T_{V}$ or overall temperature $T_{O}$ normalized by the freestream temperature $T_{\infty}$. Also, filled and empty symbols correspond to temperature distributions for axisymmetric and 2-D geometries, respectively. It is apparent from these figures that thermodynamic non-equilibrium occurs throughout the shock layer, as shown by the lack of equilibrium of the translational and internal kinetic temperatures. Thermal non-equilibrium occurs when the temperatures associated with the translational, rotational, and vibrational modes of a polyatomic gas are different from each other. In this context, an overall kinetic temperature $T_{O}$ is defined~\cite{Bird94} for a non-equilibrium gas as the weighted mean of the translational and internal temperature by the following expression,

\begin{equation}
 T_{O} =
 \frac{\zeta_{T}T_{T}+\zeta_{R}T_R+\zeta_{V}T_V}{\zeta_{T}+\zeta_{R}+\zeta_{V}}
 \label{eqn4}
\end{equation}
were $\zeta$ is the degree of freedom and subscript $T$, $R$ and $V$ stand for translation, rotation and vibration, respectively.

The overall kinetic temperature $T_{O}$ is equivalent to the thermodynamic temperature only under thermal equilibrium conditions. It should be emphasized that the ideal gas equation of state does not apply to this temperature in a non-equilibrium situation.

Referring to Figs.~\ref{BJoPP02F10} and~\ref{BJoPP02F11}, in the undisturbed freestream far from the capsule, the translational and internal temperatures have the same value and are equal to the thermodynamic temperature. Approaching the nose of the capsule, the translational temperature rises to well above the rotational and vibrational temperatures and reaches a maximum value that is a function of the altitude. Since a large number of collisions is needed to excite the molecules in the vibrational mode, from the ground state to the upper state, then the vibrational temperature increases much more slowly than rotational temperature. Still further downstream toward the nose of the capsule, the translational temperature decreases and reaches a value on the wall that is above the wall temperature ($T_w/T_\infty \approx 4$) for the 100 km and 90 km of altitude, resulting in a temperature jump. Therefore, for $x/R \approx 0$, one has $T_T \neq T_w$. In contrast, for the 80 km case, the translation, rotation and vibrational temperature are basically the same at the vicinity of the nose, indicating the the thermodynamic equilibrium is achieved.

The substantial rise in the kinetic translational temperature occurred before the density rise, as shown in Figs.~\ref{BJoPP02F06}(a-c). For instance, the translational temperature ratio, $T_{T}/T_{\infty}$, for the axisymmetric case reaches the maximum value of 119.3, 169.2, and 186.0, for altitude of 100, 90 and 80 km, respectively, around a distance of 0.40$R$, 0.18$R$, and 0.12$R$, respectively, from the nose of the capsule, while the density ratio $\rho/\rho_\infty$, for the same stations, is around 2.1, 2.3, and 2.6 for altitude of 100, 90 and 80 km, respectively. The translational kinetic temperature rise for a blunt body results from the essentially bimodal velocity distribution: the molecular sample consisting of mostly undisturbed freestream molecules with the molecules that have been affected by the shock and reflected from the capsule nose. In this scenario, the translational kinetic temperature rise is a consequence of the large velocity separation between these two classes of molecules.

\begin{figure}[t!]
 \begin{center}
   \includegraphics[width=7.0cm,height=6.0cm]{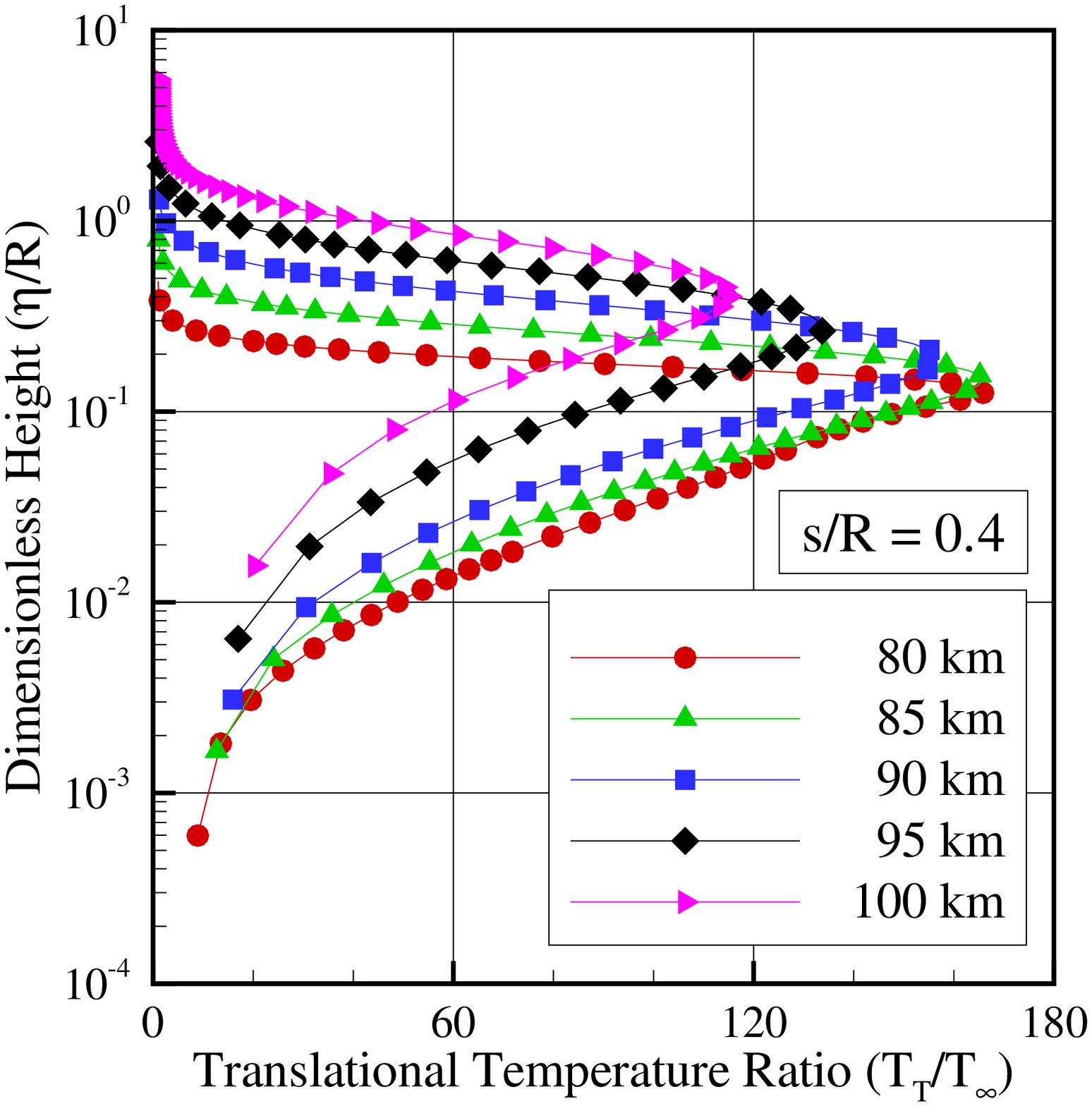}
   \includegraphics[width=7.0cm,height=6.0cm]{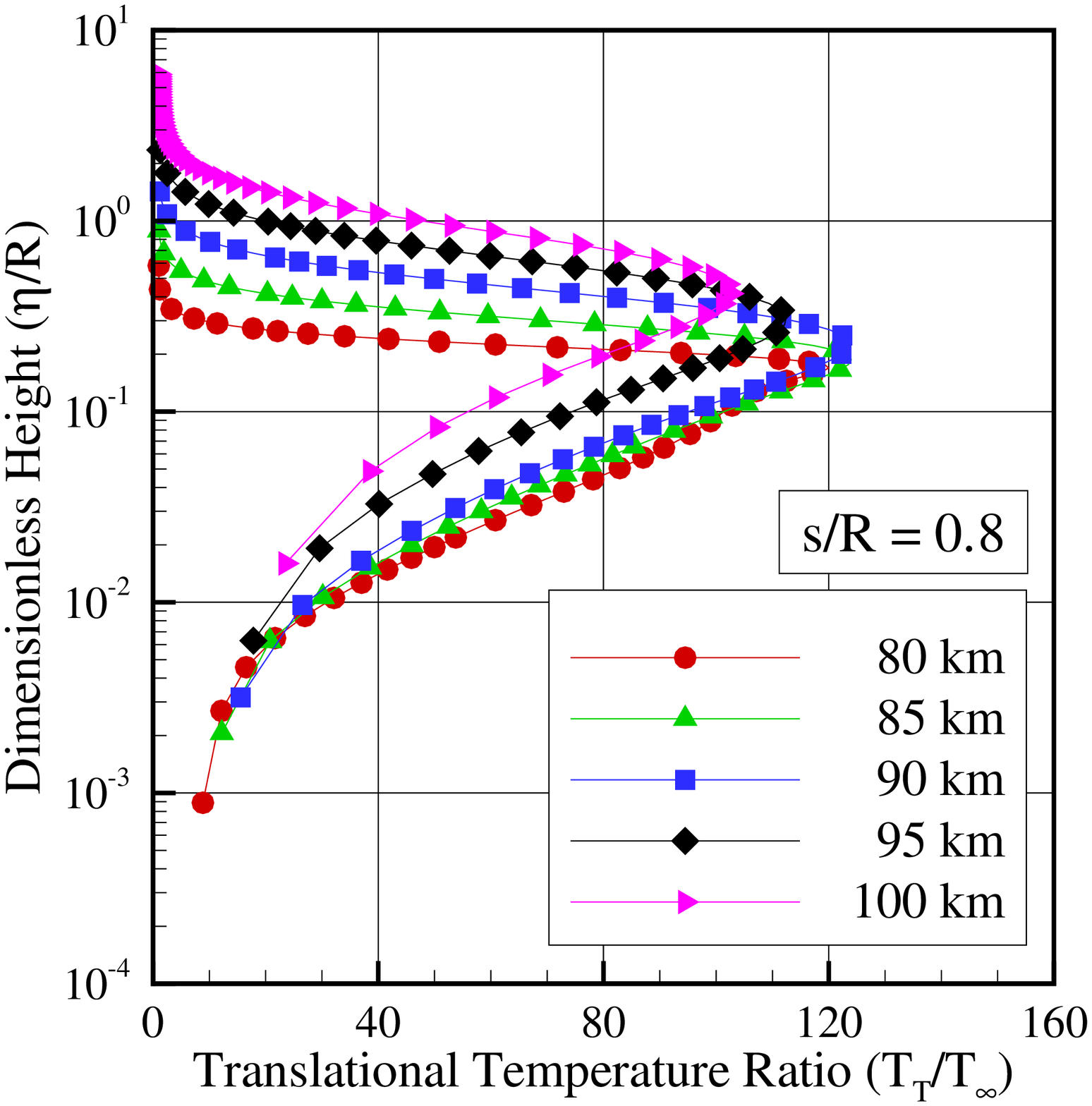}
   \includegraphics[width=7.0cm,height=6.0cm]{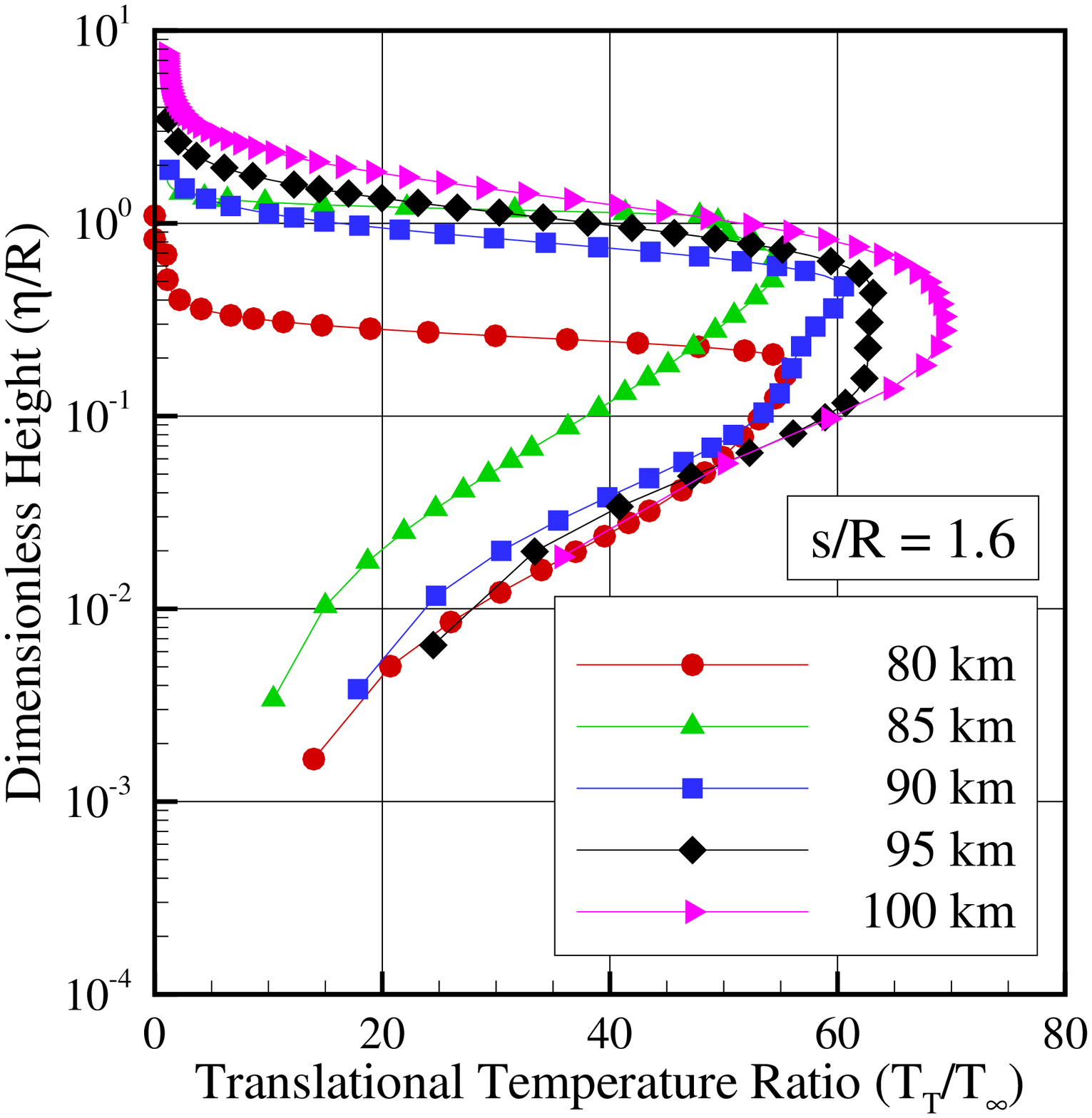}
 \end{center}
 \caption{Kinetic translational temperature ratio ($T_{T}/T_{\infty}$) profiles along the
          capsule surface for sections corresponding to $s/R$ of (a) 0.4, (b) 0.8 and (c) 1.6.}
 \label{BJoPP02F12}
\end{figure}

\begin{figure}[t!]
 \begin{center}
   \includegraphics[width=7.0cm,height=6.0cm]{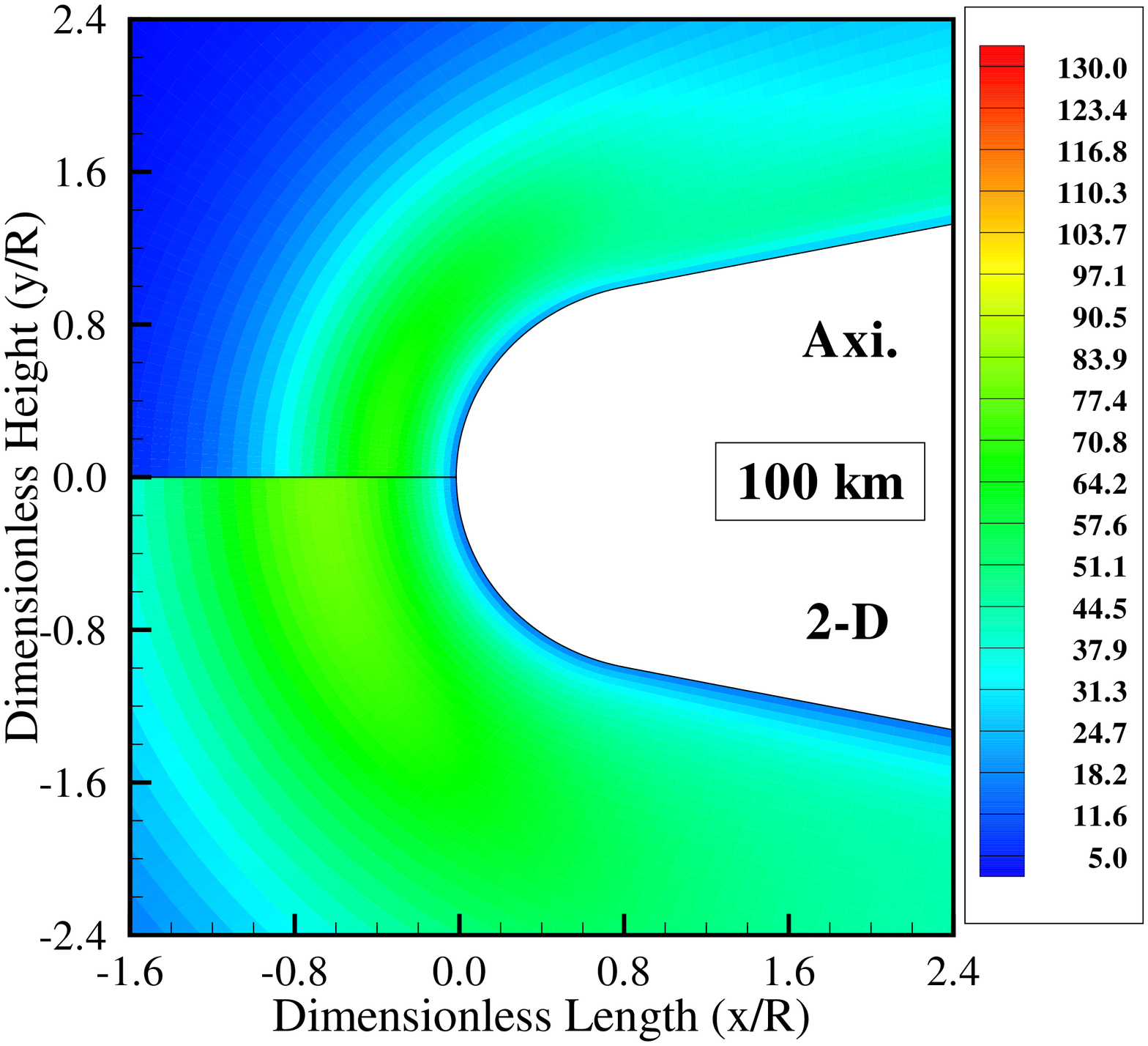}
   \includegraphics[width=7.0cm,height=6.0cm]{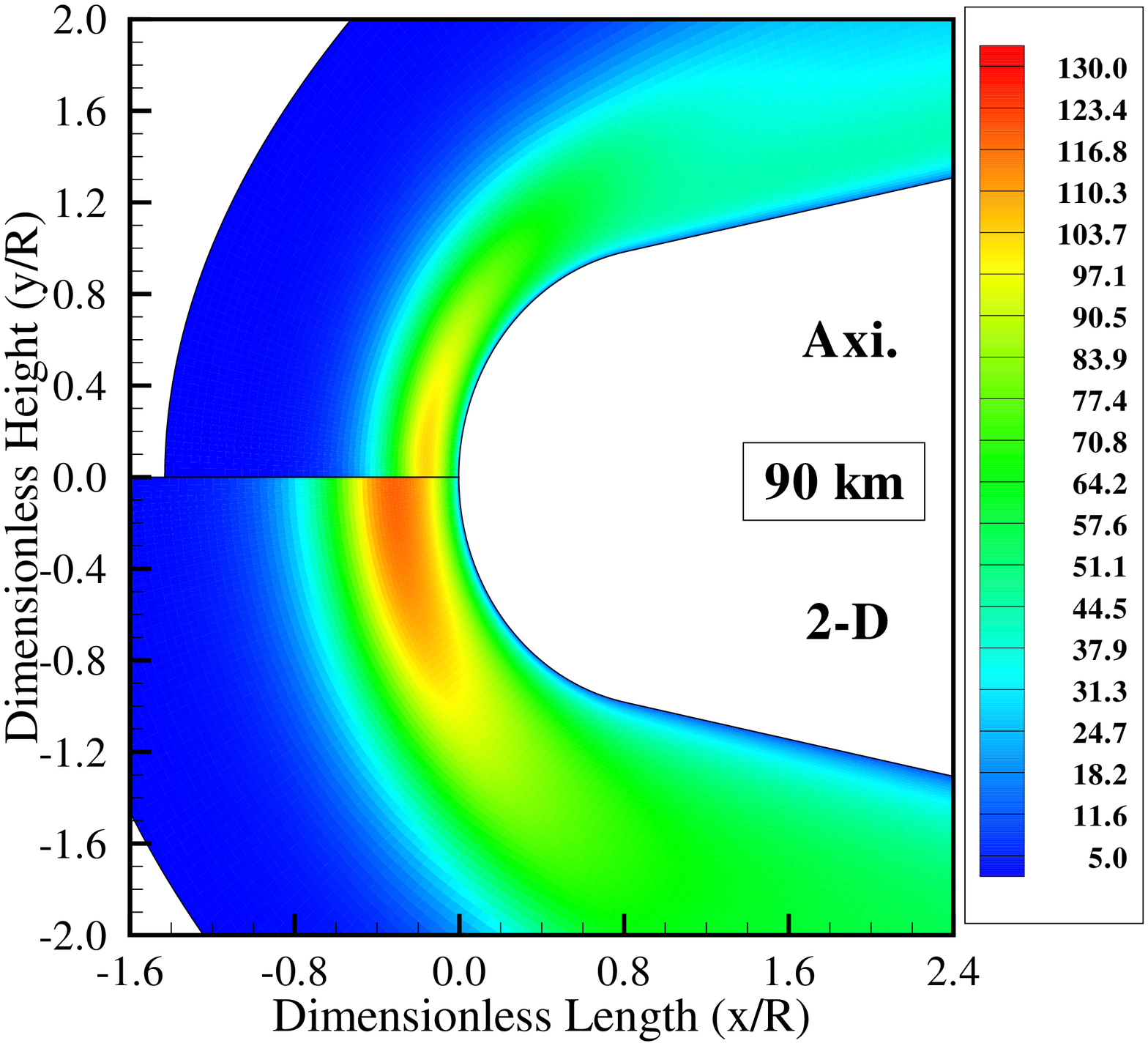}
   \includegraphics[width=7.0cm,height=6.0cm]{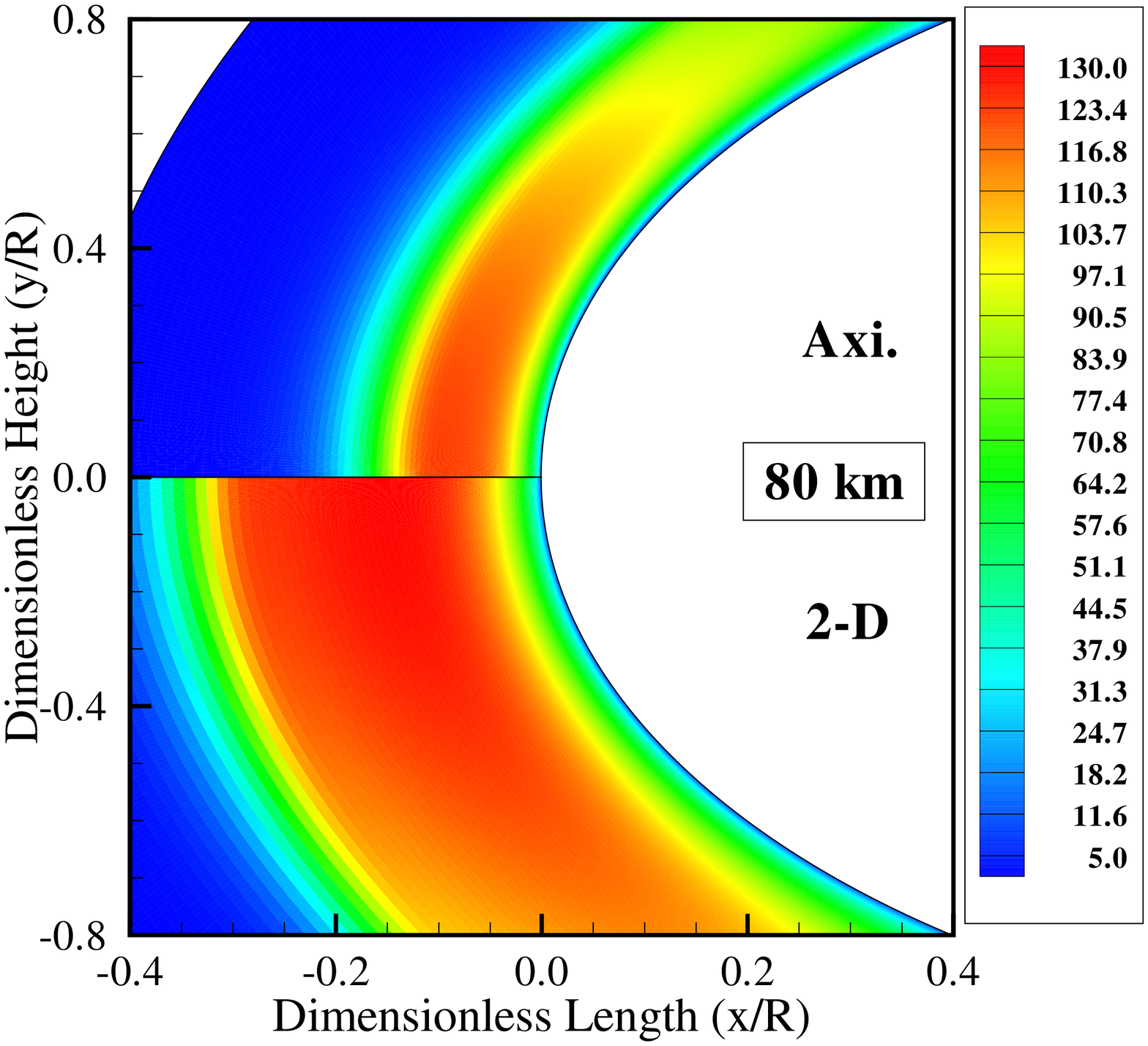}
 \end{center}
 \caption{Overall temperature ratio ($T_{O}/T_{\infty}$) contours at the vicinity of
          the capsule nose for altitude of (a) 100 km, (b) 90 km, and (c) 80 km.}
 \label{BJoPP02F13}
\end{figure}

Particular attention is paid to the kinetic translational temperature in the shock layer. In this respect, the kinetic translational temperature variation is taken normal to the capsule surface at stations corresponding to the dimensionless arc length $s/R$ of 0.4, 0.8, and 1.6. Figures~\ref{BJoPP02F12}(a-c) depict the kinetic translational temperature profiles at the considered positions normal to the capsule surface along the $\eta$-axis for the altitudes investigated. According to these figures, it is observed that the downstream evolution of the flow displays a smearing tendency of the shock wave due to the displacement of the maximum value for the kinetic translational temperature. Also, it may be recognized from the temperature distribution in Figs.~\ref{BJoPP02F12}(a-c) that significant changes in the translational temperature profiles along the spherical nose occur within a layer adjacent to the capsule surface of a few nose radius $R$ for the altitude range investigated.

In order to bring out important features of the rarefaction effects, particular attention is paid to the overall kinetic temperature at the vicinity of the capsule nose. In this scenario, overall kinetic temperature contours, normalized by the freestream temperature $T_{\infty}$, are plotted in Figs.~\ref{BJoPP02F13}(a-c) for altitude of 100, 90 and 80 km, respectively. In this set of plots, the length $x$ and height $y$ are normalized by the nose radius $R$. In addition, the upper half part of the plots represents the axisymmetric case and the lower half part stands for the 2-D case. Also, in order to emphasize points of interest, a different scale is used in the abscissa and ordinate axis of these plots. This set of plots clearly illustrates the differences in the temperature distribution produced with decreasing the altitude. Also, substantial differences in the overall kinetic temperature are observed when the axisymmetric case is compared to the 2-D case. For instance, the overall kinetic temperature core is located closer to the capsule nose for the axisymmetric case than that for the 2-D case, showing that the upstream disturbance due to the presence of the capsule becomes more pronounced for the 2-D case.

\subsection{Computational and Experimental Comparisons}
\label{sec:6.5}
\begin{figure}[t!]
 \begin{center}
   \includegraphics[width=7.0cm,height=6.0cm]{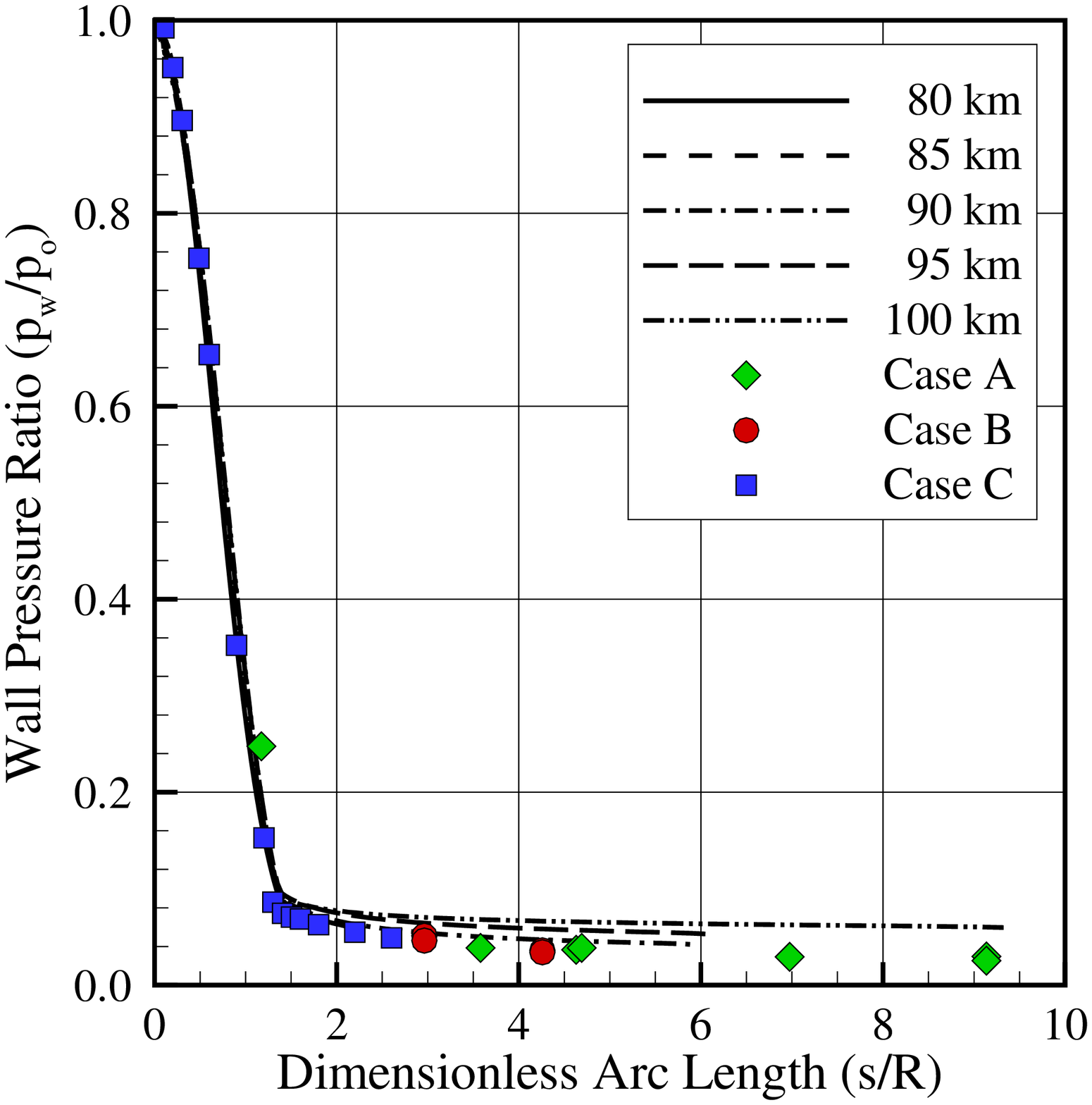}
   \includegraphics[width=7.0cm,height=6.0cm]{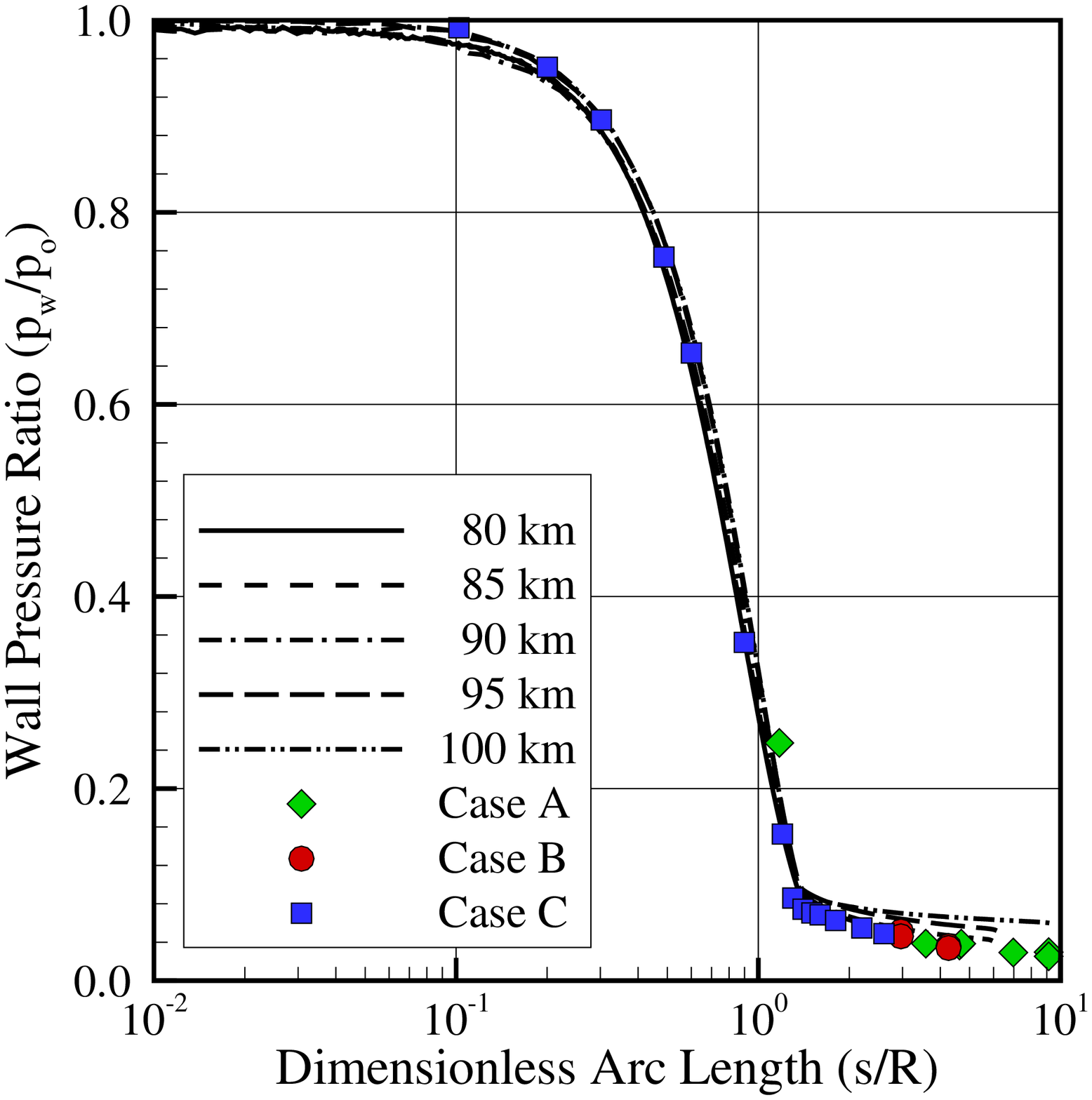}
 \end{center}
\caption{Comparison of wall pressure ratio ($p_w/p_o$) distribution along the body surface in a (a) linear scale and in a (b) logarithm scale.}
\label{BJoPP02F14}
\end{figure}

Having completed the discussion on the primary properties along the stagnation streamline and adjacent to the capsule surface, the attention now is turned to the comparison of the results. Nowadays, this is not a simple task, given the small number of studies on SARA capsule. As a result, the alternative is the comparison of the simulation data with those obtained experimentally with a similar geometry, i.e., a body defined by a spherical nose with a conical afterbody. In view of this difficult, Figs.~\ref{BJoPP02F14}(a-b) display the pressure acting on the body surface. In this set of figures, the wall pressure $p_w$ is normalized by the pressure at the stagnation point $p_o$, and the arc length $s$ along the body surface is normalized by the nose radius $R$. In addition, Fig.~\ref{BJoPP02F14}(a) illustrates the pressure ratio $p_w/p_o$ distribution with a linear scale in the abscissa, while Fig.~\ref{BJoPP02F14}(b) depicts the same pressure ratio distribution in a logarithm scale in order to emphasize the pressure ratio behavior at the vicinity of the stagnation point.

Based on this set of figures, Case A~\cite{Griffith} refers to experimental data obtained in the hypervelocity tunnel of the Arnold Engineering Development Center-von K\'{a}rm\'{a}n Facility~(AEDC-VKF) for a $9$-degree half-angle spherically blunted cone. In addition, N$_2$ was used as the working fluid, freestream Mach number near $19$ and freestream Reynolds number between 8,000 and 15,000/in. Case B~\cite{Wilkinson} corresponds to experimental data obtained from Cornell Aeronautical Laboratory~(CAL) shock tunnel for a $9$-degree half-angle spherically blunted cone, with air as a working fluid, freestream Mach number near $14.5$ and freestream Reynolds number between 5,700 and 12,000/in. Finally, Case C~\cite{Machell} represents a spherical nosed cone with semi-vertex angle of 10 degrees tested in a hypersonic wind tunnel at a Mach number of 5.8, and freestream Reynolds number in the range of 0.97--2.38$\times10^{5}$/in. Some characteristics of the models for Case A, B and C are tabulated in Table~\ref{tab4}.

\begin{table}[t!]
\caption{Flow and geometric conditions for the spherically blunted cone models}
\begin{center}
\renewcommand{\arraystretch}{1.}
\setlength\tabcolsep{3.5pt}
\begin{tabular}{cccc}
\hline\hline\noalign{\smallskip} Case & A~\cite{Griffith} & B~\cite{Wilkinson} & C~\cite{Machell} \\
\noalign{\smallskip} \hline \noalign{\smallskip}
$M_\infty$     & $\sim19$           & $\sim14.5$           & $5.8$\\
$Re_\infty$/in & $9-15\times10^{3}$ & $5.7-12\times10^{3}$ & $1.910\times10^{5}$\\
$R/R_B$        & $0.3$                & $0.3$              & $0.8$\\
$\theta$       & $9$                  & $9$                & $10$\\
\hline\hline
\end{tabular}
\label{tab4}
\end{center}
\end{table}

According to Figs.~\ref{BJoPP02F14}(a-b), it is clearly seen that the wall pressure ratio for the present work agreed reasonably well with that obtained experimentally. It is important to remark that the shape of the pressure distributions as described in nondimensional coordinates is independent of the spherical nose radius and of the Reynolds number over the range investigated.

\section{Concluding Remarks}
\label{sec:7}
Computations of a rarefied hypersonic flow over the SARA capsule have been performed by using the Direct Simulation Monte Carlo method. The calculations provided information concerning the nature of the flowfield structure about the primary properties at the vicinity of the nose and immediately adjacent to the afterbody surface by considering planar two-dimensional and axisymmetric geometries.

Effects of rarefaction on the velocity, density, pressure, and temperature for a representative range of parameters were investigated. The altitude varied from 100 to 80 km, which corresponded to Knudsen numbers $Kn_{R}$ from 0.4615 to 0.0115, Reynolds number $Re_R$ from 92 to 15249, and Mach number $M_\infty$ from 27 to 29. Therefore, cases considered in this study covered the hypersonic flow in the transitional flow regime.

It was found that changes on the altitude as well as on the capsule geometry disturbed the flowfield around the capsule, as expected. The domain of influence decreased by decreasing the altitude. In addition, the domain of influence for the axisymmetric case was smaller than that for 2-D case. Moreover, the extent of the flowfield disturbance along the stagnation streamline due to changes on the altitude was significantly different for each one of the primary flow properties. The analysis showed that the domain of influence for temperature is larger than that observed for pressure, density and velocity.

The present document has described an initial investigation of a high-altitude and low-density flow over the SARA capsule. Although this investigation has taken into account a representative range of altitudes, improvements are still desirable to a realistic capsule design. Since no database exist for such a design and since no appropriate flight data are available at such high entry velocities and conditions, the use of computational methods is essential. Since many of these issues occur in the rarefied portion of the trajectory, accurate DSMC analysis are especially important.

\section{Acknowledgments}

The author would like to thank the financial support provided by FAPESP (Funda\c{c}\~{a}o de Amparo a Pesquisa do Estado de S\~{a}o Paulo) under Grant No. 2008/03878-9.

\section*{Appendix}

This section focuses on the analysis of the influence of the cell size, the time step, and the number of molecules per computational cell on the surface properties in order to achieve grid independent solutions. Heat transfer, pressure and skin friction coefficients were selected in order to elucidate the requirements posed for the grid sensitivity study. As an illustrative example, the analysis presented in this section is limited to the capsule at an altitude of 90 km. The same procedure was employed for the other cases.

\begin{figure}[t!]
 \begin{center}
   \includegraphics[width=7.0cm,height=6.0cm]{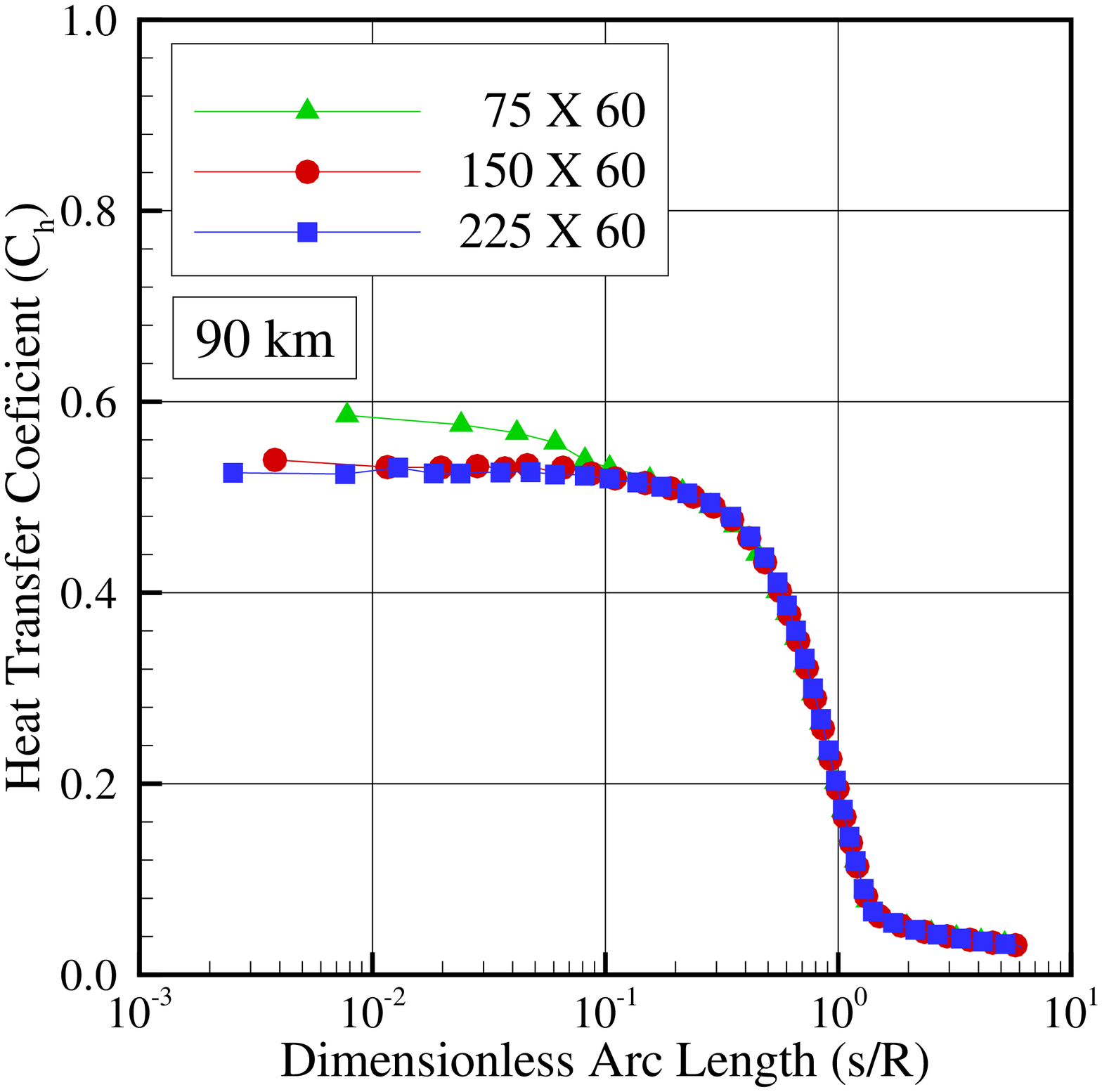}
   \includegraphics[width=7.0cm,height=6.0cm]{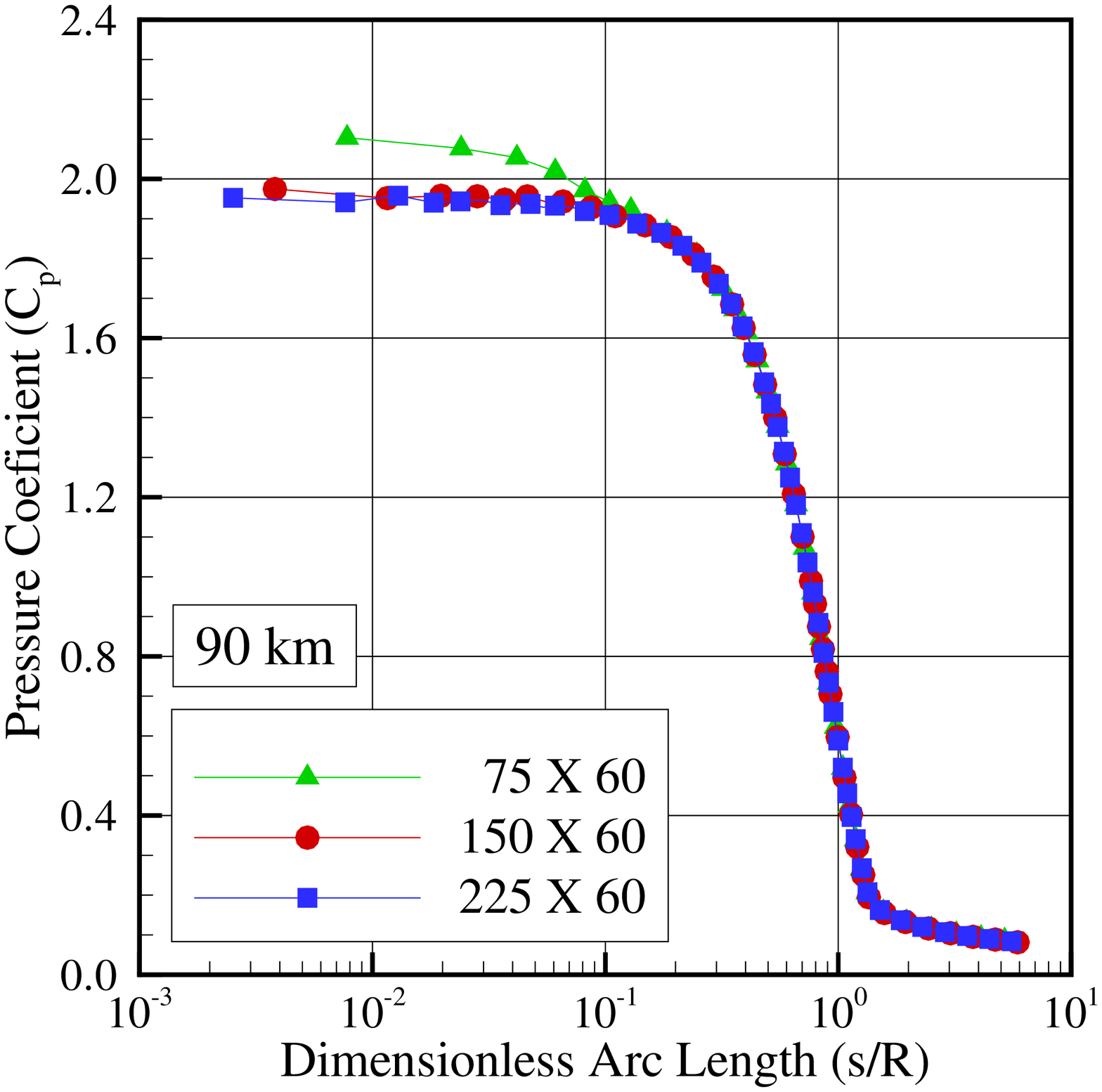}
   \includegraphics[width=7.0cm,height=6.0cm]{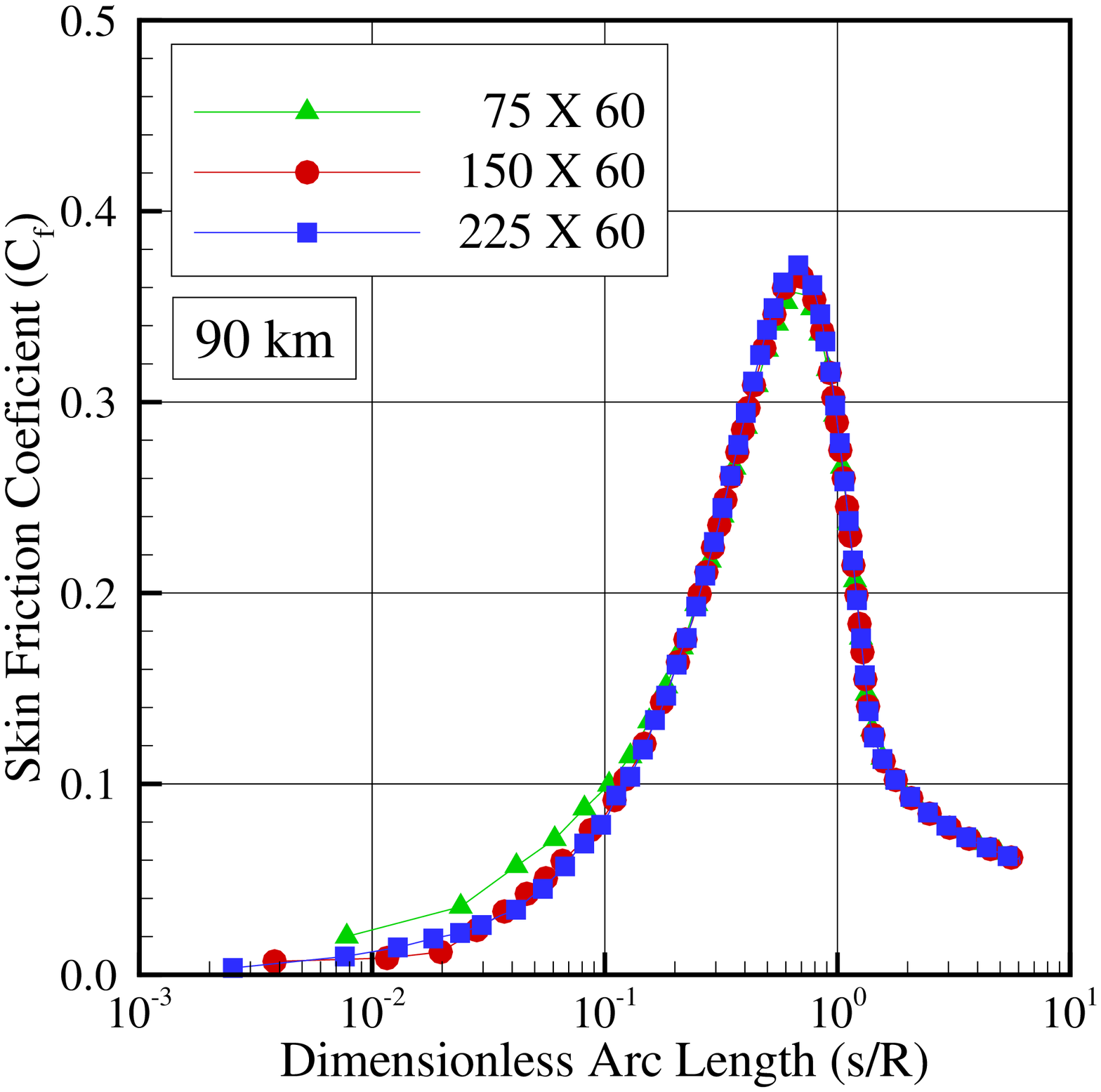}
 \end{center}
 \caption{Effect of altering the cell size along the $\xi$-direction on the coefficients of (a) heat transfer $C_h$, (b) pressure $C_p$, and skin friction $C_f$.}
 \label{BJoPP02F15}
\end{figure}

A grid independence study was made with three different structured meshes in each coordinate direction. The effect of altering the cell size in the $\xi$-direction was investigated with grids of 75 (coarse), 150 (standard) and 225 (fine) cells, and 60 cells in the $\eta$-direction. In addition, each grid was made up of non-uniform cell spacing in both directions. The effect of changing the cell size in the $\xi$-direction is illustrated in Figs.~\ref{BJoPP02F15}(a-c) as it impacts the coefficients of heat transfer $C_h$ ($\equiv 2q_w/\rho_\infty V_\infty^3$), pressure $C_p$ ($\equiv 2(p_w-p_\infty)/\rho_\infty V_\infty^2$), and skin friction $C_f$ ($\equiv 2\tau_w/\rho_\infty V_\infty^2)$. The comparison shows that the calculated results are rather insensitive to the range of cell spacing considered.

\begin{figure}[t!]
 \begin{center}
   \includegraphics[width=7.0cm,height=6.0cm]{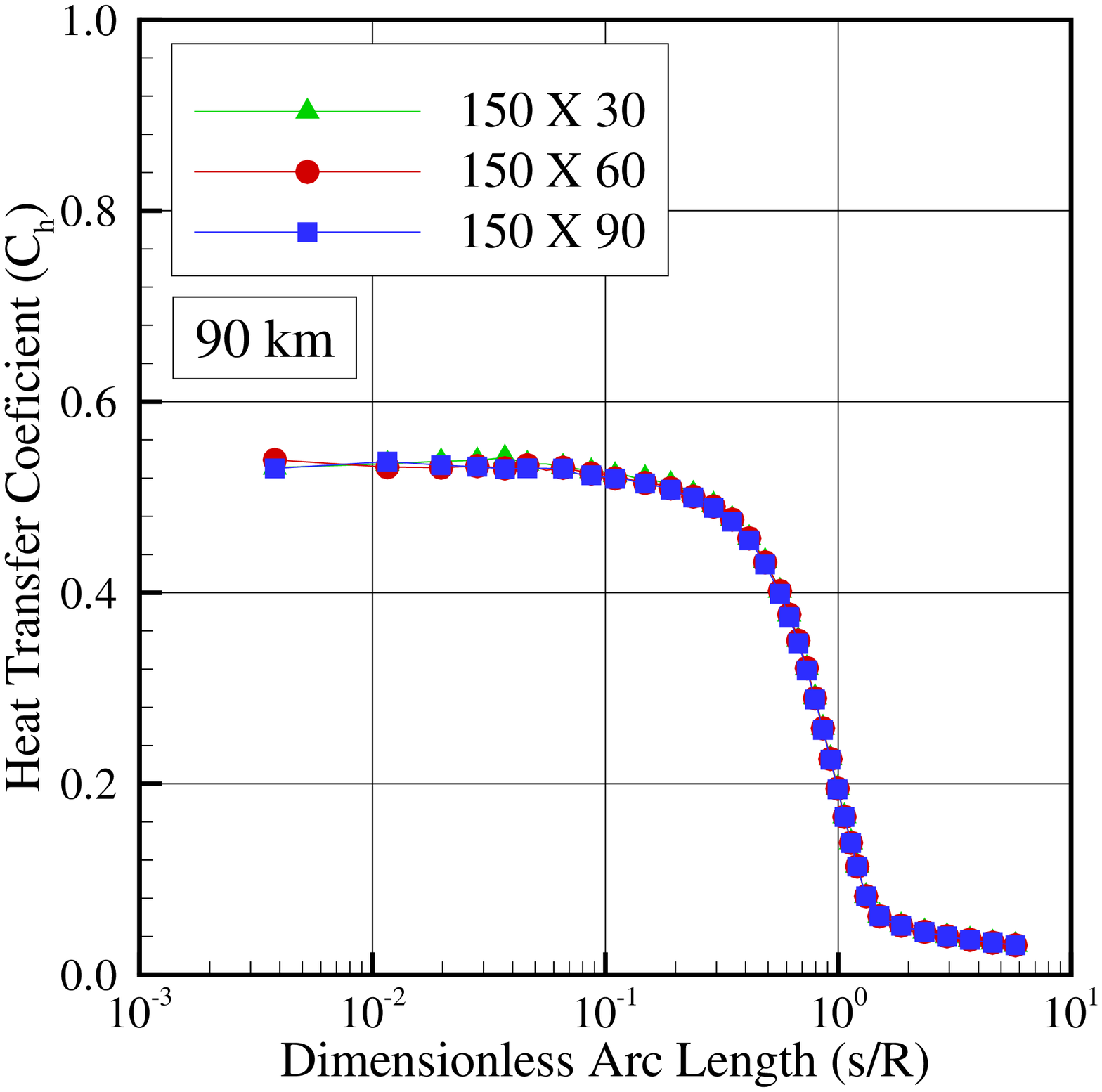}
   \includegraphics[width=7.0cm,height=6.0cm]{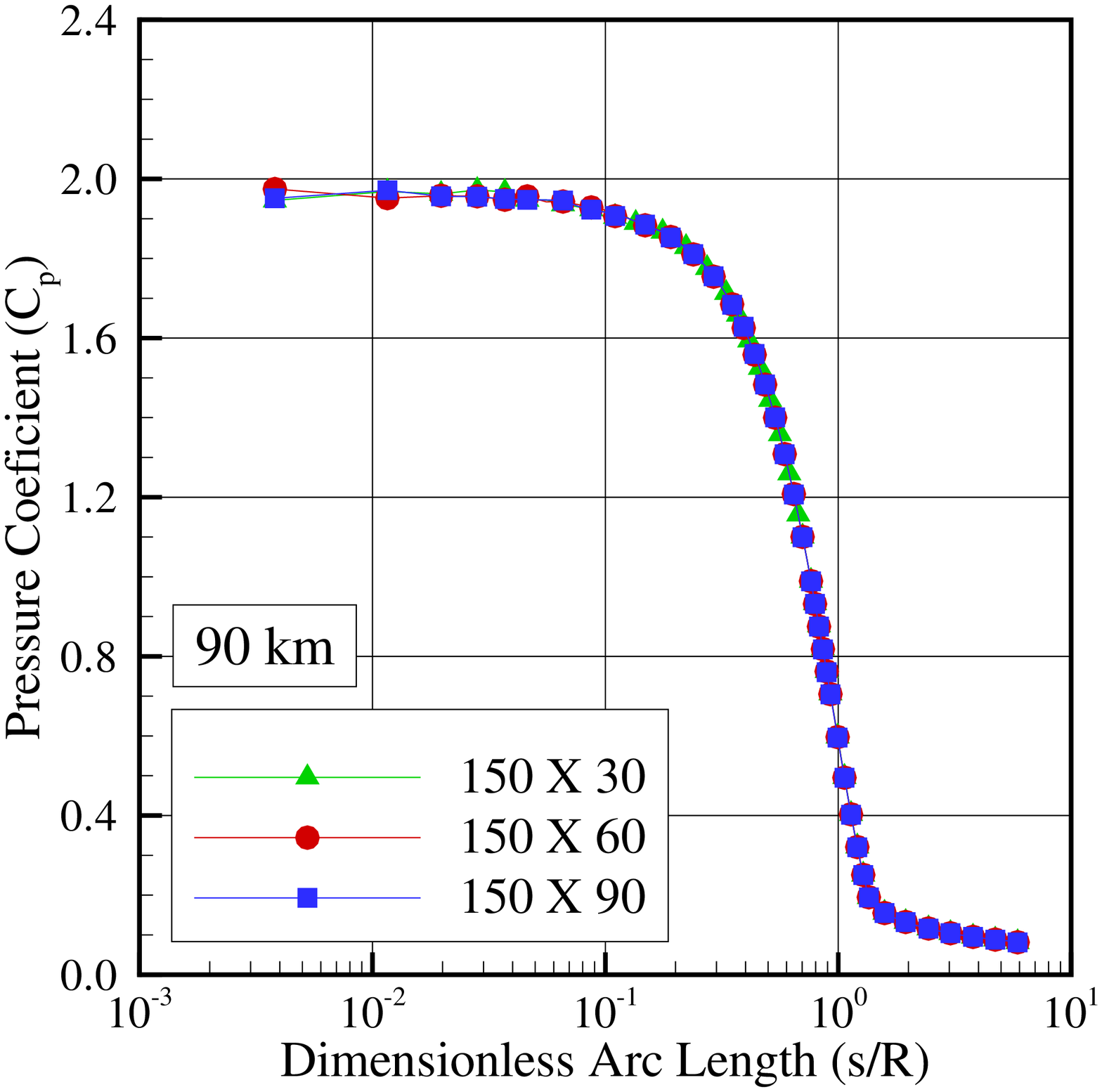}
   \includegraphics[width=7.0cm,height=6.0cm]{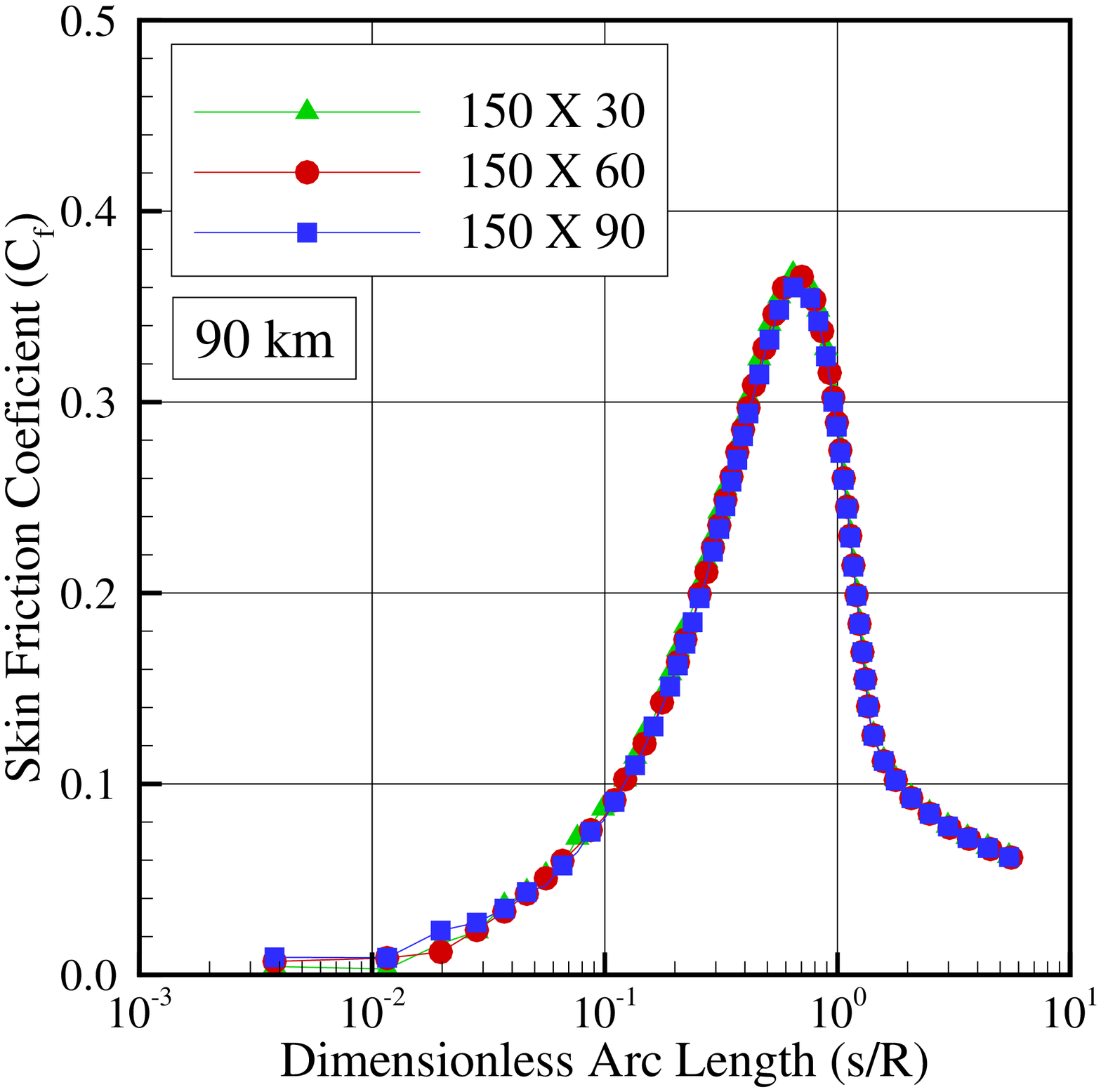}
 \end{center}
 \caption{Effect of altering the cell size along the $\eta$-direction on the coefficients of (a) heat transfer $C_h$, (b) pressure $C_p$, and skin friction $C_f$.}
 \label{BJoPP02F16}
\end{figure}

In analogous fashion, an examination was made in the $\eta$-direction with grids of 30 (coarse), 60 (standard) and 90 (fine) cells, and 150 cells in the $\xi$-direction. From the total number of cells in the $\xi$-direction, 60 cells are located along the spherical nose and 90 cells distributed along the conical afterbody surface. In addition, each grid was made up of non-uniform cell spacing in both directions. The sensitivity of the cell size variations in the $\eta$-direction on the heat transfer, pressure and skin friction coefficients is displayed in Figs.~\ref{BJoPP02F16}(a-c). Results for the three grids are basically the same, an indication that the standard grid was rather insensitive to the range of cell spacing considered. Therefore, the standard grid, 120~$\times$~110 cells, for the 90 km case, is essentially an independent grid.

One of the fundamental requirements in the DSMC method is that the time step $\Delta t$ must be smaller than the local mean collision time. This requirement is related to the important assumption employed in the DSMC method, i.e., the molecular movement can be decoupled from the intermolecular interactions in a dilute gas if a sufficient small time step is used. In addition, in order to maintain a uniform distribution of simulated particles in the entire computational domain, a different time step $\Delta t$ and scaling factor $F_N$ can be obtained for each cell. $F_N$ is the number of real particles represented by one single simulated particle. As a result, the DSMC efficiency increases, and the computational effort is balanced within the simulated domain. It is worthwhile to highlight that although the time step $\Delta t$ and scaling factor $F_N$ vary among the cells, the ratio $F_N/\Delta t$ must be the same in the entire domain. This requirement assures that the ``mass flux'' across the cell boundaries is conserved.

With this perspective in mind, the following procedure is followed: (1) a computational grid is generated based on freestream conditions; (2) $\Delta t$ and $F_N$ values are defined for each cell according to the DSMC requirements and subject to the condition that $F_N/\Delta t$ has the same value in every cell; (3) the parameters $\Delta t$ and $F_N$ are iteratively modified as the flow evolves within the simulated domain until each cell contains, on average, the desired number of simulated particles; (4) for the entire flowfield, all DSMC requirements are verified, i.e., cell size smaller than the local mean free path, the time step smaller than the time related to the local collision frequency and a number of molecules around 20-30 molecules. If within any cell these conditions are not satisfied the grid adaptation procedure, steps (1), (2) and (3), is restarted for a more appropriate spatial discretization.

In doing so, $\Delta t$ and $F_N$ will be different for coarse, standard, and fine grids. By considering the standard grid for the 90 km case, a total of 13200 cells, the time step $\Delta t$ changed from 4.5979$\times10^{-10}$ to 1.3953$\times10^{-5}$, and the scaling factor $F_N$ changed from 5.0601$\times10^{+11}$ to 1.5355$\times10^{+16}$.

\begin{figure}[t!]
 \begin{center}
   \includegraphics[width=7.0cm,height=6.0cm]{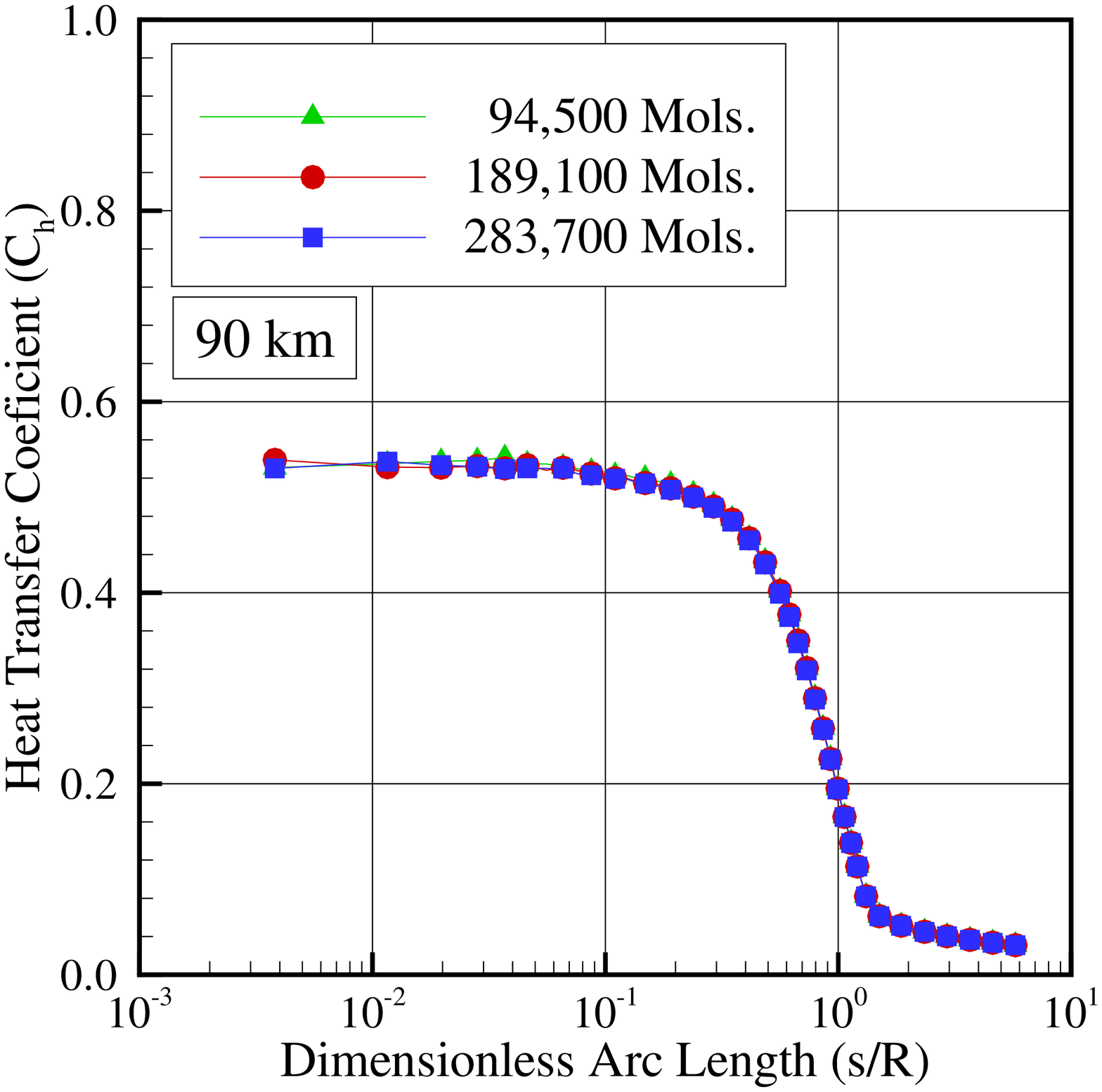}
   \includegraphics[width=7.0cm,height=6.0cm]{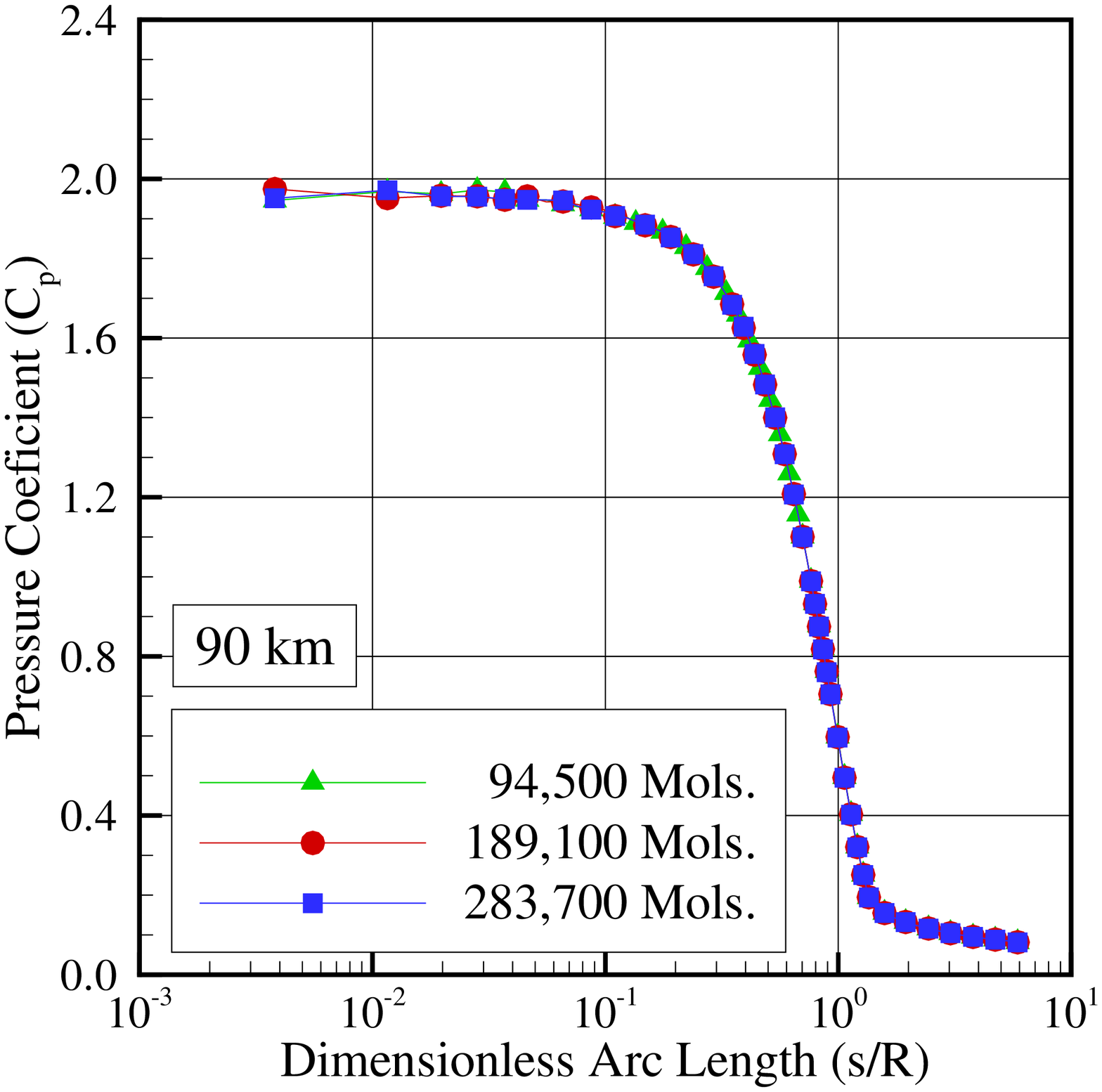}
   \includegraphics[width=7.0cm,height=6.0cm]{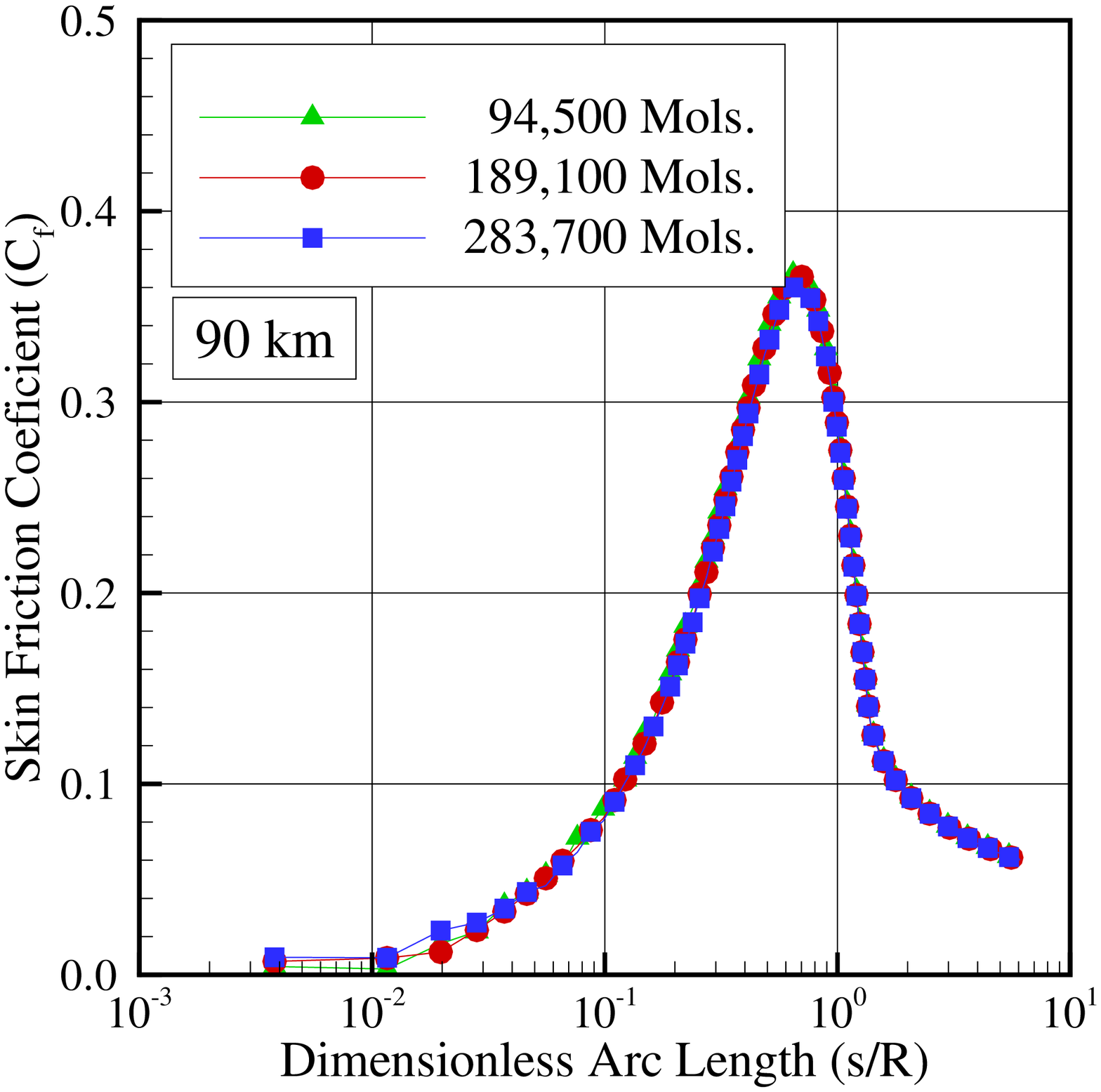}
 \end{center}
 \caption{Effect of altering the number of molecules on the coefficients of (a) heat transfer $C_h$, (b) pressure $C_p$, and skin friction $C_f$.}
 \label{BJoPP02F17}
\end{figure}

In a second stage of the grid independence investigation, a similar examination was made for the number of molecules. The standard grid for the 90 km case, 120~$\times$~110 cells, corresponds to, on average, a total of 189,100 molecules. Two new cases using the same grid were investigated. These two new cases correspond to 94,500 and 283,700 molecules in the entire computational domain. The influence on the heat transfer, pressure and skin friction coefficients due to variations in the number of cells is depicted in Figs.~\ref{BJoPP02F17}(a-c). As the three cases presented approximately the same results, hence the standard grid with a total of 189,100 molecules is considered enough for the computation of the aerodynamic surface quantities.

\begin{figure}[t!]
 \begin{center}
   \includegraphics[width=7.0cm,height=6.0cm]{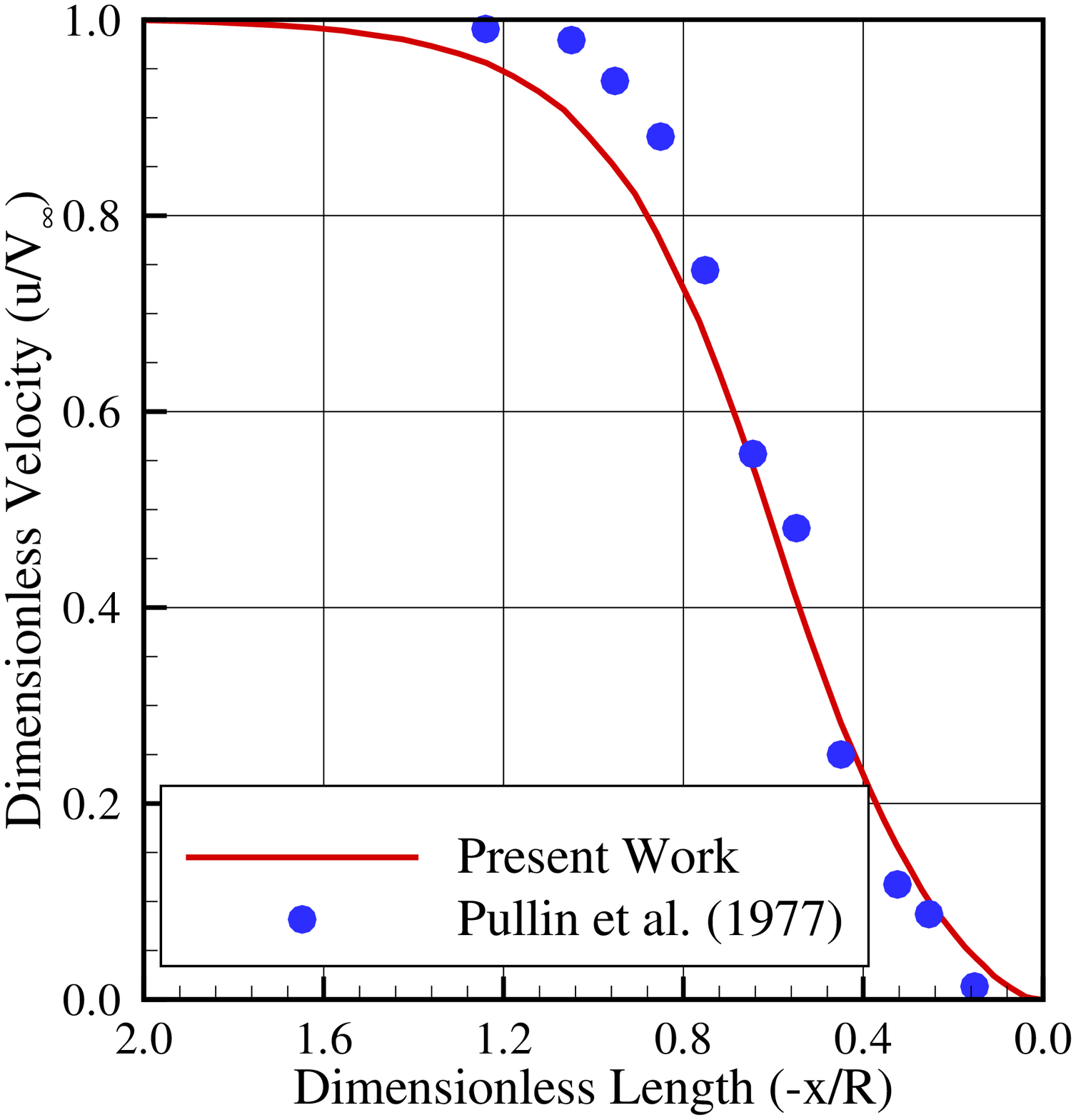}
   \includegraphics[width=7.0cm,height=6.0cm]{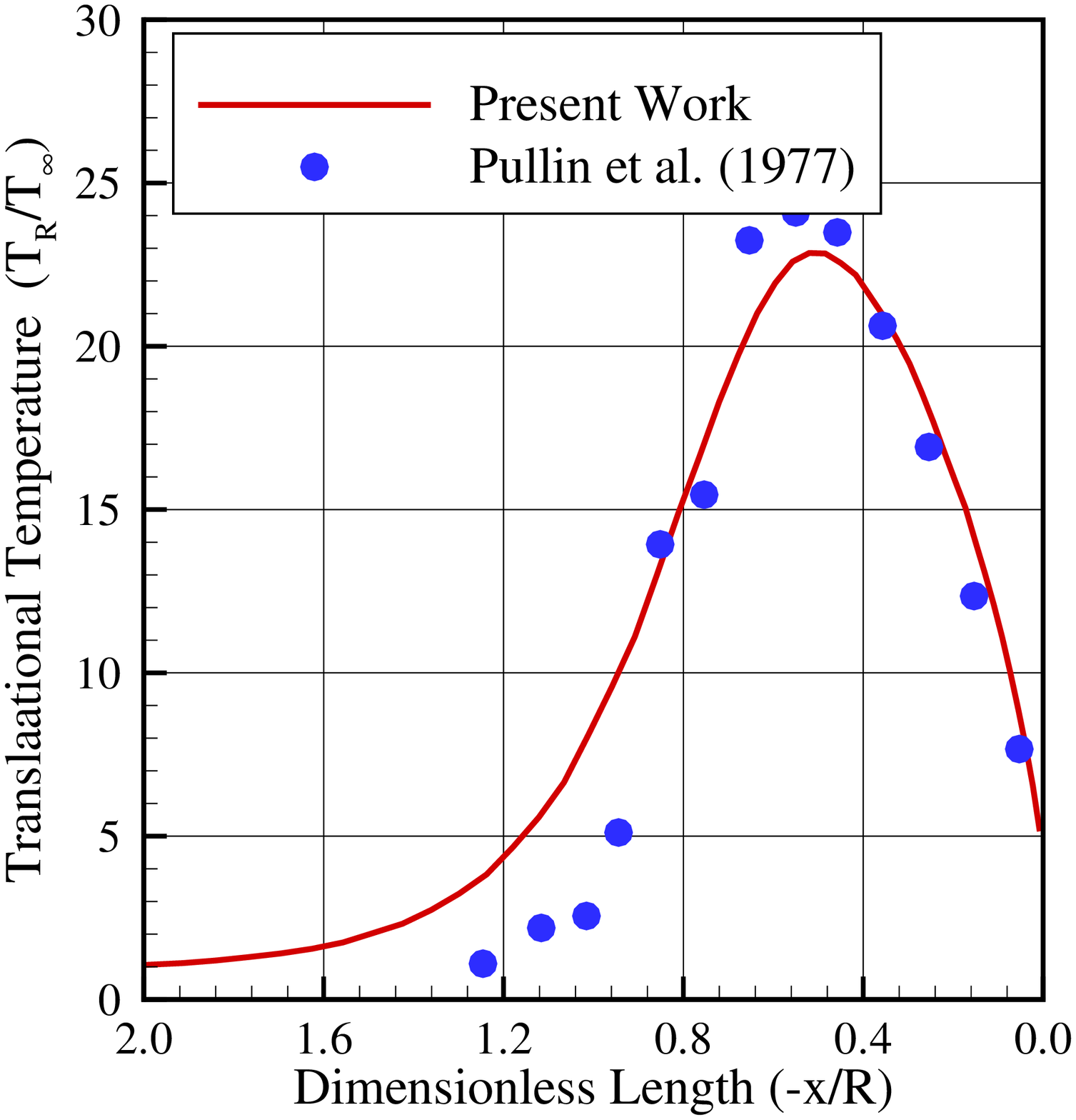}
   \includegraphics[width=7.0cm,height=6.0cm]{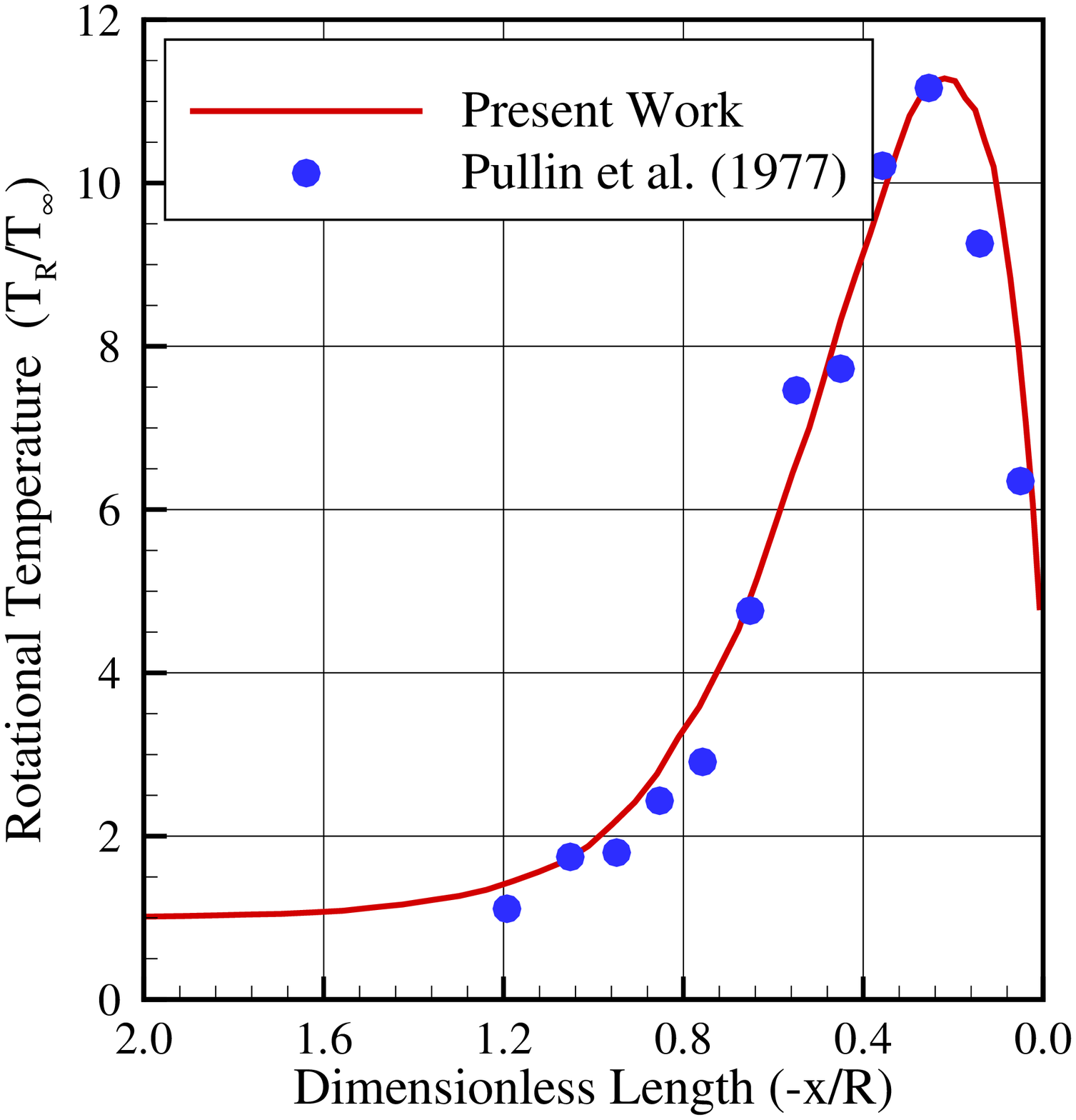}
 \end{center}
 \caption{Distribution of (a) normal velocity, (b) translational temperature, and (c) rotational temperature along the stagnation streamline for the flat-ended cylinder.}
 \label{BJoPP02F18}
\end{figure}

As part of the validation process, the axisymmetric version of this DSMC code was applied to a flat-ended circular cylinder in a rarefied hypersonic flow. Results for velocity, translational temperature, and rotational temperature distributions along the stagnation streamline were presented and compared with those obtained from another established DSMC code~\cite{Pullin} based in a code-to-code comparison. Since this comparison was published elsewhere~\cite{Santos}, details will be kept to a minimum and the discussion restricted to the significant conclusions.

In the computational solution, it was assumed that the flat-ended circular cylinder is immersed in a uniform stream flowing parallel to the cylinder itself. The cylinder is modeled with a frontal-face radius $R$ of 0.0185 m, which corresponds to a 6$\lambda_{\infty}$, and a total length $L$ of 30$\lambda_{\infty}$, where $\lambda_{\infty}~=~3.085\times10^{-3}$ stands for the freestream mean free path. The freestream velocity $V_{\infty}$ is assumed to be constant at 2694 m/s, which corresponds to a freestream Mach number $M_{\infty}$ of 10. The wall temperature $T_{w}$ is assumed constant at 570 K. This temperature is chosen in order to correspond to the temperature ratio $T_{w}/T_{\infty}$ of 3.15 assumed by~\cite{Pullin}.

Normal velocity, translational and rotational temperature profiles along the stagnation streamline are illustrated in Figs.~\ref{BJoPP02F18}(a-c), respectively. In this set of plots, the normal velocity $u$ is normalized by the velocity $V_{\infty}$, the rotational $T_R$ and translational $T_T$ temperatures are normalized by the freestream temperature $T_\infty$. In addition, the distance $x$ upstream the cylinder is normalized by the frontal-face radius $R$. Also, the solid curve represents the present DSMC simulations, and the full symbol represents the numerical data available in the literature~\cite{Pullin}. It is immediately evident from Fig.~\ref{BJoPP02F18}(a-c) that there is a close overall agreement between both DSMC simulations at the vicinity of the stagnation region.

%\begin{acknowledgements}
%\end{acknowledgements}



\end{document}